\documentclass[a4paper,twocolumn,11pt,accepted=2026-03-19]{quantumarticle}
\pdfoutput=1
\usepackage[utf8]{inputenc}
\usepackage[english]{babel}
\usepackage[T1]{fontenc}
\usepackage{amsmath}
\usepackage{hyperref}

\usepackage{tikz}
\usepackage{lipsum}

\usepackage{textcomp}
	\usepackage{amssymb}
	\usepackage{graphicx}
	\usepackage{mathtools}
	\usepackage{bm}
	\usepackage{color}
	\usepackage{xcolor}
	\usepackage{braket}
	\usepackage{comment}
	\usepackage{dsfont}
 \usepackage[numbers]{natbib}

\newcommand{\bg}{\begin{equation} \begin{gathered}}
\newcommand{\eg}{\end{gathered} \end{equation}}
\newcommand{\ba}{\begin{equation}}
\newcommand{\ea}{\end{equation}}

	\renewcommand{\i}{i}

	\newcommand{\Tr}{\mathop{\text{Tr}}\nolimits}

	\newcommand{\sign}{\text{sign}}

\begin{document}

\title{Quantum thermodynamics of the Caldeira-Leggett model with non-equilibrium Gaussian reservoirs}

\author{Vasco Cavina}
\email{vasco.cavina@sns.it}
%\affiliation{Complex Systems and Statistical Mechanics, Department of Physics and Materials Science, University of Luxembourg, 30 Avenue des Hauts-Fourneaux, L-4362 Esch-sur-Alzette, Luxembourg}
\affiliation{Scuola Normale Superiore, 56126 Pisa, Italy}

\author{Massimiliano Esposito}
\affiliation{Complex Systems and Statistical Mechanics, Department of Physics and Materials Science,
University of Luxembourg, 30 Avenue des Hauts-Fourneaux, L-4362 Esch-sur-Alzette, Luxembourg}

\maketitle

\begin{abstract}
We introduce a non-equilibrium version of the Caldeira-Leggett model in which a quantum particle is strongly coupled to a set of engineered reservoirs. The reservoirs are composed by collections of squeezed and displaced thermal modes, in contrast to the standard case in which the modes are assumed to be at equilibrium.
The model proves to be very versatile. 
Strongly displaced/squeezed reservoirs can be used to generate an effective time dependence in the system Hamiltonian and can be identified as sources of pure work. 
In the case of squeezing, the time dependence is stochastic and breaks the fluctuation-dissipation relation, this can be reconciled with the second law of thermodynamics by correctly accounting for the energy used to generate the initial non-equilibrium conditions.
To go beyond the average description and compute the full heat statistics, we treat squeezing and displacement as generalized Hamiltonians on a modified Keldysh contour. As an application of this technique, we show the quantum-classical correspondence
between the heat statistics in the non-equilibrium Caldeira-Leggett model and the statistics of a classical Langevin particle under the action of squeezed and displaced colored noises.
Finally, we discuss thermodynamic symmetries of the heat generating function, proving a fluctuation theorem for the energy balance and showing that the conservation of energy at the trajectory level emerges in the classical limit.
\end{abstract}

\section{Introduction}
The Caldeira-Leggett (CL) model \cite{caldeira1983path, leggett1987dynamics, grifoni1998driven, grabert1985quantum, fisher1985dissipative} describes a quantum particle linearly coupled to an infinite collection of harmonic oscillators initialized at thermal equilibrium and it is one of the most used microscopic models in open quantum systems theory. Its popularity arises from the fact that, despite the tractability of the degrees of freedom of the reservoirs (that can be analitically eliminated), the reduced dynamics of the system is rich enough to encompass decoherence and dissipation effects, while also reproducing the Langevin equation in the classical limit \cite{Kamenev,Altland, breuer2002theory}.  
From a thermodynamic point of view, the CL model can be used to describe the heat exchanged between a system and many reservoirs with different temperatures.
The full energy statistics of the CL model has been extensively studied \cite{Aurell2018, Aurell_2020, Funo2018, FunoQuanPRE2018, Aurell2017, CarregaNJP2014} as is the case for the Langevin equation, its classical counterpart \cite{imparato2007work, murashita2016overdamped, pucci2013entropy}.
Extending these results to non-equilibrium is interesting for several reasons. The contact with a non-equilibrium reservoir can break the time translation invariance of the system dynamics and generate an effective time dependence in the system Hamiltonian.
This idea is the basis of a self-consistent, autonomous approach to thermal machines \cite{niedenzu2019concepts, mari2015quantum, perarnau2015extractable, dann2022unification, elouard2023extending} and has been used in the context of quantum batteries \cite{andolina2018charger, mazzoncini2023optimal}, repeated interactions \cite{strasberg2017quantum} and quantum scattering setups \cite{jacob2021thermalization, jacob2022quantum, tabanera2023thermalization}.
Thus, a non-equilibrium version of the CL model is an important step to link the autonomous framework to a model that is both tractable and experimentally relevant and can represent a benchmark for future investigations.

Here we introduce the NECL (Non-Equilibrium Caldeira-Leggett), a generalization of the CL model in which the reservoirs modes are prepared in squeezed and displaced thermal states.
We discuss in detail how the contact with the Gaussian reservoirs affects the dynamics of the system: while displacement is a resource that induces effective time-dependent corrections to the system Hamiltonian (an already well-known effect in quantum optics \cite{mollow1975pure, fischer2018pulsed}), squeezing can generate stochastic forces that break the fluctuation-dissipation relation. 
Moving to thermodynamics, we use arguments based on the calculation of entropy production and entropy flows \cite{EspositoNJP2010,esposito2011second, reeb2014improved} to prove that the energy exchanged with the reservoirs in the NECL is in general a mixture of heat and work that reduces to the latter when the reservoirs are strongly out of equilibrium and the system-reservoirs coupling is weak enough.
In this context, we prove that the non-equilibrium extensions of the second law for squeezed baths \cite{niedenzu2016operation, niedenzu2018quantum, manzano2016entropy, hsiang2021fluctuation} can be fully reconciled with the equilibrium second law if the energy used to squeeze the bath is taken into account in the total energetic balance. 

Our results are not limited to average thermodynamics: to analyze the energy statistics in the NECL we combine a two-point energy measurement (TPEM) scheme \cite{Espositoreview} with Keldysh contour and non-equilibrium Green's function approaches \cite{Fei2020,Fei,Cavina2023}. Notice that similar techniques have been already applied to study the energy statistics in squeezed bosonic reservoirs \cite{yadalam2022counting, yadalam2019statistics}.
The TPEM approach allows us to derive a path integral formulation of the statistic of the heat flows,
%{\color{red} to do so, we use a new technique , contour potential etc. \cite{andreev2000counting,saito2008symmetry,utsumi2006full,utsumi2009fluctuation}, for this sake further extensions of the Keldysh contour were proposed \cite{nazarov2011flows, utsumi2015full2, Fei}}
to study its classical limit \cite{jarzynski2015quantum}, and show that it coincides with the statistics of a system coupled with squeezed and displaced classical oscillators (i.e. a non-equilibrium version of the model considered by Zwanzig \cite{Zwanzig}).
The comparison is established by showing the matching between the classical limit of the quantum path integral and the stochastic Martin-Siggia-Rose-Janssen-De~Dominicis (MSRJD) path integral \cite{martin1973statistical, DeDominicis1976, bausch1976renormalized}.
Later, we will discuss the symmetries of the energy generating functions, starting from a proof of the fluctuation theorem (FT) \cite{andrieux2009fluctuation, Gaspard, Esposito2010, campisi2011, soret2022thermodynamic, manzano2018quantum} and highlight some structural differences between the energy statistics in the quantum and classical case. For instance, the energy conservation at the trajectory level holds only in the classical scenario.
The paper is organized as follows: In Sec.~\ref{sec:dynsett} we introduce the NECL. In Sec. \ref{sec:thermo} we discuss the thermodynamics at the average level, while in \ref{sec:fullquant}
we discuss the full energy statistics.
In \ref{sec:classic} we study the thermodynamics of the classical counterpart of the NECL and the quantum-classical correspondence at the level of the heat statistics.
We conclude by analyzing the symmetries of the model, including the FT, in Sec.~\ref{sec:propsym}.

%To cite:
% \cite{Alicki1979, vinjanampathy2016quantum, Cavina2017,allahverdyan2001breakdown, allahverdyan2000extraction, aurell2015neumann, ptaszynski2019thermodynamics}  

\section{The model}
\label{sec:dynsett}
In the CL model we consider a system coupled with $N$ thermal reservoirs labeled by $\nu = 1,...N$ and initialized in Gibbs states with Hamiltonians $\hat{H}_{\nu}$:
\begin{equation} \label{eq:initial}
    \hat{\rho}(0) = \hat{\rho}_S(0)   \bigotimes_{\nu =1}^N \frac{e^{-\beta_{\nu} \hat{H}_{\nu}}}{Z_{\nu}},
\end{equation}
where $\beta_{\nu}, Z_{\nu}$ are the inverse temperature and the partition function of the reservoir $\nu$, respectively.
Each reservoir is a collection of quantum harmonic oscillators 
\begin{equation} \label{eq:reservoiram} \hat{H}_{\nu}= \sum_{k} \hat{H}_{\nu,k} = \sum_{k} \big[ \frac{\hat{P}_{\nu,k}^2}{2 m_{\nu,k}} + \frac{m_{\nu,k} \omega_{\nu,k}^2 \hat{X}_{\nu,k}^2}{2} \big],\end{equation}
where $m_{\nu,k}, \omega_{\nu,k}$ are the mass and frequency of the mode $k$ of the reservoir $\nu$ and the reservoir operators satisfy the canonical commutation relations $[\hat{X}_{\nu,k}, \hat{P}_{\nu',k'}] = i \hbar \delta_{\nu, \nu'} \delta_{k,k'}$.
For ease of notation, we leave the limits of the sums over 
$k$ in each reservoir unspecified. We emphasize, however, that we do not restrict ourselves to reservoirs with infinitely many modes, as often assumed for the Caldeira–Leggett model. Work reservoirs may for instance consist of a single mode, provided that such modes are strongly out of equilibrium, as we will discuss in Sec. III.
We assume a total Hamiltonian of the form
\begin{align} \label{eq:ham} \hat{H}(t)=\hat{H}_S+\sum_{\nu=1}^N \hat{H}_\nu + \sum_{\nu=1}^N \chi_{\nu}(t) \hat{V}_\nu,
\end{align}
where the coupling has to be linear in the reservoir variables $\hat{V}_{\nu} = \sum_{\nu,k} V_{\nu,k}(\hat{X}) \otimes \hat{X}_{\nu,k} $ where $\hat{X}$ is the position operator in the system $S$ and $\hat{H}_S$ is its free Hamiltonian.
The scalar functions $\chi_{\nu}(t)$ are the so-called {\it switching functions} and are used to couple/decouple the system to the reservoirs. 
\begin{figure} [!t]
    \centering
    \includegraphics[width=\columnwidth]{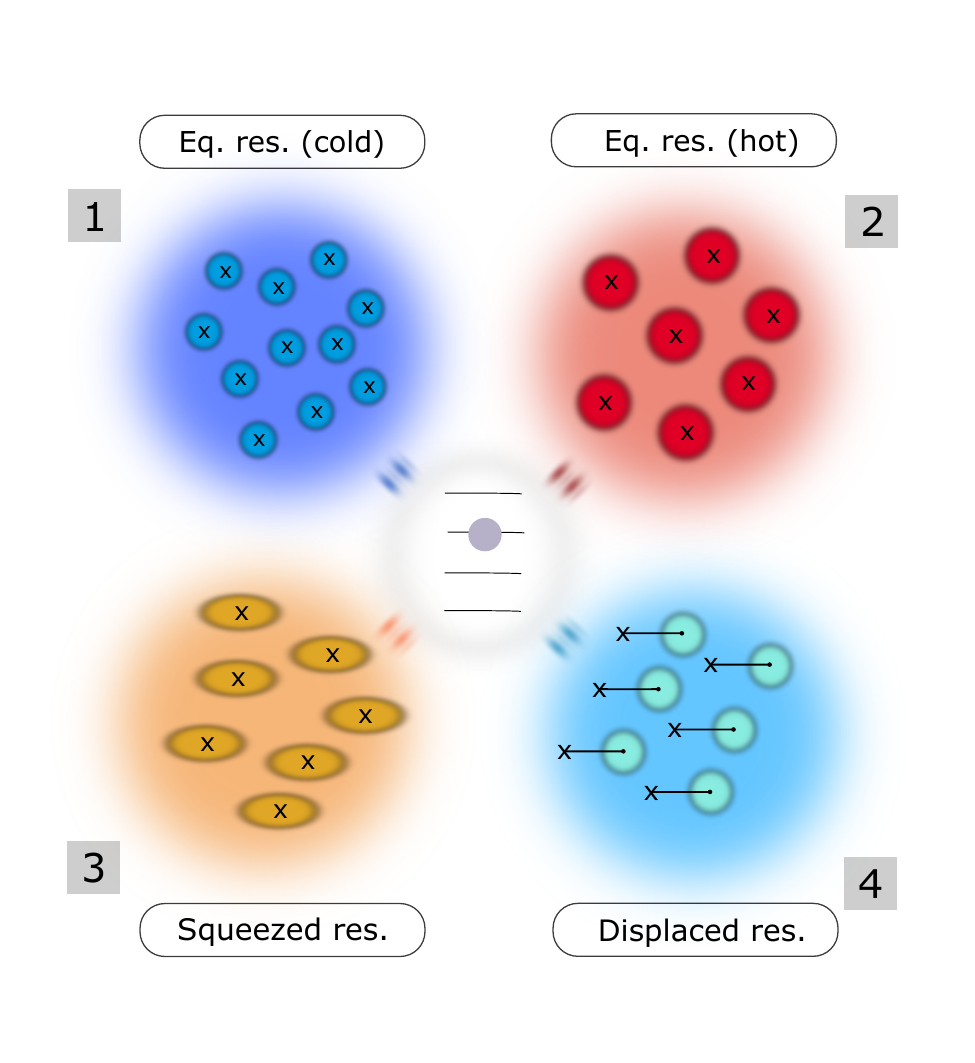}
    \caption{Pictorial representation of the example considered at the end of Section \ref{sec:dynsett}, where the system interacts with 4 reservoirs.  
    Reservoirs $\nu=1,2$ are thermal reservoirs with $\beta_1 \geq \beta_2$ \eqref{eq:therm12}.
    The modes of reservoir $\nu=3$ are squeezed \eqref{eq:squeez3}, while the modes of reservoir $\nu=4$ are displaced \eqref{eq:disp4}. 
    In this and in the following graphical representations we depict thermal, squeezed, and displaced modes in a way that is reminiscent of their phase space distributions. The squeezed modes are represented as ellipses (symbolizing the different variances in the $x$ and $p$ directions).}
    \label{fig:1}
\end{figure}
To go from the standard CL to the NECL, the reservoirs are initially moved out of equilibrium by applying a {\it squeeze operator} and a {\it displacement operator}, respectively given by
\begin{gather} \label{eq:squeeze}
    \hat{S}(r_{\nu,k}) = \exp \frac{1}{2} \big( r^*_{\nu,k} \hat{a}^2_{\nu,k} - r_{\nu,k} \hat{a}^{\dagger\, 2}_{\nu,k}\big) ,\\
    \hat{D}(\alpha_{\nu,k}) = \exp\big(\alpha_{\nu,k} \hat{a}^{\dagger}_{\nu,k} - \alpha^*_{\nu,k} \hat{a}_{\nu,k}\big), \label{eq:disp} 
\end{gather}
where $\hat{a}_{\nu,k} = \sqrt{\frac{m_{\nu,k}\omega_{\nu,k}}{2 \hbar}} ( \hat{X}_{\nu,k} + i \frac{\hat{P}_{\nu,k}}{m_{\nu,k} \omega_{\nu,k} }) $ and $\alpha_{\nu,k},r_{\nu,k}$ are arbitrary complex numbers. 
After applying both squeezing and displacement operators to all the environmental modes, the global initial state can be written as
\begin{align} \label{eq:gauss}
&\hat{\rho}(0) = \hat{\rho}_S(0) \bigotimes_{\nu,k} \hat{\rho}_{\nu,k}(0) \\
&\hat{\rho}_{\nu,k}(0) = \hat{D}^{\dag}(\alpha_{\nu,k}) \hat{S}^{\dag}(r_{\nu,k})  \frac{e^{-\beta_{\nu} \hat{H}_{\nu,k}}}{Z_{\nu,k}} \hat{S}(r_{\nu,k}) \hat{D}(\alpha_{\nu,k}) \nonumber
\end{align}
where $\hat{\rho}_{\nu,k}(0)$ represents the new initial state of the mode $k$ in the reservoir $\nu$. 
Although implicit in the notation, the displacement and squeezing operators with arguments $\alpha_{\nu,k}, r_{\nu,k}$ act on the Hilbert space of the reservoir mode identified by $\nu,k$. 
Due to the arbitrariness of $\alpha_{\nu,k}, \beta_{\nu}$ and $r_{\nu,k}$, the equation above represents the most general initial state of a system coupled to independent Gaussian modes up to phase shifts \cite{serafini2017quantum}.
In Appendix \ref{app:squeez} we summarize several useful formulas using the displacement and squeezing operators $\hat{D}$ and $\hat{S}$.

In what follows, as schematized in Fig. \ref{fig:1}, we will consider the reservoirs with $\nu=1,2$ to be standard thermal reservoirs ($\alpha_{1,k}, \alpha_{2,k} = 0$,
$r_{1,k}, r_{2,k} = 0$) with inverse temperatures $\beta_1 \geq \beta_2$:
\begin{align} \label{eq:therm12}
\hat{\rho}_{1,k}    = \frac{e^{-\beta_1 \hat{H}_{1,k}}}{Z_{1,k}}, \quad \quad 
\hat{\rho}_{2,k}    = \frac{e^{-\beta_2 \hat{H}_{2,k}}}{Z_{2,k}}.
\end{align}
Instead, the reservoir with $\nu =3 $ is squeezed, while the one with $\nu =4 $ is displaced ($\alpha_{3,k} = 0$ and
$r_{4,k} = 0$):
\begin{align} \label{eq:squeez3}
\hat{\rho}_{3,k}   & = \hat{S}^{\dag}(r_{3,k})\frac{e^{-\beta_3 \hat{H}_{3,k}}}{Z_{3,k}} \hat{S}(r_{3,k}), \\ \label{eq:disp4}
\hat{\rho}_{4,k}   & =   \hat{D}^{\dag}(\alpha_{4,k})  \frac{e^{-\beta_4 \hat{H}_{4,k}}}{Z_{4,k}}  \hat{D}(\alpha_{4,k}) .
\end{align}

\section{Average thermodynamics}
\label{sec:thermo}
We assume that the switching functions in Eq. \eqref{eq:ham} are set to zero outside the time interval $[0,\tau]$. 
The total unitary evolution reads
\begin{equation} \label{eq:unipro}
    \hat{U}(\tau,0) = \mathcal{T}\{e^{- \frac{i}{\hbar} \int_0^{\tau} \hat{H}(t) dt} \},
\end{equation}
where $\mathcal{T}$ is the time-ordering operator and $\hat{H}(t)$ is given in Eq. \eqref{eq:ham}. 
As a result, the final state is
\begin{equation}
    \hat{\rho}(\tau) = \hat{U}(\tau,0) \hat{\rho}(0) \hat{U}^{\dag}(\tau,0)
\end{equation}
where $\hat{\rho}(0)$ is given by \eqref{eq:gauss}.
In this section, we will discuss the general properties of the energy and entropy flows between the system and its reservoirs during the unitary process \eqref{eq:unipro}. 
\subsection{First and second laws of thermodynamics} \label{sec:avgtherm}

We build on the conventional formulation of the second law for open quantum systems
\cite{EspositoNJP2010, esposito2011second, reeb2014improved,strasberg2017quantum}.
The main difference is that a quench, consisting in the squeeze and displacement operations in Eq.~\eqref{eq:gauss}, is initially performed on the reservoirs. 

As a preliminary, we define the average energy, entropy, and free energy in the system $X=S$ (resp. reservoir $X=\nu$) as:  
\begin{align}
&\langle E_X \rangle(\tau) = Tr[\hat{H}_X \hat{\rho}_X(\tau)], \nonumber\\
&S_X(\tau) =- Tr[\hat{\rho}_X(\tau) \ln \hat{\rho}_X(\tau) \big] ] ,\label{eq:defenergy} \\
&F_X(\tau)=\langle E_X \rangle(\tau)-\beta^{-1}_{X} S_X(\tau)
,\nonumber
\end{align}
where $\hat{\rho}_{X}$ denotes the reduced density matrix. The changes in these quantities between time $0$ (right after the quench) and time $\tau$  are denoted respectively $\langle \Delta E_X \rangle$, $\Delta S_X $, and $\Delta F_X$. Here, $\hat{\rho}_{\nu}(0)$ is the initial state of the reservoir $\nu$ taken from Eq. \eqref{eq:gauss}.
The initial quench is unitary so it does not produce any change in the entropy of the reservoirs, $S_{\nu}(0)=S_{\nu}^{\rm{eq}}$, while their average energy changes by an amount
\begin{equation} \label{eq:quenchwork}
\langle W_{\nu} \rangle = Tr[\hat{H}_{\nu} \big[\hat{\rho}_{\nu}(0) -  \frac{e^{-\beta_{\nu} \hat{H}_{\nu}}}{Z_{\nu}}\big] ],
\end{equation}
which can be thus seen as a source of work.
The variation in the average energy of reservoir $\nu$ that occurs subsequently up to $\tau$ is
\begin{align}
\langle \Delta E_{\nu} \rangle = Tr[\hat{H}_{\nu} \big[\hat{\rho}_{\nu}(\tau) - \hat{\rho}_{\nu}(0) \big] ] \;,\label{eq:defenergyRes}
\end{align}
and is solely due to the interaction with the system.
Therefore, the overall variation in the average energy in the reservoir $\nu$ is
\begin{align} \label{eq:defheat}
\hspace{-0.2cm} \langle Q_{\nu} \rangle =  \langle W_{\nu} \rangle + \langle \Delta E_{\nu} \rangle 
= Tr[\hat{H}_{\nu} \big[\hat{\rho}_{\nu}(\tau) - \frac{e^{-\beta_{\nu} \hat{H}_{\nu}}}{Z_{\nu}}\big]]. 
\end{align}
We can derive a set of entropic inequalities for the reservoirs:
\begin{align} \label{eq:correntropy}
D_{\nu} &\equiv D\big(\hat{\rho}_{\nu}(\tau)|| \frac{e^{-\beta_{\nu} \hat{H}_{\nu}}}{Z_{\nu}} \big) = \beta_{\nu} \langle Q_{\nu} \rangle - \Delta S_{\nu} \\
&= \beta_{\nu} (F_{\nu}(\tau) - F_{\nu}^{\rm{eq}}) = \beta_{\nu} (\langle W_{\nu} \rangle + \Delta F_{\nu}) \geq 0, \nonumber
\end{align}
where $F_{\nu}^{\rm{eq}}=-\beta_{\nu}^{-1} \ln Z_{\nu}$ and $D$ denotes the quantum relative entropy. From the last two equalities, we see that the overall change in free energy in a given reservoir consists of a gain initially supplied as work by the squeezing and displacement operations, followed by a change in free energy due to the subsequent interaction with the system. %$D_\nu \approx 0$ would mean that the free energy initially put in as work has been fully released into the system at the end of the process.

We now give a closer look on what happens in the system. 
The interaction between the system and each reservoir $\nu$ is only turned on after the squeezing and displacement operations and the energy supplied as work for switching on and off such interaction reads 
\begin{align} \label{SwitchWork} 
\langle W_{\nu}^{\chi} \rangle= \int_{0}^{\tau} dt \; \frac{d}{dt} [\chi_{\nu}(t)]  \; Tr[\hat{V}_\nu \hat{\rho}(t)].
\end{align}
The sum of the contribution in Eq. \eqref{SwitchWork} and the work supplied by the initial quench must be equal to the overall energy variation in the system and reservoirs. The following first law thus holds
\begin{align}
\langle \Delta E_S \rangle
&= \sum_{\nu} \langle W_{\nu}^{\chi} \rangle  + \sum_{\nu} \langle W_{\nu} \rangle - \sum_{\nu} \langle Q_{\nu}\rangle \nonumber \\
& = \sum_{\nu} \langle W_{\nu}^{\chi} \rangle - \sum_{\nu} \langle \Delta E_{\nu}\rangle .
\label{eq:1stlaw}
\end{align}
Furthermore, a second law inequality can be derived by introducing the {\it entropy production}:
\begin{align} \notag
\Sigma 
&\equiv D\big( \hat{\rho}(\tau) || \hat{\rho}_S(\tau) \bigotimes_{\nu =1}^N \frac{e^{-\beta_{\nu} \hat{H}_{\nu}}}{Z_{\nu}} \big) 
\\ & = \Delta S_S + \sum_{\nu} \beta_{\nu} \langle Q_{\nu} \rangle  \nonumber \\
&= \beta_{1} [\sum_{\nu} \langle W_{\nu}^{\chi} \rangle + \sum_{\nu} \langle W_{\nu} \rangle - \Delta F_{S}^{1} ] \nonumber\\
&\hspace{0.3cm} - \sum_{\nu} (\beta_{1} - \beta_{\nu}) \langle Q_{\nu} \rangle \geq 0  ,\label{eq:modprod2}
\end{align}
where defining $\Delta F_S^1 = \langle \Delta E_S \rangle - \beta_{1}^{-1}\Delta S_S $ we singled out the contribution of the cold bath $\nu=1$ and used \eqref{eq:1stlaw} to obtain the last equality.
This reveals that besides the non-conservative forces resulting from thermal gradients, which vanish in isothermal situations, $\langle W_{\nu}^{\chi}\rangle $ and $\Delta F_{\nu}$ can now be used as resources to increase the free energy of the system.
We can further split $\Sigma$ as \cite{esposito2011second, reeb2014improved} 
\begin{equation} \label{eq:EPsplit}
\Sigma =  I + \sum_{\nu} D_{\nu} \;,
\end{equation}
where $I$ is the mutual information between the system and all the reservoirs established during the time evolution.
Using \eqref{eq:correntropy} and \eqref{eq:1stlaw},
we find that $I$ can be written as  
\begin{align} \notag
I &=
D\big(\hat{\rho}(\tau)||\hat{\rho}_S(\tau) \bigotimes_{\nu=1}^N \hat{\rho}_{\nu}(\tau)\big)=\Delta S_{S} + \sum_{\nu} \Delta S_{\nu} \\
&= \beta_{1} [\sum_{\nu} \langle W_{\nu}^{\chi} \rangle - \sum_{\nu} \Delta F_{\nu} - \Delta F_{S}^{1}] \nonumber \\
& \; \; \; - \sum_{\nu} (\beta_1 -\beta_{\nu}) \beta_{\nu}^{-1} \Delta S_{\nu} \geq 0. \label{eq:mutalinfo}
\end{align} 
Putting again non-isothermal effects aside, we observe that an increment of $I$ goes at the expense of the free energy of the system (at parity of resources $\langle 
 W_{\nu}^{\chi} \rangle, \,\Delta F_{\nu}$).
As we will see in Section \ref{sec:worksour} the quantities $I,\Sigma$ can be used to discriminate deterministic from stochastic work reservoirs and understand how their performances are related. 
\subsection{Work reservoirs in the NECL}  \label{sec:worksour}
If we want a reservoir $\nu$ to transfer a positive amount of energy to the system we need $ \langle \Delta E_{\nu} \rangle \leq 0$. 
In addition, to interpret this energy as work, it should not be associated with any outgoing entropy from $\nu$, i.e., we have to impose $\Delta S_{\nu} = 0$. 
Using these two restrictions in the bound \eqref{eq:correntropy} results in $ | \langle  \Delta E_{\nu}\rangle| \leq \langle W_{\nu} \rangle $, that is, the non-equilibrium nature of the reservoir is essential to pump a positive amount of energy in $S$. 
Based on these premises, we define a {\it deterministic work reservoir} as one whose influence does not affect the entropy of the system. In other words, if this reservoir were the only one coupled to $S$ we would have $\Delta S_S = 0$ and (from Eq.~\eqref{eq:mutalinfo}) $ I = 0$.

We naturally extend this concept by exploring a scenario where the mutual information $I = \Delta S_S$ does not vanish. 
This scenario, considering again the case in which the reservoir of interest $\nu$ is the sole coupled to the system, is compatible with the requirement $\Delta S_{\nu} = 0$ made at the beginning of this section.
We call this case {\it stochastic work reservoir}
since it represents a situation in which the reservoir does not change its entropy but its randomness leads to a stochastic transfer of energy to the system.
A comparison of the performances of deterministic and stochastic work reservoirs is offered by the second line of Eq. \eqref{eq:modprod2}. The stochastic reservoirs allow to transfer more of the initial energetic resources $\langle W_{\nu}\rangle$,  $\langle W^{\chi}_{\nu}\rangle$ to the system, at the expense of increasing its entropy (such increment of entropy would decrease $\Delta F_S^1$ and make room for a larger increment in $\langle \Delta E_{S} \rangle$).
In the following we will see that a strongly displaced reservoir behaves like a deterministic work reservoir, while a strongly squeezed one behaves like a stochastic work reservoir.

{\it Deterministic work from displaced reservoirs -} \label{sec:dsres}
For ease of calculation, let us assume that the system interacts with a single displaced mode. The results can be easily generalized to the case with many reservoir modes since the different modes are independent one from the other. 
In the language of Eqs. \eqref{eq:therm12}, \eqref{eq:squeez3} and \eqref{eq:disp4}, we now ignore the coupling with the reservoirs 1,2,3 and consider the reservoir 4 with $V_{4,k}(\hat{X}) = \delta_{1,k} V_{4,1}(\hat{X})$, that is, only the first mode of the reservoir $4$ is put in contact with the system.
In this case the total Hamiltonian \eqref{eq:ham} is given by
\begin{align} \notag
\hat{H}(t) & = \hat{H}_S + \frac{1}{2}m_{4,1} \omega_{4,1}^2 \hat{X}_{4,1}^2 \\ & + \frac{1}{2m_{4,1}} \hat{P}_{4,1}^2  + \chi_4 \hat{V}_{4,1}(\hat{X}) \hat{X}_{4,1} \label{eq:hampart}
\end{align}
in which $\hat{X}$ is the system position operator and $\hat{X}_{4,1}, \hat{P}_{4,1}$ are the single mode position and momentum operators, according to the notation introduced in Sec.~\ref{sec:dynsett}.
In Eq. \eqref{eq:hampart} we also assumed $\chi_4(t) = \chi_4$ for simplicity.
We look for a regime in which the fluctuations are negligible and the effect of the reservoir is only to induce a time-dependent deterministic modification of the system Hamiltonian.
%A way to ensure this is to work in a regime in which the typical length scale of fluctuations is much smaller than the initial displacement position of the reservoir}  
%To ensure that noisy effects are irrelevant, we rely on the intuition that the only source of noise, the uncertainty in the position of the reservoir $4$, should be small in some sense.
%In this way, the interaction of the system with the reservoir mode is dominated by and interaction with the "center of mass" of the Gaussian state and fluctuatons are negligible.  
This can be achieved by assuming that the typical length scale of environmental fluctuations is much smaller than the initial displacement of the reservoir position
$L_{4,1} = - \sqrt{\frac{2 \hbar}{m_{4,1} \omega_{4,1}}} \mathrm{Re} \alpha_{4,1}$.
The typical length scale of fluctuations is measured by the standard deviation of the observable $\hat{X}_{4,1}$ in the initial state of reservoir $4$, so that we impose
\begin{equation} \label{eq:bigbig}
|L_{4,1}| \gg
\sqrt{\frac{\hbar \coth \big( \frac{\beta_4 \hbar \omega_{4,1}}{2}\big)} {2 m_{4,1} \omega_{4,1} }} .
\end{equation}
We immediately see that the condition above can be realized by choosing $|L_{4,1}| \rightarrow \infty$ and consequently renormalizing the coupling strength to ensure that the force acting on $S$ remains finite, that is, working in the limit $\chi_{4} \rightarrow 0$, while keeping the product $\chi_4 L_{4,1}$ constant. Under these assumptions, we obtain the following equation of motion for the reduced density matrix of the system:
\begin{equation}
\dot{\hat{\rho}}_S(t) 
= - \frac{i}{\hbar} [ \hat{H}_S + \chi_4 L_{4,1}  
 V_{4,1}(\hat{X})  \cos \omega_{4,1} t,  \hat{\rho}_S(t)] .\label{eq:motionnew}
\end{equation}
Since the reduced dynamics of the system expressed by Eq. \eqref{eq:motionnew} is unitary, in this regime $\Delta S_S = 0$. 
In Appendix \ref{app:balreservoir} we give more details on the derivation of Eqs. \eqref{eq:bigbig}, \eqref{eq:motionnew} and we also prove that the increase in entropy of the reservoir is negligible, thus concluding that the entropic quantities introduced in Sec. \ref{sec:avgtherm} satisfy $\Delta S_B = I \approx 0$ and coincide with the case of an ideal deterministic work reservoir. In the appendix the rate of energy flowing into the reservoir is also computed, and reads \begin{equation} \label{eq:phenpow}
\langle \dot{W} \rangle = \langle \dot{E}_{4,1} \rangle  
= Tr[ \dot{\hat{H}}_{eff}(t) \hat{\rho}_S(t)],
\end{equation}
where $\hat{H}_{eff}(t)$ is the time dependent Hamiltonian in Eq. \eqref{eq:motionnew}.
This coincides with the phenomenological equation typically used to describe the power in the context of driven heat engines \cite{Alicki1979, esposito2010quantum, Cavina2017, vinjanampathy2016quantum, cavina2018optimal,cavina2018variational}.

\begin{figure}
    \centering
\includegraphics[width=1\columnwidth]{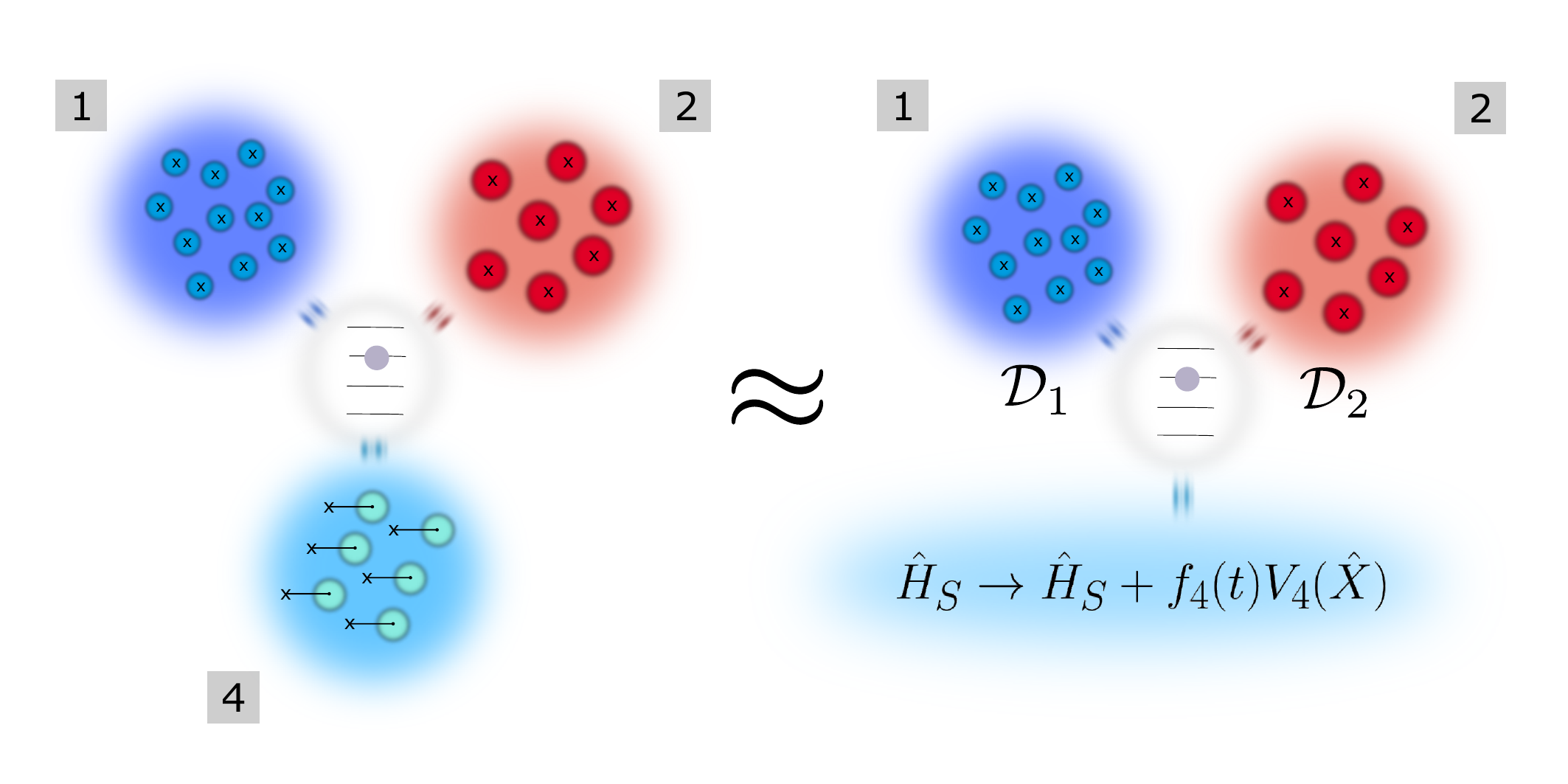}
    \caption{In the limit in which the coupling is small ($\chi_4 \approx 0$) and the reservoir is strongly displaced ($L_{4,k} \rightarrow \infty$) the reservoir behaves as a deterministic work source, causing an effective time dependent Hamiltonian that can be used to drive the system. A generic time dependent profile can be obtained using a suitable engineering of the reservoir (see Section \ref{sec:engineer}).}
    \label{fig:3}
\end{figure}

{\it Squeezed reservoirs as stochastic work sources -} \label{sec:sqres}
A similar approach can be used to determine the effect of the contact with a squeezed reservoir mode.
Starting again from Eqs. \eqref{eq:therm12}, \eqref{eq:squeez3} and \eqref{eq:disp4}, we assume that the only mode coupled with the system is the first mode of the reservoir with $\nu=3$, that is, $V_{3,k}(\hat{X}) = \delta_{k,1} V_{3,1}(\hat{X})$.
In line with our previous discussion regarding the displaced case, we seek a regime where the fluctuations in the system's dynamics caused by the initial squeezing are significantly greater than the thermal fluctuations, yet remain finite.
This scenario is realized again in the weak coupling limit, in which $\chi_3 \rightarrow 0$ and $r_{3,1} \rightarrow \infty$ while keeping the product $\chi_3 e^{r_{3,1}}$ finite (see Appendix \ref{app:squeeztherm}).
Under these assumptions, the system evolves under an equation of the form
\begin{equation} \label{eq:squeezmaster}
    \dot{\hat{\rho}}^{(c)}_S(t) = - \frac{i}{\hbar} [\hat{H}_S + \xi \hat{\Gamma}(t),\hat{\rho}_S^{(c)}(t)],
\end{equation}
where $\hat{\Gamma}(t) =  \chi_3(t)  V_{3,1}[\hat{X}(t)] \frac{e^{r_{3,1}}}{m_{3,1} \omega_{3,1}} \cos \omega_{3,1} t$ and $\xi$ is a Gaussian random variable such that $\langle \xi \rangle = 0$ and
$ \langle \xi^2 \rangle = \frac{\hbar}{2 m_{3,1} \omega_{3,1}} \coth \big(\frac{\beta_3 \hbar \omega_{3,1}}{2} \big) $.
Equation~\eqref{eq:squeezmaster} is a particular case of a stochastic von Neumann equation~\cite{tanimura2006stochastic, stockburger2002exact, cavina2025unifying}. Such equations hold under the fundamental assumptions that the reservoirs are Gaussian and linearly coupled to the system, and they generally give rise to rich non-Markovian features. In the specific case of Eq.~\eqref{eq:squeezmaster}, these features are absent due to the single-mode assumption and to the weak-coupling condition, which makes the back-action of the system on the mode negligible.
As is customary when the generator depends on a classical stochastic variable, the standard density matrix can be recovered by averaging over the possible realizations of the noise, such that $\langle \hat{\rho}_S^{(c)}(t) \rangle = \hat{\rho}_S(t) $, in analogy with stochastic Master Equations \cite{gardiner2004quantum, carmichael2013statistical,albarelli2024pedagogical}. %where  $\hat{\rho}_S^{(c)}(t)$.
Stochastic Hamiltonians of the form \eqref{eq:squeezmaster} have also been used to generate squeezing and improve the thermalization speed of the system \cite{dupays2021shortcuts, dupays2020superadiabatic}.
In App. \ref{app:squeeztherm} we show that also in this case the entropy variation in the reservoir, multiplied by its temperature, is negligible compared to the energy one, so that this squeezed mode behaves like a stochastic work source.
% as defined at the beginning of this section.
From a physical point of view, this can be understood by noting that equation \eqref{eq:squeezmaster} could also be obtained by assuming that an external agent randomly picks a coupling constant $\xi$ between a system and a driving Hamiltonian $\hat{\Gamma}(t)$, and then evolves the system. 
The stochastic reservoir thus implements a form of effective stochastic driving on the system \cite{verley2014work}.
\subsection{Multimode work reservoirs}
\label{sec:engineer}
By combining the results of Section \ref{sec:dsres} we can obtain a broad class of effective system Hamiltonians.
We can also extend the results by considering many different modes in the reservoir 4. This allows us to engineer the phase of the cosine in Eq.~\eqref{eq:motionnew} 
so that we can perform the following replacement 
\begin{align} \notag
 & L_{4,1} V_{4,1}(\hat{X}) \cos \omega_{4,1} t \rightarrow \\ 
& \rightarrow 
 \sum_k  A_{4,k} V_{4,k}(\hat{X}) \cos{(\omega_{4,k} t + \phi_{k})}, \notag
\end{align}
where $A_{4,k} = \sqrt{ 2 \hbar  m_{4,k}^{-1} \omega_{4,k}^{-1}} |\alpha_{4,k} |$ and $\phi_{k} = \arctan[Im(\alpha_{4,k})/ Re(\alpha_{4,k})]$. 
If we now assume that the functional dependence of $V_{4,k}$ by $\hat{X}$ is uniform over the modes, i.e. $V_{4,k}(\hat{X}) =  c_{4,k} V_{\nu}(\hat{X})$, 
we can  write the equation of motion for the density matrix of the system as
\begin{align}  \label{eq:motionnew3}
\dot{\hat{\rho}}_S(t) 
 = - \frac{i}{\hbar}  [ \hat{H}_S    + f_4(t) V_{4}(\hat{X}) ,   \hat{\rho}_S(t)],
\end{align}
where $f_4(t) = \chi_4 \sum_k c_{4,k} A_{4,k} \cos [\omega_{4,k} t + \phi_k] $.
%If the frequencies of the modes are uniformly spaced \footnote{A possible example of a collection of harmonic modes with equally spaced frequencies is represented by an optical cavity.}, the expression of $f_4(t)$ is the phase amplitude representation of a Fourier series. In this case, by wisely engineering the frequencies and phases of the reservoir modes  we can obtain any desired function in the time interval $[0,t]$. 
Eqs.~\eqref{eq:motionnew3} and \eqref{eq:phenpow} tell us that, in the regime considered, the reservoir is not only a work reservoir but it is also Markovian, i.e. the energy transferred to the system and its dynamics are described by equations that are local in time.
Similarly, we can depict the dynamics induced by equilibrium reservoirs using a time-local master equation, based on appropriate assumptions regarding weak coupling and timescale separation \cite{davies1974markovian, dumcke1979proper, breuer2002theory, d2022time, trushechkin2021unified, Nathan2020,di2023time}. 
Under such assumptions the dynamics of a particle coupled to the displaced reservoir with $\nu=4$ and two equilibrium reservoirs with $\nu=1,2$ can be written as
\begin{align}  \notag
\dot{\hat{\rho}}_S(t) 
 = & - \frac{i}{\hbar}  [ \hat{H}'(t)    + f_4(t) V_{4}(\hat{X}) ,   \hat{\rho}_S(t)] \\ & + \sum_{\nu=1,2} \mathcal{D}_{\nu}(t)[\hat{\rho}_S(t)]  \label{eq:dynmaster}\end{align}
where $\hat{H}'(t)$ is the original system Hamiltonian plus a Lamb shift contribution induced by the reservoirs \cite{breuer2002theory} and $\mathcal{D}_{\nu}$ is the dissipator due to the action of the reservoir $\nu$ on the system (see Fig.~\ref{fig:3}). According to the approximation scheme used, the dissipators can depend on the eigenvectors and eigenvalues of the driven Hamiltonian \cite{Nathan2020, dann2018time,yamaguchi2017markovian}.
Equations of the form \eqref{eq:dynmaster} have been extensively used in quantum thermodynamics as a paradigmatic theoretical model to describe heat engines. Their suitability for this purpose has been thoroughly investigated, also starting from consistent microscopic descriptions \cite{soret2022thermodynamic,dann2022unification,soret2024optical}.
In the setting of Eq. \eqref{eq:dynmaster}, the work performed by, or extracted from, the system can be of two distinct types: thermoelectric work, arising from the presence of a bias (i.e., a chemical potential difference) between the two leads, and work generated by an explicit time dependence of the system Hamiltonian, which is more relevant from our perspective. Our approach provides a microscopic and self-consistent justification of the latter contribution [see Eq. (25)] and is closely connected to experimental implementations in which quantum systems—such as qubits—are directly driven by out-of-equilibrium photons, as in realizations of the Dicke or Tavis–Cummings models \cite{canzio2025single}.

The displaced case corresponds to a regime in which the cavity undergoes lasing. In this regime, coherent photons are generated and can be interpreted as a form of extractable work, closely related to recent proposals and experimental realizations of quantum batteries based on light–matter systems \cite{campaioli2024colloquium, quach2022superabsorption}. More generally, the possibility not only to drive a system but also to harvest the extracted work by storing energy in a harmonic “piston mode” has been thoroughly discussed in the literature on autonomous quantum engines \cite{niedenzu2019concepts}.
In the forthcoming sections, when studying the full energy statistics in the NECL, we will abstain from doing specific approximations, even though we confine our analysis to the case of Gaussian reservoirs.
\section{Full energy statistics in the quantum regime}
\label{sec:fullquant}
In this section we study the fluctuating thermodynamics of the NECL by computing the moment generating function (MGF) of the energy flows in the reservoirs.
The MGF can be expressed as a functional of the sole variables of the system if we manage to trace away the degrees of freedom of the reservoirs. A possible way to approach this problem is to write a path integral expression of the MGF using Keldysh techniques, as we will detail in the sections below.
\label{sec:fullen}

\subsection{Fluctuating thermodynamics}
\label{sec:fluct}
To go beyond the average description of the preceding section and study fluctuations, we apply the {\it two point energy measurement (TPEM) } technique \cite{Espositoreview}, in which the system and reservoirs Hamiltonians are measured at the initial and final times and the difference of the outcomes is used to quantify the variation of internal energy and heat.
The Hamiltonians of the system and reservoirs can be measured simultaneously since $[\hat{H}_S, \hat{H}_{\nu}] = 0$ for all $\nu =1,...N$, this will produce a set of outcomes $\epsilon_S(\tau_0), \epsilon_S(\tau)$ for the initial and final energies of the system and
$\epsilon_{\nu}(\tau_0), \epsilon_{\nu}(\tau)$ for the initial and final energies of the reservoirs. The results of the measurement can be gathered in a joint probability distribution $P[\epsilon_S(\tau)-\epsilon_S(\tau_0), \boldsymbol{\epsilon_B}(\tau) - \boldsymbol{\epsilon_B}(\tau_0)]$, where $\boldsymbol{\epsilon_B} = (\epsilon_{1},...\epsilon_N)$,
and used to build the MGF for the total energy balance
\begin{equation}
M(\lambda_S, \boldsymbol{\lambda_B},\tau) = 
\sum_{\epsilon_S, \boldsymbol{\epsilon_B}}  P[\Delta_S, \boldsymbol{\Delta_B}] e^{\lambda_S \Delta_S + \boldsymbol{\lambda_B}\cdot \boldsymbol{\Delta_B}},
\end{equation}
where $\sum_{\epsilon_S, \boldsymbol{\epsilon_B}} $ is a shorthand notation for $\sum_{\epsilon_S(\tau_0), \epsilon_S(\tau)} \sum_{\boldsymbol{\epsilon_B}(\tau_0), \boldsymbol{\epsilon_B}(\tau)}$ and $\Delta_S =\epsilon_S(\tau)-\epsilon_S(\tau_0) $, $\boldsymbol{\Delta_B} =\boldsymbol{\epsilon_B}(\tau) - \boldsymbol{\epsilon_B}(\tau_0) $ (see also \cite{soret2022thermodynamic}).
If the initial Hamiltonian commutes with the initial state, the MGF assumes a rather simple form:
\begin{align} \notag
M(\lambda_S, \boldsymbol{\lambda_B},\tau) & = 
Tr\big[\hat{U}_{\lambda_S,\boldsymbol{\lambda_B}}(\tau,\tau_0) \hat{\rho}(\tau_0) \\ & \times \hat{U}^{\dagger}_{\lambda_S,\boldsymbol{\lambda_B}}(\tau,\tau_0)\big] \label{eq:genmom}
\end{align}
in which $\hat{U}(\tau, \tau_0)$ is the evolution operator of the process and we introduced the tilted evolution operator as
\begin{align}  \notag
\hat{U}_{\lambda_S,\boldsymbol{\lambda_B}} & =
e^{(\lambda_S \hat{H}_S + \boldsymbol{\lambda_B}\cdot\boldsymbol{\hat{H}_B})/2} \hat{U}(\tau,\tau_0) \\ & \times e^{-(\lambda_S \hat{H}_S+ \boldsymbol{\lambda_B}\cdot\boldsymbol{\hat{H}_B})/2} 
\label{eq:Utilt} \end{align}
where $\boldsymbol{\hat{H}_B} = (\hat{H}_1,..,\hat{H}_N)$ is a vector containing the Hamiltonians of the reservoirs.
We now apply the TPEM to the setting described in Secs.~\ref{sec:dynsett} and \ref{sec:avgtherm} in which a system evolves after being coupled to non-equilibrium reservoirs.
First, we need to define at which times $\tau_0, \tau$ the measurements are performed.
The non-stationarity of the squeezed and displaced reservoirs is crucial if we want to observe any non-equilibrium reservoir-induced effect on the dynamics of the system. 
\begin{figure*}
    \centering
    \includegraphics[width=1\textwidth]{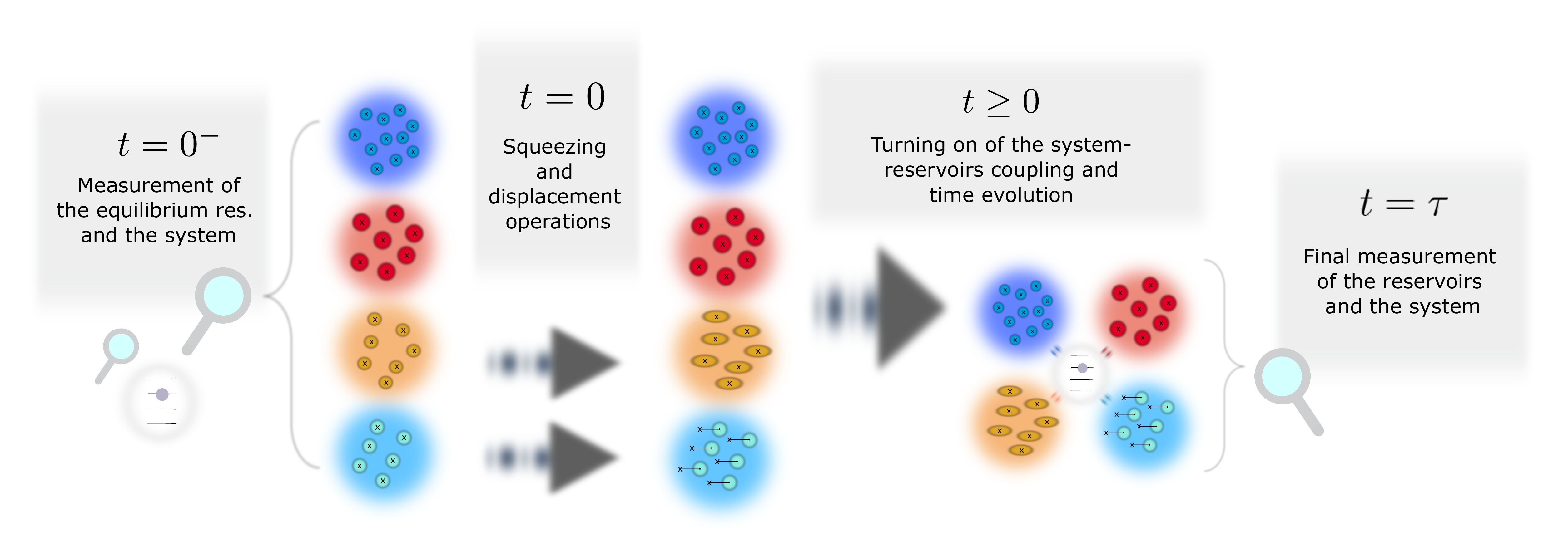}
\caption{Schematization of the TPEM for the four reservoirs example introduced at the end of sec \ref{sec:dynsett}. 
The reservoirs, prepared at equilibrium, are measured at time $t=0^-$ and collapse to eigenvectors of their respective Hamiltonians. The system energy is measured as well. After the collapse, at time $0$, some of the reservoirs, in this case the ones with $\nu=3,4$, are quenched out of equilibrium. At the same time, the coupling with the system is switched on. Subsequently, the time evolution of the system and reservoirs occurs up to time $\tau$, when the second energy measurement is performed.}
    \label{fig:2}
\end{figure*}
Performing the projective measurement of $\hat{H}_{\nu}$ after the squeeze and displacement operations would cause the reservoir $\nu$ to collapse into an eigenstate of $\hat{H}_{\nu}$. As a result, the reservoir would become stationary and the coherent effects induced by initial displacement and squeezing would be lost.
To bypass the problem of the destruction of initial coherences, alternative approaches to evaluate the energy statistics that do not rely on TPEMs have been proposed \cite{allahverdyan2005fluctuations,solinas2015full,talkner2016aspects, talkner2020colloquium, watanabe2014generalized, perarnau2017no}.
However, one can dodge the issue by assuming that the reservoirs are initially prepared at equilibrium and the
first measurement is performed {\it before} the squeezing and displacement operations.
In this way the TPEM scheme will fully capture the coherences produced by these initial operations.
From now on we assume that the first measurement happens at time $\tau_0 = 0^-$, when the modes of the reservoirs are still described by equilibrium states, while squeezing and displacement are performed immediately after, at time $0$.

This idea is schematized in Fig. \ref{fig:2}.
 Under these assumptions the energy flows $\boldsymbol{\Delta_B}$ represent the total energy variation in the reservoirs and
 constitute a fluctuating version of the average heat flows $\langle Q_{\nu} \rangle$ introduced in Section \ref{sec:avgtherm}.
 We identify them as {\it fluctuating heat}, that is, $\boldsymbol{\Delta_B} = \boldsymbol{Q_B}$ where $\boldsymbol{Q_B} = (Q_1,...,Q_N)$.
The energy variation of the system is the fluctuating version of the one discussed in Section \ref{sec:avgtherm}, that is, $ \Delta_S = \Delta E_S$.

The MGF associated to the heat and system energy statistics in the TPEM above is equal to \eqref{eq:genmom} with $\tau_0 = 0^-$ and 
\begin{align}  \label{eq:replaceU}
    \hat{U}(\tau,0^-) & \equiv   \hat{U}_{\boldsymbol{\alpha}, \boldsymbol{r}}(\tau,0^-) \\ & =   \hat{U}^{}(\tau,0) \prod_{\nu,k} \hat{D}^{\dagger}(\alpha_{\nu,k}) \hat{S}^{\dagger}(r_{\nu,k}) ,  \notag
\end{align}
 where we introduced $\boldsymbol{\alpha} = \{\alpha_{\nu,k}\}_{\nu,k}$ and $\boldsymbol{r}=\{r_{\nu,k}\}_{\nu,k}$ i.e. vectors containing all the displacement and squeezing parameters.
 The equation above takes into account the fact that if we perform the first measurement at time $0^-$, the evolution between the initial and final measurement is a composition of the initial quench with the evolution between times $0$ and $\tau$ generated by \eqref{eq:ham}.
Finally, we obtain 
\begin{align} \notag
& M(\lambda_S, \boldsymbol{\lambda_B},\tau) = \frac{1}{Z} Tr [ e^{\lambda_S \hat{H}_S + \boldsymbol{\lambda_B} \cdot \boldsymbol{\hat{H}_B}} \hat{U}_{\boldsymbol{\alpha}, \boldsymbol{r}}(\tau,0^-)   \\ &  \times e^{-\lambda_S \hat{H}_S -(\boldsymbol{\lambda_B} +\boldsymbol{\beta_B}) \cdot \boldsymbol{\hat{H}_B} } \hat{\rho}_S(0) 
\hat{U}^{\dagger}_{\boldsymbol{\alpha}, \boldsymbol{r}}(\tau,0^-) ],\label{eq:finalgen}
\end{align}
with $Z = \prod_{\nu} Z_{\nu}$.
Note that by computing the first derivatives, we can recover the average thermodynamic quantities introduced in Sec. \ref{sec:avgtherm} that is
$ \frac{\partial}{\partial{\lambda_{\nu}}} M = \langle Q_{\nu}\rangle $ and $ \frac{\partial}{\partial{\lambda_{S}}} M = \langle \Delta E_S\rangle $. The statistics of the fluctuating work can be obtained by setting $\lambda_{\nu} =\lambda_S = \lambda_{W}$ for all values of $\nu$, that is, by computing $M(\lambda_{W}, \lambda_W \boldsymbol{1},\tau)$ where $\boldsymbol{1} = (1,....1)$ (similarly to what happens for equilibrium reservoirs \cite{soret2022thermodynamic}).
\subsection{Contour integrals and heat statistics}
\label{sec:contourstat}
A convenient way to perform an exact elimination of the degrees of freedom of the reservoirs is to use a path integral formulation. 
We start by computing the path integral representation of the amplitude associated to the process
\eqref{eq:unipro}
\begin{equation} \label{eq:quantampl}
    \bra{x_i} \hat{U}(\tau,0)\ket{x_f}  = \int_{x(0)= x_i}^{x(\tau)=x_f} \mathcal{D} x  \mathcal{D} \boldsymbol{x_B}  e^{\frac{i}{\hbar} \mathcal{S}[\{x(t),\boldsymbol{x}_B(t)\}_0^{\tau}]}
\end{equation}
in which $\boldsymbol{x_B}$ is a vector of the coordinates of all the modes of the reservoirs and $\mathcal{D} x \mathcal{D} \boldsymbol{x_B} $ is the measure of the path integral in the configuration space \cite{kleinert2009path}. In the equation above we already integrated over the momentum variables to write the path integral in terms of the Lagrangian form of the action \cite{feynman2018statistical}
\begin{align} \notag
 \mathcal{S}  & =  \int_0^{\tau} \big\{ \frac{m}{2} \dot{x}(t)^2 -V[x(t)] \\  & - \sum_{\nu,k} \chi_{\nu}(t) x_{\nu,k}(t) V_{\nu,k}[x(t)]  \\ & +  \sum_{\nu,k} \big[\frac{m_{\nu,k}}{2} \dot{x}_{\nu,k}(t)^2 - \frac{m_{\nu,k} \omega_{\nu,k}^2}{2} x_{\nu,k}(t)^2\big]  \big\} dt. \notag
\end{align}
To grasp dissipative effects, we need to work in the density matrix formalism, and the amplitude \eqref{eq:quantampl} is not sufficient for such a task.
However, the path integral formalism can be extended to offer a comprehensive description of dissipative effects \cite{caldeira1983path, leggett1987dynamics, smith1987generalized, ingold1992charge}. 
Instrumental in this sense is the introduction of the {\it Keldysh contour} \cite{keldysh1965diagram, schwinger1961brownian,konstantinov1961diagram,baym1961conservation}.
The idea behind the Keldysh contour is that, 
in the same way a unitary evolution operator $\hat{U}(\tau,0)$ can be written as a time-ordered product by introducing a time-ordering on the real axis (between $0$ and $\tau$) the same can be done for two evolution operators acting jointly as in $\hat{U}(\tau,0) \hat{\rho}(0) \hat{U}^{\dag}(\tau,0)$, provided that we define a new suitable ordering domain.
The solution is to duplicate the real axis in two branches, one for $\hat{U}(\tau,0)$ and one for $\hat{U}^{\dag}(\tau,0)$, respectively called the {\it forward} ($\gamma_-$) and {\it backward} ($\gamma_+$) branch, the Keldysh contour is the union of the two \cite{Stefanucci,rammer2007quantum}.
After the time-ordered products are formulated in terms of the new contour, a path integral expression of the density matrix can be obtained with the usual textbook approach \cite{Kamenev}.

\begin{figure}
   \hspace{-1cm}
\includegraphics[width=\columnwidth]{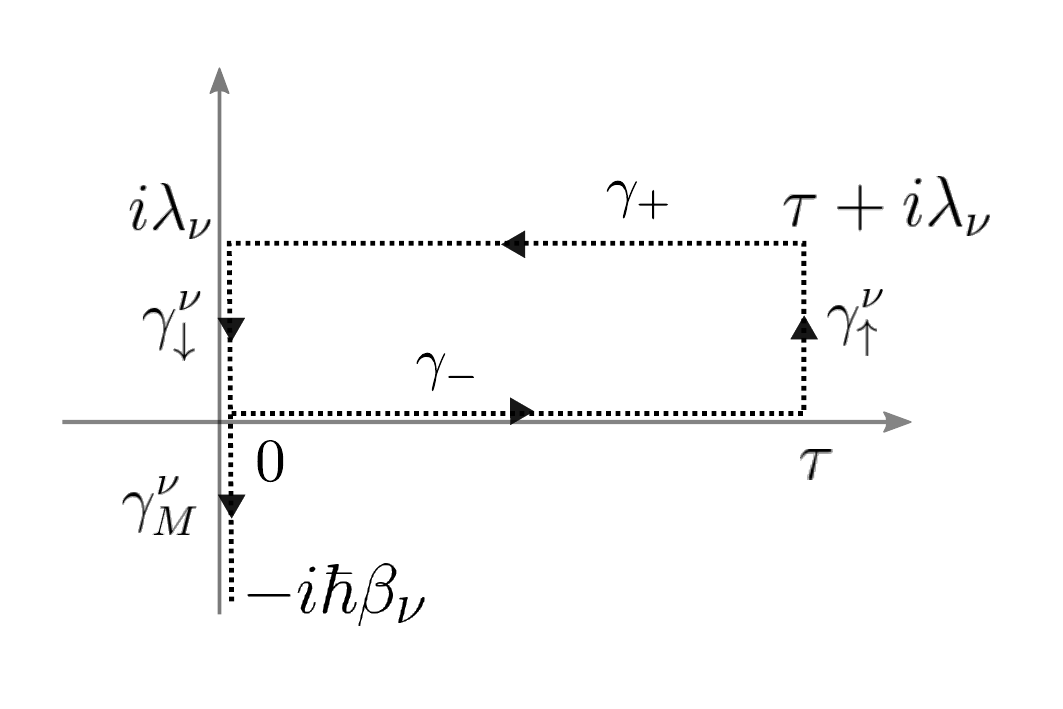}
    \caption{The contour of integration for the MGF in Eq. (\ref{eq:quantampl2}). The horizontal branches $\gamma_{-}, \gamma_{+}$ are already present in the standard Keldysh contour. In agreement with the ordering of the exponential operators in Eq. \eqref{eq:GFequi}, in the new contour there are new vertical tracks $\gamma_{\uparrow}^{\nu}$, $\gamma_{\downarrow}^{\nu}$ and $\gamma_M^{\nu}$ respectively associated to the exponentials of $\lambda_{\nu} \hat{H}_{\nu}$, $-\lambda_{\nu} \hat{H}_{\nu}$ and $-\beta_{\nu} \hat{H}_{\nu}$. The contour starts in $0$ (at the beginning of the forward branch) and ends in $- \i \hbar \beta_{\nu}$.
}
    \label{fig:4}
\end{figure}

We follow recent literature \cite{Funo2018,FunoQuanPRE2018, Cavina2023} to adapt the ideas mentioned above to the calculation of the MGF in a TPEM scheme. To make the procedure clear, we start by considering the MGF of the heat statistics \eqref{eq:finalgen} for a system coupled to a single equilibrium reservoir $\nu$:
\begin{align} \notag
    M^{eq}_{\nu}&  =\frac{1}{Z_{\nu}} Tr[e^{\lambda_{\nu} \hat{H}_{\nu}} \hat{U}(\tau,0) \hat{\rho}_S(0) \\  & \otimes e^{-(\beta_{\nu} + \lambda_{\nu})\hat{H}_{\nu}} \hat{U}^{\dag}(\tau,0)], \label{eq:GFequi}
\end{align}
in which we used the shortened notation $M_{\nu} \equiv M(0,0, ..., \lambda_{\nu}, ..., 0,\tau)$.
We can see Eq.~\eqref{eq:GFequi} as a single ordered exponential of the form
\begin{align} M^{eq}_{\nu} = \frac{1}{Z_{\nu}} \Tr\left[ \mathcal{T}_{\gamma^{\nu}} \big\{ e^{- \frac{i}{\hbar} \int_{\gamma^{\nu}} \hat{H}_{\gamma^{\nu}}(z) dz} \hat{\rho}_S(0)\big\}\right] ,  \label{genfin} \end{align}
where we introduced a new contour $\gamma^{\nu}$ (represented in Fig.~\ref{fig:4}) with its ordering operator $\mathcal{T}_{\gamma^{\nu}}$, and defined the contour Hamiltonian
\begin{align}  \hat{H}_{\gamma^{\nu}}(z) = \begin{cases}
   \hat{H}(t) & \text{for } z = t \in \gamma_{+}, \gamma_-, \\
  \hat{H}_{\nu} & \text{for } z\in \gamma^{\nu}_{\uparrow}, \gamma^{\nu}_{\downarrow}, \gamma^{\nu}_M,
  \end{cases}   \label{piecewise}    
\end{align}
where $\gamma_{\uparrow}^{\nu}, \gamma_{\downarrow}^{\nu} ,\gamma_{M}^{\nu}$ are the branches of the new contour respectively associated to the initial and final measurements and to the initial state.
A step-by-step derivation of the equations above is contained in \cite{Cavina2023}.
Using a Trotter decomposition of Eq.~\eqref{genfin}, the MGF for the energy flows in the reservoir $\nu$ can be written as a path integral over the modified contour $\gamma^{\nu}$:
\begin{align} \notag
   & M_{\nu}  = \frac{1}{Z_{\nu}} \int_{\mathcal{B}} \mathcal{D} x   \mathcal{D} \boldsymbol{x_B} \\ & \times \exp\big(\frac{i}{\hbar}  \mathcal{S}[\{x(z), \boldsymbol{x_B}(z)\}_{\gamma^{\nu}}] \big) 
  \bra{x_-} \hat{\rho}_S(0) \ket{x_+}  \label{eq:quantampl2}
 \end{align}
where $\mathcal{B}$ is the boundary condition imposing that the system field is equal to $x_-,x_+$ at the beginning of the forward and backward branch respectively, while the reservoir fields satisfy periodic boundary conditions on $\gamma^{\nu}$.
The action reads
 \begin{align} \notag
 \mathcal{S}  =  \int_{\gamma_{\pm}} & \big\{ \frac{m}{2} \dot{x}^2(z) -V[x(z)] \big\} dz + \sum_k  \mathcal{S}_{\nu,k} \\ \mathcal{S}_{\nu,k}  = \int_{\gamma^{\nu}} & \big\{  \frac{m_{\nu,k}}{2}\dot{x}_{\nu,k}(z)^2 - \frac{m_{\nu,k}  \omega_{\nu,k}^2}{2} x_{\nu,k}^2(z)  \notag \\
 & -\chi^{(\gamma)}_{\nu}(z) x_{\nu,k}(z) V_{\nu,k}[x(z)] \big\} dz, \label{eq:quantact}
 \end{align}
and we extended the switching function $\chi_{\nu}$ to a function with arguments on $\gamma^{\nu}$
    \begin{align}
\chi^{(\gamma)}_{\nu}(z) = \begin{cases}
   \chi_{\nu}(t) &  \text{for } z =t \in \gamma_{\pm} , \\
  0  &  \text{for } z \in \gamma^{\nu}_{M,\downarrow, \uparrow}.
  \end{cases}    \label{eq:piecewise3}
\end{align}
Note that $\chi^{(\gamma)}_{\nu}(z) = 0$ on $\gamma^{\nu}_{\uparrow,\downarrow}$ since when the measurements are performed the system and the reservoir are decoupled (see Section \ref{sec:fluct}).
In Eq.~\eqref{eq:quantampl2} we dropped the superscript $eq$ because, as we will show in the next section, the equations ~\eqref{eq:quantampl2} and \eqref{eq:quantact} hold also out of equilibrium, provided that we suitably redefine the quantities $m_{\nu,k}, \chi_{\nu}^{\gamma}$ and $V_{\nu,k}$.

%The path integrals appearing in Eqs. \eqref{eq:quantampl} and \eqref{eq:quantampl2} are formally the same, but the domains of definition of the fields are different, $[0,t]$ for Eq. \eqref{eq:quantampl} and $\gamma^{\nu}$ for \eqref{eq:quantampl2}. 
 \subsection{Out of equilibrium reservoirs in the contour}
 \label{sec:otresc}
The formalism with the modified contour can be easily extended to the case in which the initial state of the reservoir is sent out of equilibrium by performing additional squeezing/displacement operations.
For ease of notation, we start by considering a single reservoir $\nu$ composed of a single mode $k$ (the generalization to many reservoirs and many modes is straightforward, since the different modes are independent).
Following the protocol described in Section \ref{sec:fluct},
performing an initial displacement corresponds to using
\begin{equation} \label{eq:repcont}
   \hat{U}_{\boldsymbol{\alpha}, 0}(\tau,0^-) = \hat{U}(\tau,0) \hat{D}^{\dagger}(\alpha_{\nu,k}),
\end{equation}
inside equation  \eqref{eq:GFequi}.
We obtain
\begin{align} \nonumber
    M_{\nu} & =\frac{1}{Z_{\nu}} Tr[e^{\lambda_{\nu}  \hat{H}_{\nu}  }\hat{U}(\tau,0) \hat{\rho}_S(0) \\ 
    & \otimes e^{-(\beta_{\nu} + \lambda_{\nu}) \hat{D}^{\dag}(\alpha_{\nu,k})\hat{H}_{\nu} \hat{D}(\alpha_{\nu,k})}   \hat{U}^{\dag}(\tau,0)].
    \label{eq:GFequi2} 
\end{align}
The effect of a displacement with real parameter reduces to a translation of the initial position of the harmonic mode $\hat{X}_{\nu,k} \rightarrow \hat{X}_{\nu,k} + L_{\nu,k}$ where $L_{\nu,k} = - \sqrt{\frac{2 \hbar}{m_{\nu,k} \omega_{\nu,k}}} \alpha_{\nu,k} $.
We have
\begin{equation} \notag
\hat{D}^{\dagger}(\alpha_{\nu,k}) \hat{H}_{\nu} \hat{D}(\alpha_{\nu,k}) = \hat{H}_{\nu} - m_{\nu, k} \omega^2_{\nu,k} L_{\nu,k} \hat{X}_{\nu,k} + N_{\nu,k}
,
\end{equation}
with $N_{\nu,k} = \frac{ m_{\nu,k} \omega_{\nu,k}^2 L_{\nu}^2}{2}$.
Up to the constant $N_{\nu,k}$, this shift in the Hamiltonian corresponds to an equal shift in the action on the branches $\gamma^{\nu}_{\downarrow}$, $\gamma^{\nu}_M$ and it can be absorbed in a redefinition of the switching function 
    \begin{align}
\chi^{(\gamma)}_{\nu}(z) = \begin{cases}
   \chi_{\nu}(t) &  \text{for } z =t \in \gamma_{\pm} , \\
  1  &  \text{for } z \in \gamma^{\nu}_{M,\downarrow, \uparrow},
  \end{cases}    \label{eq:piecewisquee}
\end{align}
and a modification of the potential
    \begin{align}
V^{(\gamma)}_{\nu,k}(z) = \begin{cases}
  V_{\nu,k}(t) &  \text{for } z =t \in \gamma_{\pm} , \\
 - m_{\nu,k} \omega^2_{\nu,k} L_{\nu,k} &  \text{for } z \in \gamma^{\nu}_{M}, \gamma^{\nu}_{\downarrow}, \\ 
 0  & \text{for } z \in \gamma^{\nu}_{\uparrow}.
  \end{cases}    \label{eq:piecewisquee2}
\end{align}
A single squeezed reservoir mode can be treated in the same way, after substituting the operator $\hat{D}(\alpha_{\nu,k})$ with the squeeze operators $\hat{S}(r_{\nu,k})$. Since for a real squeeze parameter $r_{\nu,k}$ we have
   \begin{align}\label{eq:appsqueezham}
     \hat{S}^{\dag}(r_{\nu,k}) \hat{H}_{\nu,k} \hat{S}(r_{\nu,k}) & = \frac{ e^{2 r_{\nu,k}}}{2 m_{\nu,k}} \hat{P}^2_{\nu,k} \\ & +  \frac{e^{-2r_{\nu,k}} m_{\nu,k} \omega^2_{\nu,k}}{2} \hat{X}^2_{\nu,k} ,  \notag 
\end{align} 
the squeeze operator acts as a quench of the mass $m_{\nu,k}$ operated at time equal to $0$. The masses appearing inside $\hat{U},\hat{U}^{\dag}$ remain untouched. This translates, in the language of the modified contour, into a dependence of the form
  \begin{align}
m_{\nu,k}(z) = \begin{cases}
   m_{\nu,k} &  \text{for } z =t \in \gamma_{\pm}, \gamma^{\nu}_{\uparrow} , \\
  e^{-2 r_{\nu,k}} m_{\nu,k}  &  \text{for } z \in \gamma^{\nu}_{M}, \gamma^{\nu}_{\downarrow}.
  \end{cases}    \label{eq:piecewisquee3}
\end{align}
Note that, if squeezing and displacement are both present, the redefinition \eqref{eq:piecewisquee3} affects also the mass appearing in Eq.~\eqref{eq:piecewisquee2} and in the constant $N_{\nu,k}$.
Equations \eqref{eq:piecewisquee}, \eqref{eq:piecewisquee2} and \eqref{eq:piecewisquee3} allow us to encode the non-equilibrium properties of the initial state of the reservoir in suitable redefinitions of the physical quantities, without altering the quadratic structure of the action \eqref{eq:quantact}, hence, it will be possible to perform an exact elimination of the reservoirs degrees of freedom through standard Gaussian integration methods.
 \subsection{Effective system action and heat statistics}
 \label{sec:disp}
 \label{sec:greenreservoir}
\begin{figure}
    \centering
    \includegraphics[width=\columnwidth]{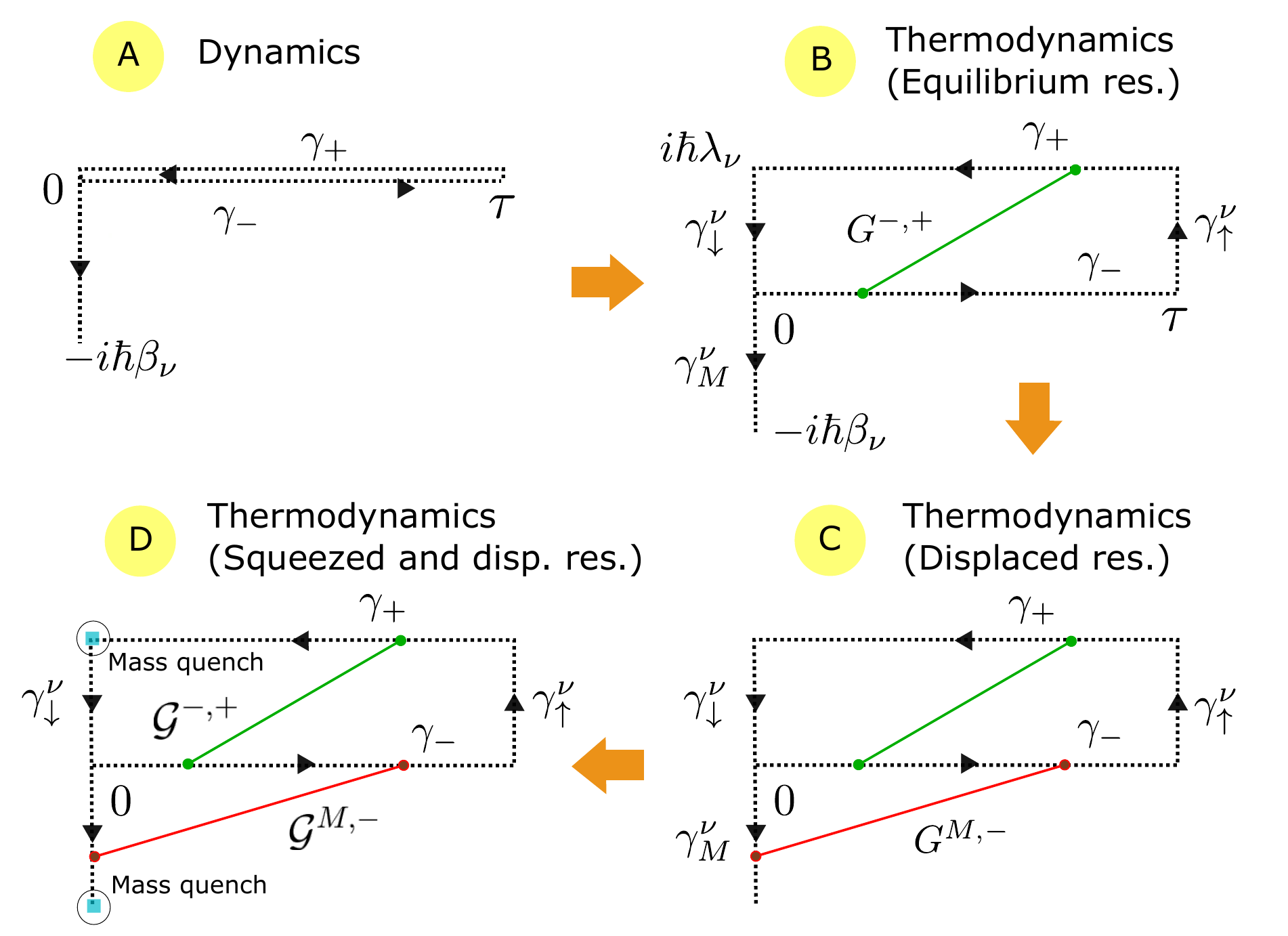}
    \caption{Schematic representation of the application of the modified Keldysh contour to the calculation of the energy statistics for the NECL.
    Panel (A): the standard Keldysh contour with the Matsubara branch for the initial state \cite{konstantinov1961diagram,Stefanucci}.
    Panel (B): the modified contour $\gamma^{\nu}$ for computing the energy statistics \cite{Funo2018,FunoQuanPRE2018,Cavina2023} The green line that connects $\gamma_+$ and $\gamma_-$ indicates that the components of the GF \eqref{eq:defcompmain} that contribute are only those with $i,j=\pm$.
    Panel (C): In the displaced case, other components of the GF appear (e.g. $G^{M,-}$ as represented by the red line).
    Panel (D): In the squeezed case, an additional quench of the reservoir masses has to be performed in the instants indicated by the blue dots (see Eq. \eqref{eq:piecewisquee3}).
    }
    \label{fig:5}
\end{figure}

For equilibrium reservoirs the action is given by Eq. \eqref{eq:quantact} which after an integration by parts reduces to the sum over $\nu,k$ of the differential operators $x_{\nu,k}(z)(\partial_z^2 + \omega_{\nu,k}^2 ) x_{\nu,k}(z) $, which can be eliminated by a Gaussian integration of the reservoirs degrees of freedom in \eqref{eq:quantampl2}.
Using the Gaussian integral formula, the result is an exponential of the inverse of the differential operator \cite{Kamenev}, i.e. the function $G_{\nu,k}$ satisfying
\begin{equation}
(\partial_z^2 + \omega_{\nu,k}^2) G_{\nu,k}(z,z') = \omega_{\nu,k} \delta(z-z'),
\end{equation}
with $z,z' \in \gamma^{\nu}$ and periodic boundary conditions
$ G_{\nu,k}(0,z')= G_{\nu,k}(-i \hbar \beta, z') $, $G_{\nu,k}'(0,z') =  G'_{\nu,k}(-i \hbar \beta, z')  $. 
This solution is the contour {\it Green's function }(GF) \cite{Kamenev, Stefanucci,Cavina2023}
\begin{align} \notag
    G_{\nu,k}(z,z') = & \frac{i}{2}\coth\big(\frac{\hbar \omega_{\nu,k} \beta_{\nu}}{2}\big) \cos\omega_{\nu,k}(z-z') \\+ & \big[\Theta(z-z') -\frac{1}{2} \big] \sin \omega_{\nu,k}(z-z'), \label{eq:GFmain}
\end{align}
where $\Theta(z-z')$ is the Heaviside function on the contour $\gamma^{\nu}$, valued $1$ if $z'$ precedes $z$ in $\gamma^{\nu}$ in Fig. \ref{fig:4} and $0$ otherwise.
In the case of a squeezed reservoir, the mass of the  modes in the contour is not constant (see Eq. \eqref{eq:piecewisquee3}) so that the equation of the GF has to be replaced with the one of an oscillator with a time-dependent mass:
 \begin{equation} \label{eq:squeeqg}
   [\partial_z \frac{m_{\nu,k}(z)}{m_{\nu,k}} \partial_z  + \omega_{\nu,k}^2 \frac{m_{\nu,k}(z)}{m_{\nu,k}} ] \mathcal{G}_{\nu,k}(z,z') = 
  \omega_{\nu,k} \delta(z-z'), \end{equation}
  where $m_{\nu,k}(z)$ satisfies Eq. \eqref{eq:piecewisquee3}. 
Summarizing, if there is no initial squeezing, the reduced action of the system after eliminating the degrees of freedom of all the reservoirs is (see also App.~\ref{app:disp})
  \begin{align} \label{eq:actquant}
 & \bar{\mathcal{S} }  = \int_{\gamma_{\pm}} \big\{\frac{m}{2}\dot{x}^2(z) - V[x(z)] \big\} dz + \sum_{\nu,k}  \mathcal{S}_{\nu,k} \\   \notag &  \bar{\mathcal{S}}_{\nu,k} =  \iint_{\gamma_{\nu}} \frac{dz  dz'}{2 m_{\nu,k} \omega_{\nu,k}} G_{\nu,k}(z,z') \times \\ & \quad \quad \times  \chi^{(\gamma)}_{\nu}(z) \chi^{(\gamma)}_{\nu}(z') V^{(\gamma)}_{\nu,k}[z] V^{(\gamma)}_{\nu,k}[z'] \label{eq:dissipact} \\   & \quad \quad + i \hbar(\beta_{\nu} + \lambda_{\nu})N_{\nu,k},  \notag
\end{align}
where $\chi_{\nu}^{\gamma}(z)$ is given by Eq.~\eqref{eq:piecewise3} for equilibrium reservoirs and by Eq.~\eqref{eq:piecewisquee} for displaced ones and $N_{\nu,k}$ is defined above Eq.~\eqref{eq:piecewisquee}.
In the presence of squeezing we replace the GF with the result of Eq.~\eqref{eq:squeeqg} obtaining
  \begin{align}     \notag &  \bar{\mathcal{S}}_{\nu,k} =  \iint_{\gamma_{\nu}} \frac{dz  dz'}{2 m_{\nu,k} \omega_{\nu,k}}  
\mathcal{G}_{\nu,k}(z,z')  \\ \label{eq:dissipactsquee} & \quad \quad \times  \chi^{(\gamma)}_{\nu}(z) \chi^{(\gamma)}_{\nu}(z') V^{(\gamma)}_{\nu,k}[z] V^{(\gamma)}_{\nu,k}[z'] 
 \\ & \quad \quad + i \hbar(\beta_{\nu} + \lambda_{\nu})N'_{\nu,k} ,  \notag
\end{align}
with $N'_{\nu,k} = \frac{m_{\nu,k} e^{- 2 r_{\nu,k}} \omega_{\nu,k}^2 L_{\nu,k}}{2}$.
The dependence of \eqref{eq:dissipact} and \eqref{eq:dissipactsquee} on the counting fields $\lambda_{\nu}$ is not evident at first sight, but is "hidden" in the geometry of the contour $\gamma^{\nu}$.
This dependence becomes clear when selecting specific values of $z,z'$, thus introducing the {\it components} \cite{langreth1976linear,Stefanucci} of the Green's function on the contour.
This can be done by adopting a function that selects the real or imaginary parts of $z$ according to the branch of the contour, $ \Psi^j (t+i \zeta) = t $ for $j = -,+$ and $ \Psi^j (t+i \zeta) = \zeta $ for $j = \uparrow,\downarrow, M$ and defining 
 \begin{align} \notag 
 & {\rm if} \; z\in \gamma^{\nu}_{i}, \, z'\in\gamma^{\nu}_{j} \quad \: \\ & G_{\nu,k}^{i,j}(\Psi^j(z),\Psi^j(z')) \equiv G_{\nu,k}(z,z').\label{eq:defcompmain}
 \end{align}
Analogous definitions work for $\mathcal{G}_{\nu,k}$ and for any function with arguments on the contour e.g.~$x_{\pm}(t) \equiv x(t+ i \zeta)$ for $t+ i \zeta \in \gamma_{\pm}$.
 With these definitions we immediately recover the lesser, greater, time-ordered, anti time-ordered and Matsubara components \cite{Stefanucci, Kamenev} of the Green's function respectively by selecting $(i,j) = (-,+), (+,-), (-,-), (+,+), (M,M)$.
In the usual Keldysh contour, the greater and lesser components 
are associated with emission and absorption of excitations, the same is true for the contour $\gamma^{\nu}$ in which they bear an additional dependence by $\lambda_{\nu}$ (see the Appendix \ref{app:Green} for a list of all the GF components). 
To write Eq.~\eqref{eq:dissipactsquee} (or, analogously, Eq. \eqref{eq:dissipact}) in terms of the components \eqref{eq:defcompmain} it is sufficient to split the integrals on the contour $\gamma^{\nu}$ as a sum of integrals over $\gamma_{-}, \gamma_-, \gamma^{\nu}_M, \gamma^{\nu}_{\downarrow}$.
After dividing the integral in components
and assuming that the coupling between the system and the modes of a given reservoir are functionally the same apart from a prefactor $v_{\nu,k}(t) =  c_{\nu,k} v_{\nu}(t) $ where $ v_{\nu}(t) = \chi_{\nu}(t) V_{\nu}(t)$
we have (see also App.~\ref{app:disp})
\begin{align}  \notag
   & \bar{\mathcal{S}}_{\nu,k}  =    \sum_{l,j=\pm} \iint_{0}^{\tau}   \frac{ c^2_{\nu,k} v^l_{\nu}(t) v_{\nu}^{j}(t') }{2 m_{\nu,k}  \omega_{\nu,k}}  j l \mathcal{G}^{l,j}_{\nu,k}(t,t') dt dt'  \\      \label{eq:bigactdisp} &
    +  i \sum_{l=\pm}  \int_{0}^{\tau}   \, l   \omega_{\nu,k}c_{\nu,k} v^l_{\nu}(t) e^{- 2 r_{\nu,k}} L_{\nu,k}   \\ &  \times  \big[ \int_{0}^{-\hbar \beta} \mathcal{G}^{l,M}_{\nu,k}(t,\zeta)   d\zeta
    %+   \int_0^{\hbar \lambda_{\nu}} ds'G^{l,\uparrow}_{\nu,k}(s,\zeta)  
    + \int_{\hbar \lambda_{\nu}}^{0} \mathcal{G}^{l,\downarrow}_{\nu,k}(t,\zeta) d\zeta]  dt+ \mathcal{N}_{\nu,k},
\notag 
 \end{align}
where $\mathcal{N}_{\nu,k}$ is equal to $ i \hbar (\beta_{\nu} + \lambda_{\nu}) N'_{\nu,k}$ plus the contribution of  $\mathcal{G}_{\nu,k}^{M, \downarrow}$, $\mathcal{G}_{\nu,k}^{\downarrow,\downarrow}$, $\mathcal{G}_{\nu,k}^{M,M}$ and does not depend on the system fields (see Appendix \ref{app:disp}), and $v^{l}_{\nu}(t)$ is a shorthand notation for $v_{\nu}[x_{l}(t)]$ with $l = \pm$.
To obtain the statistics of the heat flows in the bath $\nu$ we have to add the contributions of all the modes; 
the sum over $k$ is often replaced by an integral
over the {\it spectral density} of the reservoir $\nu$, defined as
\begin{align} \label{spectraldensitymain}
    J_{\nu}(\omega)\equiv \frac{\pi}{2} \sum_k \frac{c_{\nu,k}^2}{m_{\nu,k} \omega_{\nu,k}} \delta\left(\omega-\omega_{\nu,k} \right).
\end{align}
The spectral density at a specific frequency $\omega$ measures how much the modes of a given reservoir are concentrated around $\omega$ and how strong the coupling between these modes and the system is.
In an analogous way, we can introduce 
\begin{equation} \label{eq:Kdensity}
    K_{\nu}(\omega)\equiv \frac{\pi}{2} \sum_k c_{\nu, k} L_{\nu,k} e^{-2r_{\nu,k}} \delta\left(\omega-\omega_{\nu,k} \right).
    \end{equation}
    The function $ K_{\nu}(\omega)$ measures how the magnitude of the force induced by the displacement of the reservoir modes is distributed over their spectrum.
We can now sum over $k$ in  Eq. \eqref{eq:bigactdisp} and use the definitions \eqref{spectraldensitymain}, \eqref{eq:Kdensity} to obtain $\bar{\mathcal{S}}_{\nu} \equiv \sum_k \bar{\mathcal{S}}_{\nu,k}$, that reads
 \begin{align}  \notag
 &  \bar{\mathcal{S}}_{\nu} =   \sum_{l,j=\pm} \int d\omega \iint_{0}^{\tau}  \frac{J_{\nu}(\omega)}{\pi} v_{\nu}^l v_{\nu}^j  j l\mathcal{G}^{l,j}_{\nu}(\omega)   dt dt'  \\  \notag &
    +  i \sum_{l=\pm} \int d \omega  \: \frac{2 K_{\nu}(\omega) \omega}{\pi}  \int_{0}^{\tau} \,   l     \big[ \int_{0}^{-\hbar \beta_{\nu}}  \mathcal{G}^{l,M}_{\nu,k} (\omega) d\zeta \\ & 
    %+   \int_0^{\hbar \lambda_{\nu}} ds'G^{l,\uparrow}_{\nu}(\omega)  
    + \int_{\hbar \lambda_{\nu}}^{0}   \mathcal{G}^{l,\downarrow}_{\nu}(\omega) d\zeta ]  v_{\nu}^l   dt + \mathcal{N}_{\nu}, \label{eq:bigactdisp22}
 \end{align}
 where $\mathcal{N}_{\nu} = \sum_{k} \mathcal{N}_{\nu,k}$ and we did not write the time dependences for ease of notation (they are the same as in Eq. \eqref{eq:bigactdisp}).
 The equation above is the effective system action appearing in the path integral expression for the calculation of the heat flows.
 If $\lambda_{\nu} = 0$ and with no squeezing and displacement, the action reduces to the standard action of the CL model (see App. \ref{app:classqdyn}) and the contour reduces to the standard Keldysh contour augmented with the Matsubara branch for initial states (see Fig.~\ref{fig:5} A). The other panels of Fig.~\ref{fig:5} contain a recap of the effects of initial displacement and squeezing of the bath $\nu$ on the effective action of the system.
\section{Fluctuating thermodynamics in the Zwanzig model and classical limit of the quantum action} \label{sec:classic}
In this chapter we obtain the MGF of the heat flows  in the classical analog of the NECL and show the quantum/classical correspondence at the level of generating functions. We compute all the relevant quantities in the generic framework in which both squeezing and displacement are present, but show the full quantum/classical correspondence only in the displaced case.
 
\subsection{Thermodynamics in the classical regime}
\label{sec:thermclass}
We now turn to the classical counterpart of the quantum NECL.
Our intention is to later recover it as the classical limit $\hbar \rightarrow 0$ of the NECL. 

The classical analog of the quenches in Fig. \ref{fig:2} are the classical squeezing and displacement operations, i.e. a quench of the masses $m_{\nu,k}$ and a change of the center of the average initial harmonic potential. 
The analog of the Hamiltonian \eqref{eq:ham} is the Zwanzig \cite{Zwanzig} model
\begin{align}  \notag  H(t) & = \frac{p(t)^2}{2 m} + V[x(t)] +\sum_{\nu,k} \chi_{\nu}(t) x_{\nu,k}(t) V_{\nu,k}[x(t)] 
\\ 
& + \sum_{\nu,k} \big[ \frac{p_{\nu,k}(t)^2}{2 m_{\nu,k}} + \frac{m_{\nu,k} \omega_{\nu,k}^2 x_{\nu,k}(t)^2}{2} \big] \label{eq:hamclass}
,
\end{align}
where $x(t),p(t),x_{\nu,k}(t), p_{\nu,k}(t)$ are position and momentum variables of the system and of all the modes of the reservoirs. From \eqref{eq:hamclass} we derive the force exerted by all reservoir modes on the system $F^{(S)}_{\nu,k}[x,\boldsymbol{x_B}] $ and the force exerted by the system on each reservoir mode $F_{\nu,k}[x,\boldsymbol{x_B}]$.
The energy variation in the system is defined as the difference between the final and initial energies i.e. the integral of the associated power
\begin{align}
    \Delta E_S(\tau_0,\tau) &= \int_{\tau_0}^{\tau} \frac{d}{dt} H_S(t) dt \nonumber\\ \label{eq:powDer} & \hspace{-1cm}= \int_{\tau_0}^{\tau} \big\{ \frac{p(t) \dot{p}(t)}{m} + V'[x(t) \big\} \dot{x}(t) \big] dt \\ 
    &\hspace{-1.5cm}=  \int_{\tau_0}^{\tau} \big\{ \frac{p(t) (-V' +\sum_{\nu,k}F^{(S)}_{\nu,k})}{m} + V'[x(t) \big\} \dot{x}(t) \big] dt. \nonumber
\end{align}

The terms containing $V'$ cancel out, leaving the energy variation of the system and each reservoir respectively equal to
\begin{align} \notag
\Delta E_S(\tau_0,\tau) & = \sum_{\nu,k} \int_{\tau_0}^{\tau} F^{(S)}_{\nu,k}[\bar{x}(t), \boldsymbol{\bar{x}_B}(t)] \dot{x}(t) dt, 
\\ \Delta E_{\nu}(\tau_0,\tau) & = \sum_k \int_{\tau_0}^{\tau} F_{\nu,k}[\bar{x}(t), \boldsymbol{\bar{x}_B}(t)] \dot{\bar{x}}_{\nu,k}(t) dt, \label{eq:heatclass}
\end{align}
where the latter can be obtained by following analogous steps to those in Eq. \eqref{eq:powDer}, the trajectories $\bar{x}(t)$, $\bar{x}_{\nu,k}(t)$ satisfy the Hamilton equations derived from \eqref{eq:hamclass} and 
\begin{align} 
 F^{(S)}_{\nu,k}[\bar{x}(t),\boldsymbol{\bar{x}_B}(t)] & = -  \chi_{\nu}(t)  V'_{\nu,k}[\bar{x}(t)] \bar{x}_{\nu,k}(t),
 \\
 F_{\nu,k}[\bar{x}(t)] & = - \chi_{\nu}(t) V_{\nu,k}[\bar{x}(t)].
\end{align}
Eq. \eqref{eq:heatclass} assumes that $\tau_0$ is greater of equal to $0$, so that the energy variation in the reservoirs contains only the contribution due to the contact with the system.
As in Sec. \ref{sec:thermo}, we distinguish this energy variation from the total heat $Q_{\nu}$ that can be written as the sum of $\Delta E_{\nu}(0,\tau)$ and the stochastic energy initially pumped in the reservoirs by the initial quench. 
The contributions of the squeezing and displacement are respectively given by 
\begin{align} \notag
\Delta E^{0,sq}_{\nu} & = \sum_k \big[ \frac{1}{2} (1-e^{-2 r_{\nu,k}}) m_{\nu,k} \omega_{\nu,k}^2 \bar{x}_{\nu,k}(0)^2 \\ 
& + \frac{1}{2m_{\nu,k} } (1-e^{+2r_{\nu,k}})  \bar{p}_{\nu,k}^2(0) \big], \label{eq:ensquein} \\ \notag
\Delta E_{0,\nu}^{dp} & = 
   e^{-2r_{\nu,k}} m_{\nu,k} \omega_{\nu,k}^2 \\ & \times  \big[ \bar{x}_{\nu,k}(0) L_{\nu,k} -\frac{1}{2} L_{\nu,k}^2  \big],  \label{eq:endispin}
\end{align}
that are expressed in terms of the values $\bar{x}_{\nu,k}(0), \bar{p}_{\nu,k}(0)$ of the position and momentum after the quench, as we show in App.~\ref{app:coordispsque}.
We introduce two generating functions, one for $\Delta E_{\nu}$ and one for $Q_{\nu} = \Delta E_{\nu} + \Delta E_{\nu}^{0,sq} + \Delta E_{\nu}^{0,dp}$. In the first case, we have
\begin{equation} \label{eq:momclass}
M^{ex}_{cl}(\lambda_S, \boldsymbol{\lambda_B},\tau) = \langle e^{\lambda_S \Delta E_S(0,\tau) + \boldsymbol{\lambda_B}\cdot \boldsymbol{\Delta E_{\nu}}(0, \tau)}  \rangle,
\end{equation}
where $\langle...\rangle$ is the average over all the initial preparations of the system and the reservoirs, and ${\bf \Delta E_B} = (\{ \Delta E_{\nu} \}_{\nu=1}^N)$.
Instead, the classical analog of the moment generating function \eqref{eq:finalgen} is given by
\begin{equation} \label{eq:momclass2}
M^{tot}_{cl}(\lambda_S, \boldsymbol{\lambda_B},\tau) = \langle e^{\lambda_S \Delta E_S(0,\tau) + \boldsymbol{\lambda_B}\cdot \boldsymbol{Q}}  \rangle.
\end{equation}
Note that in the TPEM for the quantum case, there is no analog of the MGF \eqref{eq:momclass}, since every measurement performed after the quenches inevitably destroys the initial coherences and affects the subsequent dynamics.
\subsection{Stochastic dynamics for system variables}
\label{sec:redusist}
Our goal in the next two sections is to express Eqs. \eqref{eq:momclass} and \eqref{eq:momclass2} in a path integral form and compare the associated actions with Eq. \eqref{eq:bigactdisp22}.
We start by showing that both the dynamics generated by \eqref{eq:hamclass} and the thermodynamic quantities \eqref{eq:heatclass} can be expressed in terms of noise variables associated to the reservoirs.
The solution of the Hamilton equations generated by \eqref{eq:hamclass} is $\bar{x}_{\nu,k}(t) = y_{\nu,k}(t) + w_{\nu,k}(t) $ where 
\begin{align} \notag
y_{\nu,k}(t)  & =  \sum_{k}\big[\bar{x}_{\nu,k}(0) \cos \omega_{\nu,k}t + \frac{\bar{p}_{\nu,k}(0)}{m_{\nu,k} \omega_{\nu,k}} \sin \omega_{\nu,k}t \big], \\ 
w_{\nu,k}(t)&  =-\int_{0}^{t} dt' \frac{\chi_{\nu}(s) c_{\nu,k}}{m_{\nu,k} \omega_{\nu,k}} V_{\nu}[x(s)] \sin{\omega_{\nu,k} (t-s) }. \label{eq:reservoirvar}
\end{align}
In the equation above, we assumed that the coupling between the system and each mode of a specified reservoir has the same form for all the modes up to a constant, i.e. $V_{\nu,k}(x) = c_{\nu,k} V_{\nu}(x)$.
The external force acting on the system can be divided into a fluctuating and a deterministic contribution, obtaining the following generalized Langevin equation: 
\begin{equation} \label{eq:genlang}
m \ddot{x} = -V'(x) - \sum_{\nu} \big[ \chi_{\nu} V'_{\nu}(x)\xi_{\nu} +\chi_{\nu} V'_{\nu}(x)R_{\nu}\big] 
\end{equation}
where $\xi_{\nu} = \sum_{k} c_{\nu,k} y_{\nu,k}$, $R_{\nu} = \sum_{k} c_{\nu,k} w_{\nu,k}$. A step by step derivation of \eqref{eq:genlang} is given in App.~\ref{app:GLEQ}.
For reservoirs that are initially at equilibrium, it is possible to show that $\langle \xi_{\nu} \rangle =0$ and by defining $\langle \xi_{\nu}(t) \xi_{\nu'}(t') \rangle \equiv C^{(2)}_{\nu}(t-t')$ and $R_{\nu}(t) =  - \int_{0}^t \chi_{\nu}(s) C_{\nu}^{(1)}(t-s) V_{\nu}[x(s)] $ we have 
\begin{align} \label{eq:corr1}
C^{(2)}_{\nu}(t-t') =  \int \frac{ 2 J_{\nu}(\omega) \cos \omega (t-t')}{\pi \beta_{\nu} \omega} d \omega, \\ \label{eq:corr2}
C^{(1)}_{\nu}(t-t') = \frac{2}{\pi} \int J_{\nu}(\omega) \sin \omega (t-t')  d \omega ,
\end{align}
where we introduced the spectral density of the reservoir $\nu$, $J_{\nu}(\omega)$.
The variation in system energy and the energy flow in the reservoir $\nu$ can be expressed as
\begin{align} \notag 
\Delta E_S(0, \tau) & = - \int_0^{\tau}  
 \chi_{\nu}(t) V'_{\nu}[x(t)] \dot{x} \\ &\times (\xi_{\nu}(t) + R_{\nu}(t)) dt, \label{eq:genen}
\\ \notag
   \Delta E_{\nu}(0,\tau) & =  \int_0^{\tau}  
  \frac{d}{dt} \{\chi_{\nu}(t) V_{\nu}[x(t)] \} \\ & \times (\xi_{\nu}(t) + R_{\nu}(t)) dt. \label{eq:genheat}
\end{align}
We highlight that there is a term proportional to $\dot{\chi}_{\nu}$ in both equations \eqref{eq:genheat} and \eqref{eq:genlang}, in the latter, this can be explicitly obtained by integrating by parts in $s$ in the definition of $R_{\nu}(t)$.
This term is the so-called {\it Sekimoto force} \cite{sekimoto1997complementarity, sekimoto1998langevin}. It disappears when the coupling is not driven.
If we allow for the possibility of an initial displacement $L_{\nu,k}$ in the reservoir modes, we have $ \langle \bar{x}_{\nu,k} (0) \rangle_d = L_{\nu,k} $ which in turn gives
$\langle y_{\nu,k}(t)\rangle_d = L_{\nu,k} \cos \omega_{\nu,k} t$, where the subscript $d$ is to make explicit that we are averaging on a displaced mode,
thus the random force in Eq. 
\eqref{eq:genlang} satisfies
\begin{equation} \label{eq:avgforce}
    \langle \xi_{\nu}(t) \rangle_{d} \equiv - F_{\nu,d}(t) = \frac{2}{\pi}
    \int^{\infty}_{-\infty} K_{\nu}(\omega) \cos \omega t d\omega,
 \end{equation}
with $K_{\nu}(\omega)$ given by Eq. \eqref{eq:Kdensity} (with $r_{\nu,k} =0$).
In the case of initial quench of the reservoir masses, the structure of the correlation function \eqref{eq:corr1} changes, becoming
\begin{align}\notag
\mathcal{C}_{\nu}^{(2)}(t,t') & = 
\frac{2}{\pi} \int_{-\infty}^{\infty} \frac{1}{\beta_{\nu} \omega} \big[  \mathcal{J}_{\nu}(\omega) \cos \omega (t-t') \\ & +  \Delta \mathcal{J}_{\nu}(\omega) \cos \omega t \cos \omega t' \big] d\omega \label{eq:modcorr2}
\end{align}
where $\mathcal{J}_{\nu}$ and $\Delta \mathcal{J}_{\nu}$ are two squeezed versions of the spectral density, that respectively reduce to $J_{\nu}$ and $0$ in the absence of squeezing (see App. \ref{app:quencma}).
It is not surprising that using squeezed reservoirs can lead to a violation of the second law since the correlation function \eqref{eq:modcorr2} violates the fluctuation/dissipation theorem. We will see how this relates with the second law and the fluctuation theorem in Sec. \ref{sec:propsym}.
\subsection{MSRJD path integral and thermodynamics}
\label{sec:MSRJDfinal}
%To show the quantum/classical correspondence we look for a uniformed strategy that allows us to treat both classical (stochastic) and quantum thermodynamics in the same ground.
%The path integral is a good choice, since the quantum and stochastic path integrals share the same structure, and the differences and similarities between the two can be spotted at the level of the action. As a first step,
As demonstrated in the preceding section, the average on the initial conditions in Eq.~\eqref{eq:momclass}  can be written as an average of the exponential of the stochastic energetic functionals \eqref{eq:genen} and \eqref{eq:genheat} over the noise variables.
Using the MSRJD approach \cite{martin1973statistical,DeDominicis1976,Altland}, this can be written as an integral over stochastic paths.
For any stochastic functional $f[\{x(t), \boldsymbol{\xi_B}(t)\}^{\tau}_{t=0}]$, where $x(t)$ is constrained to satisfy a set of equations of motion like \eqref{eq:genlang}, we can produce an integral over the paths by expressing the dynamical constraint as a set of delta functions (one for each infinitiesimal time interval), and representing each delta function as a phase integral over an auxiliary variable $\eta_x(t)$ (see Appendix \ref{app:classpath}).
Denoting with $\langle ... \rangle$ the average over the noises $\boldsymbol{\xi_B}= \{ \xi_{\nu}\}_{\nu=1,..,N}$ we have 
\begin{align} \notag
\langle  f  [...] \rangle & = \langle \int \mathcal{D}x e^{i \mathcal{S}^{(cl)} } f[\{x(t), \boldsymbol{\xi_B}(t)\}^{\tau}_{t=0}] \\ & \times P[x(0),\dot{x}(0)] \rangle
%\\  = & \int \mathcal{D}x \mathcal{D}\xi e^{i S'_{sys}[x,\eta,\xi]} f[\{x(t)\}^{\tau}_0] P[x(0),\dot{x}(0)],
\label{eq:avclasspath}
\end{align}
where $P$ is the initial probability distribution in the system configuration space, $\mathcal{D}x$ is the path integral measure and $\mathcal{S}^{(cl)}$ is the stochastic dynamical action
\begin{align} \label{eq:claclos}
     \mathcal{S}^{(cl)} = \int_0^{\tau} \eta_x(t) \{m \ddot{x}(t) - \mathcal{F}[x(t),\boldsymbol{\xi_B}(t)]\} dt,
\end{align}
with $\mathcal{F}$ being the total force acting on the system (r.h.s. of Eq. \eqref{eq:genlang}).
Due to the specific form of the functional inside Eq. \eqref{eq:momclass}, to compute this MGF it is sufficient to replace the dynamical action \eqref{eq:claclos} with
\begin{equation} \label{eq:slambdasys}
     \mathcal{S}'  =
        \mathcal{S}^{(cl)} - i \lambda_S \Delta E_S(0,\tau) - i \boldsymbol{\lambda_B}\cdot \boldsymbol{\Delta E_B}(0,\tau), 
\end{equation}
in which we dropped the arguments of $\mathcal{S}^{(cl)}$ for ease of notation and $\Delta E_S$ and $\Delta E_{\nu}$ are given by Eqs. \eqref{eq:genen} and \eqref{eq:genheat}.
We conclude by performing the average in Eq. \eqref{eq:avclasspath} explicitly (see App. \ref{app:classpath}).
In a completely general setting, in which the average of the noise can be different from $0$, like in Equation \eqref{eq:avgforce}, and the correlation function is squeezed as in \eqref{eq:modcorr2}, we obtain a MSRJD form of the action as
$\mathcal{S}' = \mathcal{S}_{lin} + \mathcal{S}_{quad} $ where 
\begin{align} \notag
    \mathcal{S}_{lin} &= \int_0^{\tau} \eta_x(t) \big[m \ddot{x} + V'(x) + \sum_{\nu} \chi_{\nu} V_{\nu}'(x)  R'_{\nu} \big] dt \\   \notag
& - \int_0^{\tau}\sum_{\nu} \big\{ \big[ i \lambda_{\nu} \frac{d}{dt} (\chi_{\nu} V_{\nu})   - i \lambda_S \chi_{\nu} V_{\nu}'(x) \dot{x}  \big] R'_{\nu} \big\} dt
\\
\mathcal{S}&_{quad} = \frac{i}{2} \sum_{\nu} \iint_{0}^{\tau} \Xi_{\nu}(t) \Xi_{\nu}(t') \mathcal{C}^{(2)}_{\nu,s}(t,t') dt dt', \label{eq:finalclass}
\end{align}
in which, for ease of notation, we suppressed the time arguments (when there is no ambiguity) and introduced the auxiliary quantities
\begin{equation}
    \Xi = \chi_{\nu} V_{\nu}'[x][\eta_x+ i \lambda_S \dot{x}] - i \lambda_{\nu} \frac{d}{dt}[\chi_{\nu} V_{\nu}(x)],
\end{equation}
and the total average force induced by the contact with the reservoir $R'_{\nu} = (R_{\nu}- F_{\nu,d}) $.
We can do the same calculations in the case of the MGF \eqref{eq:momclass2}, for which we have to replace $\Delta E_{\nu}$ with $Q_{\nu}$. %The difference between the two is the energy $\Delta E^0_{\nu}$ that is the energy spent to create the non-equilibrium reservoir starting from an equilibrium configuration.
To compute the MGF \eqref{eq:momclass2} we simply have to add $\Delta E^0_{\nu} = \Delta E^{0,sq}_{\nu} + \Delta E^{0,dp}_{\nu}$ to each $\Delta E_{\nu}$ in Eq. \eqref{eq:slambdasys}, to obtain the action
\begin{equation}
\mathcal{S}'' = \mathcal{S}^{(cl)} - i \lambda_S \Delta E_S(0,\tau) - i \boldsymbol{\lambda_B} \cdot \boldsymbol{Q}. \label{eq:momclassexp2}
\end{equation}
The summation over the noises can be done exactly also in this case, obtaining an equation similar to Eq. \eqref{eq:finalclass}. 
The explicit calculation (limited to the case without squeezing) is done in Appendix
\ref{app:coordispsque} and results in a correction $\Delta \mathcal{S}_{lin}$ to the first of Eqs.
\eqref{eq:finalclass} 
\begin{align} \notag
    \Delta \mathcal{S}_{lin} &=  \int_0^{\tau} \eta_x(t)  \big[ \sum_{\nu}\frac{\lambda_{\nu}}{\beta_{\nu}}\chi_{\nu} V_{\nu}'(x) F_{\nu,d} \big]dt \\ \label{eq:finalclass2}
    & -  \int_0^{\tau}\big\{ \sum_{\nu} \frac{\lambda_{\nu}}{\beta_{\nu}} \big[ i \lambda_{\nu} \frac{d}{dt} (\chi_{\nu} V_{\nu})   \\ & - i \lambda_S \chi_{\nu} V_{\nu}'(x) \dot{x}  
    \big ]    F_{\nu,d} \big\}dt 
    - i\sum_{\nu}\mathcal{M}_{\nu} ,
\notag
\end{align}
with $\mathcal{M}_{\nu} = \sum_{k}\frac{m_{\nu,k} \omega_{\nu,k}^2 L_{\nu,k}^2 }{2} \big( \lambda_{\nu} + \frac{ \lambda^2_{\nu}}{ \beta_{\nu}}\big)$. 
\begin{figure*}[!t]
    \centering    \includegraphics[width=1\textwidth]{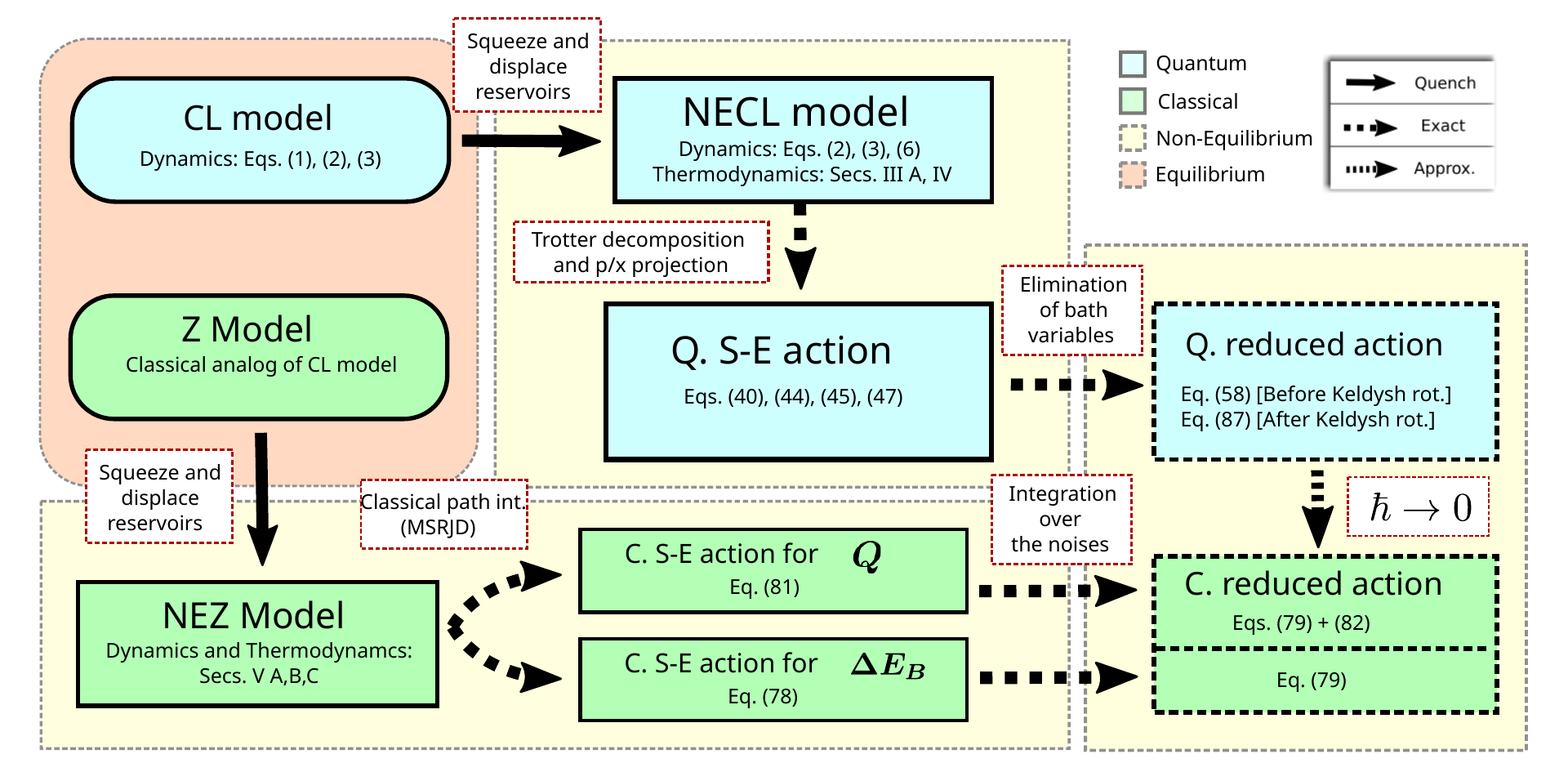}
    \caption{A summary of the results of Secs. \ref{sec:fullquant} and \ref{sec:classic}.
    The boxes colored in light blue represent a roadmap of Sec. \ref{sec:fullquant}. We start from a microscopic description of the system in contact with equilibrium reservoirs (CL model) and squeeze and displace the reservoirs, obtaining the NECL. In Sec. \ref{sec:otresc} we express the MGF in terms of a path integral, thus obtaining a quantum action for the system and environmental variables ({\bf Q. S-E action}). After eliminating the variables of the reservoirs, we obtain a reduced action in terms of the system fields ({\bf Q. reduced action}). 
    The light green boxes represent an outline of Sec. \ref{sec:classic}. We use Z and NEZ as abbreviations for Zwanzig and Non-Equilibrium Zwanzig respectively, the classical analogues of the CL and NECL. In Sec. \ref{sec:MSRJDfinal} we introduce the MGF of the energy variation of the reservoirs during the system evolution ({\bf C. S-E action for } $ \boldsymbol{\Delta E_B}$) and of the total energy variation of the reservoirs (heat), including the work in the initial quench ({\bf C. S-E action for }$\boldsymbol{Q_B}$).
     To connect the two paths, we prove that the quantum action reduces to the classical action for $\boldsymbol{Q_B}$ in the classical limit $\hbar \rightarrow 0$ (the matching is explicitly shown in the displaced case).  }
    \label{fig:6}
\end{figure*}
\subsection{Classical limit of the quantum action} \label{sec:semiclq}
The two path integrals in \eqref{eq:actquant} and \eqref{eq:finalclass} are formally different, indeed
\begin{enumerate}
\item The action \eqref{eq:finalclass} depends on 2 fields, the physical field $x(t)$ and the auxiliary field $\eta_x(t)$, while the quantum action depends on a single field $x(z)$.
\item The domain of integration of the classical action is $[0,\tau]$, while every component of the quantum action containing $\nu$ lives on a modified contour, like the one in Fig. \ref{fig:1}.
\end{enumerate}
For the standard dissipative action problem, based on the Keldysh contour, there is a way to transform the quantum action that solves both the problems mentioned above, that consists in introducing the components and performing the {\it Keldysh rotation} \cite{Kamenev, Polkovnikovreview, polkovnikov2003quantum}.
First, after introducing the components as in \eqref{eq:defcompmain}, the double integral on the contour in Eq. \eqref{eq:actquant} reduces to a set of integrals on $\gamma^{\nu}_{\uparrow, \downarrow,M}$ and $\gamma_{\pm}$ (see Eq. \eqref{eq:bigactdisp22}). The integrals on $\gamma_{\pm}$ can be transformed into integrals between $0$ and $\tau$ since $\int_{\gamma_+} dt = -\int_{\gamma_-} dt = -\int_0^{\tau} dt$.
In this way, we go from a description based on a single field in a ``doubled" contour ($\gamma_+ \oplus \gamma_-$) to a two-field description on the real segment $[0,\tau]$.
The Keldysh rotation is now a redefinition of the fields that ensure the convergence between the quantum and classical dynamical actions for  $\boldsymbol{\lambda_B}=0$, $\hbar \rightarrow 0$. 
This transformation is 
\begin{equation}\label{eq:keldishrot}
    x_-(t) = x_{cl}(t) + \frac{x_q(t)}{2}, \quad   x_+(t) = x_{cl}(t) - \frac{x_q(t)}{2}.
\end{equation}
The "quantum field" $x_q$ measures the asymmetry between the forward and backward quantum trajectories, 
as such, is a purely quantum term and it is customary to assume that it is small in the classical limit, with a scaling linear in $\hbar$ \cite{polkovnikov2003quantum, Polkovnikovreview, Kamenev}. 
The same rotation can be applied to other quantities defined on the contour, for instance the interaction potential
\begin{equation} \label{eq:potrot}
    v_{\nu}^-(t) = v_{\nu}^{cl}(t) + \frac{v_{\nu}^q(t)}{2}, \quad   v_{\nu}^+(t) = v_{\nu}^{cl}(t) - \frac{v_{\nu}^q(t)}{2}.  
\end{equation}
The linear transformations \eqref{eq:keldishrot} and \eqref{eq:potrot} make it convenient to introduce the Keldysh components of the contour Green's function, defined as (see also \cite{Kamenev})
\begin{align} \label{eq:rotgf}
 \notag \mathcal{G}_{\nu}^{cl,cl} & = \mathcal{G}_{\nu}^{-,-} + \mathcal{G}_{\nu}^{+,+} - \mathcal{G}_{\nu}^{+,-} - \mathcal{G}_{\nu}^{-,+}, \\
  \notag \mathcal{G}_{\nu}^{cl,q} & = \frac{1}{2} \big[\mathcal{G}_{\nu}^{-,-} - \mathcal{G}_{\nu}^{+,+} - \mathcal{G}_{\nu}^{+,-} + \mathcal{G}_{\nu}^{-,+}\big], \\
   \notag \mathcal{G}_{\nu}^{q,cl} & = \frac{1}{2} \big[ \mathcal{G}_{\nu}^{-,-} - \mathcal{G}_{\nu}^{+,+} + \mathcal{G}_{\nu}^{+,-} - \mathcal{G}_{\nu}^{-,+} \big], \\
     \mathcal{G}_{\nu}^{q,q} & = \frac{1}{4} \big[\mathcal{G}_{\nu}^{-,-} + \mathcal{G}_{\nu}^{+,+} + \mathcal{G}_{\nu}^{+,-} + \mathcal{G}_{\nu}^{-,+} \big], 
\end{align}
where we omitted the time arguments for ease of notation, and the transformation is the same both in the discrete and continuum case (in which there is an additional dependence on $k$ and $\omega$, respectively, i.e. $\mathcal{G}_{\nu}^{i,j} \equiv \mathcal{G}_{\nu,k}^{i,j} $ and $\mathcal{G}_{\nu}^{i,j} \equiv \mathcal{G}_{\nu}^{i,j}(\omega) $). Note that in \eqref{eq:rotgf} we are using a notation in which the indexes $q,cl$ are exchanged with respect to the common approach used in the literature \footnote{$G^{cl,cl}$ as defined in Eq. \eqref{eq:rotgf} is associated with the quantum fluctuations of the reservoir, for this reason the choice to call this quantity $G^{q,q}$ is preferred in the literature. However, the quantum fluctuations in the reservoir couple to classical fluctuations in the system, so that our $G^{cl,cl}$ is naturally multiplied by $x_{cl}$ in the non-equilibrium action.}, for instance see Eq. 2.41 in \cite{Kamenev}. This choice allows for a notational simplification when writing scalar products of the form $x_i \mathcal{G}^{i,j} x_j$. 
In an analogous way, we can introduce the Keldysh-rotated version of the other relevant components on the contour
\begin{equation}
    \mathcal{G}_{\nu}^{cl,j} = \mathcal{G}_{\nu}^{-,j} - \mathcal{G}_{\nu}^{+,j}; \quad  \mathcal{G}_{\nu}^{q,j} = \frac{1}{2} \big[\mathcal{G}_{\nu}^{-,j} + \mathcal{G}^{+,j}_{\nu} \big],
\end{equation}
where $j=\uparrow, \downarrow, M$. After writing Eq. \eqref{eq:bigactdisp22} in terms of the potential \eqref{eq:potrot} and of the Keldysh components \eqref{eq:rotgf}, we have (see Appendix \ref{app:keld} for the explicit calculations)
\begin{align}  \notag
  \bar{\mathcal{S}}_{\nu} & =  \sum_{l,j=cl,q} \int d\omega \int_{0}^{\tau}  \int_{0}^{\tau}   \frac{J_{\nu}(\omega)}{\pi} v^l_{\nu}  v^{j}_{\nu} \mathcal{G}^{l,j}_{\nu}(\omega)   dt dt' 
   \notag \\ & - i  \sum_{l=cl,q}  \int d\omega \frac{2 K_{\nu}(\omega) \omega}{\pi}   \int_{0}^{\tau} ds \big[ \int_{\hbar \lambda_{\nu}}^{0} d\zeta       \mathcal{G}^{l,\downarrow}_{\nu}(\omega)
  \notag \\  
    & +      \int_{0}^{-\hbar \beta_{\nu}} d\zeta    
       \mathcal{G}^{l,M}_{\nu}(\omega)  \big] v^{l}_{\nu} + \mathcal{N}_{\nu}
  % +   \int_{0}^{\hbar \lambda_{\nu}} d\zeta'         G^{l,\uparrow}_{\nu}(\omega)    \big] v^l_{\nu}(s) 
  ,\label{eq:bigactdispcontmain}
    \end{align}
where we omitted the time arguments in the GFs and in $v^{l}_{\nu}(t)$ for ease of notation (they are the same as in Eq. \eqref{eq:bigactdisp}).
The advantage of the definitions \eqref{eq:rotgf} is clear from the structure of \eqref{eq:bigactdispcontmain}, which is the same as the one of \eqref{eq:finalclass} after we do the customary replacement $x_q \rightarrow  - \hbar \eta_x $.
We will not write all the calculations in the main text, but we will focus on some terms of the action in which the correspondence between the classical limit of the quantum model and the classical model can be seen easily.
We start by assuming that there is no squeezing so that $\mathcal{G}_{\nu} = G_{\nu}$. The components appearing in the first line of Eq. \eqref{eq:bigactdispcontmain}
 are $G^{q,q}_{\nu}, G^{cl,q}_{\nu}, G^{q,cl}_{\nu}$ and $G^{cl,cl}_{\nu}$, the first reduces, in the classical limit, to the classical noise kernel (see App. \ref{app:classqdyn})
 \begin{equation}
G^{q,q}_{\nu}(\omega,t,t') = \frac{i}{ \hbar \omega \beta} 
\cos \omega(t-t') + O(\hbar)
 \end{equation}
that after integrating over $\omega$
in \eqref{eq:bigactdispcontmain} takes exactly the form in Eq. \eqref{eq:corr1}.
Using this result, we obtain that the terms proportional to $\eta_x^2$ in Eq. \eqref{eq:finalclass} and in the classical limit of the first line in Eq. \eqref{eq:bigactdispcontmain} (with $x_q \rightarrow -  \hbar \eta_x$) coincide, as expected \cite{Kamenev}.
A similar discussion holds for the component $G^{cl,cl}_{\nu}$ for which we have
\begin{align}  \notag
     G^{cl,cl}_{\nu}&(\omega,t,t')   = i \coth\big( \frac{\hbar \omega \beta_{\nu}}{2} \big)
\cos \omega (t-t') \times \\ & \times \big[ 1 - \cos i \omega \hbar \lambda_{\nu}  \big] -  \cos \omega (t-t') \sin i \omega \hbar \lambda_{\nu}, \label{eq:classiGcl}
\end{align}
that at the first non-zero order in $\hbar$ becomes
\begin{align} \notag 
   G^{cl,cl}_{\nu}(\omega,t,t') & = - i \hbar \omega \big(    \beta  \lambda_{\nu}^2 -  \omega  \lambda_{\nu} \big) \\ & \times \cos \omega (t-t') + O(\hbar^3) \label{eq:clclmainsem}
\end{align}
that corresponds to the term quadratic in $\lambda_{\nu}$ in the second of Eqs. \eqref{eq:finalclass}.
Note that while the effect of the additional force $F_{\nu,d}$ (see Eq. \eqref{eq:avgforce}) comes from the last term of Eq. \eqref{eq:bigactdispcontmain},
the second to last term gives a contribution proportional to $K_{\nu} \frac{\lambda_{\nu}}{\beta_{\nu}}$ that is absent in Eq. \eqref{eq:finalclass}, indeed it coincides with the first line of Eq. \eqref{eq:finalclass2}. 
The mismach between the classical limit of Eq. \eqref{eq:bigactdispcontmain} and \eqref{eq:finalclass} is expected, since the latter do not contain the energetic contributions pumped in the reservoir by the initial quench.
In Appendix \ref{app:worksource} we conclude the calculations in the displaced case and show the full correspondence between \eqref{eq:finalclass} + \eqref{eq:finalclass2} and the classical limit of Eq. \eqref{eq:bigactdispcontmain}. In the case of squeezing, we show the calculations of the quantum/classical correspondence only of a part of the action (the noise kernel) in App. \ref{sec:squeezKeld}.
 A recap of all the main results of the last two sections (Secs. \ref{sec:fullquant} and \ref{sec:classic}) is given in Fig. \ref{fig:6}.
%7
\section{Non-equilibrium symmetries of the MGF}
\label{sec:propsym}
In this chapter we discuss the symmetries of the MGF, in particular the ones related to thermodynamic consistency, such as the fluctuation dissipation relation and the fluctuation theorem \cite{kurchan2001, bochkov1977general, Gaspard, Espositoreview, campisi2011}. 
\subsection{Classical vs quantum action and energy conservation}
In Eq.~\eqref{eq:finalclass} the dependence on the counting fields ($\boldsymbol{\lambda_B}$, $\lambda_S$) is quadratic, as opposed to Eq. \eqref{eq:bigactdispcontmain} in which the action contains terms proportional to higher powers of $\lambda_{\nu}$ that disappear only in the classical limit.
There is a symmetry behind the simple structure of \eqref{eq:finalclass}, the {\it conservation of energy} at the trajectory level.
Indeed, when we can ignore the contribution to the action due to the explicit time derivatives (i.e. when the coupling is weak, or sufficiently slowly driven, $\dot{\chi}_{\nu} \approx 0$) the action \eqref{eq:finalclass}
becomes invariant under the following transformation
\begin{align} \label{eq:shift}
\lambda_{\nu} \rightarrow \lambda_{\nu} + \bar{\lambda} , \quad \; \lambda_S \rightarrow \lambda_S + \bar{\lambda}.
\end{align}
A translation of all the counting fields by the same amount $\bar{\lambda}$ corresponds to a shift of the action proportional to a factor $\Delta E_S (0,\tau) + \sum_{\nu} \Delta E_{\nu}(0,\tau)$
(see Eq. \eqref{eq:slambdasys}), hence, the invariance under Eq.~\eqref{eq:shift} corresponds to an energy conservation requirement at the level of the single trajectories.
The symmetry is broken if we consider the contribution of the initial displacement and squeezing (see Eq. \eqref{eq:finalclass2}).
In the quantum case the symmetry is broken anyway (even without initial quench), this can be checked by considering also $\hat{H}_S$ in the TPEM scheme and generalizing Eq. \eqref{eq:quantampl2} to the case in which $\lambda_S \neq 0$.
This is a consequence of the fact that energy is not defined at the level of the single quantum path integral trajectories. There is a way to define energy at the trajectory level only in certain cases, for instance in the secular regime for master equations (see \cite{soret2022thermodynamic}, \cite{kosloff2021}).
\subsection{Fluctuation theorem}
\label{sec:fluctuationtheo}
The FT is a general symmetry that holds in the hypothesis that the system and the reservoirs are initialized in a Gibbs state as in Eq. \eqref{eq:initial} with $\hat{\rho}_S = \frac{e^{-\beta_S \hat{H}_S}}{Z_S(0)}$.
When the Hamiltonian is time reversal invariant the theorem can be expressed as a relation between the MGF and its time-reversed counterpart \cite{andrieux2009fluctuation, andrieux2007fluctuation,Esposito2010, soret2022thermodynamic}
\begin{equation} \label{eq:fluctgen}
    M(\lambda_S, \boldsymbol{\lambda_B}, \tau) = M^R(-\lambda_S- \beta_S, -\boldsymbol{\lambda_B}- \boldsymbol{\beta_B}, \tau) \frac{Z_S(\tau)}{Z_S(0)},
\end{equation}
where $Z_S(\tau)$ is the equilibrium partition function of the system at the final time $\tau$.
The time-reversed generating function can be obtained, in the quantum case, by reversing the explicit time dependences $t\rightarrow \tau-t$ and by applying the time reversal operator $\Theta$ to the initial state (see, for example, \cite{andrieux2009fluctuation}).
In the setting in which the reservoirs are squeezed and displaced, it is straightforward to prove a fluctuation theorem if we adopt the point of view of Sec. \ref{sec:fluct}, where the squeezing and displacement are seen as a preliminary time evolution that affects the reservoirs, initially prepared at equilibrium.
Given this, a small difference from the usual \eqref{eq:fluctgen} is due to the squeeze/displacement Hamiltonians not being time reversal invariant. This property can be taken into account by properly inverting the driving parameters,
in complete analogy to what happens when proving the FT in presence of magnetic fields
\cite{andrieux2009fluctuation, barbier2018microreversibility}.
After explicitly specifying the parametric dependence by the squeezing and displacement parameters, we can finally write (see Appendix \ref{app:fluctsqdisp})
\begin{align}   \nonumber
\frac{Z_S(0)}{Z_S(\tau)} M(\lambda_S, \boldsymbol{\lambda_B},\tau, \{\alpha_{\nu,k}, r_{\nu,k}\}_{\nu,k}) =
 \\  M^R(-\lambda_S- \beta_S, -\boldsymbol{\lambda_B}- \boldsymbol{\beta_B}, \tau, \{-\alpha^*_{\nu,k}, -r^*_{\nu,k}\}_{\nu,k}), \label{eq:ftsqdis}
\end{align}
where we explicited the squeezing and displacement parameters that define the initial quench.
We can discuss the FT also at the level of the path integral action, i.e. at the level of single trajectories, similarly to what is done in the classical case where FT can be derived at the level of stochastic trajectories using detailed balance \cite{majewski1984detailed,alicki1976detailed,fagnola2008detailed,agarwal1973open} or (when many sources of noise are present) local detailed balance~\cite{Rao2018} conditions.
To compute the time-reversed generating function, we should use, for each reservoir, a contour that is obtained by reflecting $\gamma^{\nu}$ in Fig.~\ref{fig:4} around the $t = \frac{\tau}{2}$ axis \cite{Cavina2023}.
After doing this reflection, the lower branch becomes anti time-ordered, while the upper branch becomes time-ordered. Thus, the reflection causes an inversion between the forward branch $\gamma_-$ and the backward branch $\gamma_+$ (see also \cite{Whitney2018}).
At the level of the action \eqref{eq:bigactdisp22}, in the absence of squeezing, the replacement $\gamma_- \leftrightarrow \gamma_+$ produces an exchange of the GF components $G^{+,-}_{\nu}$ and $G^{-,+}_{\nu}$, that using Eq. \eqref{eq:rotgf} leads to $G^{q,cl}_{\nu} \leftrightarrow G^{cl,q}_{\nu}$ in Eq. \eqref{eq:bigactdispcontmain}.
The original action can be restored by transforming the counting field $\lambda_{\nu} \rightarrow - \lambda_{\nu} - \beta_{\nu}$, this is easy to prove by direct calculation (by replacing $\lambda \rightarrow - \lambda -\beta $ in Eq. \eqref{eq:4GFs}).
When squeezing and displacement are present, it is more difficult to reason at the level of the action since explicit time dependences on the contour appear (see Eqs. \eqref{eq:piecewisquee2} and \eqref{eq:piecewisquee3}). In this case, to prove FT at the level of the action it would be crucial to include the contributions of $\mathcal{G}^{M, \downarrow}_{\nu}, \mathcal{G}^{\downarrow, \downarrow}_{\nu}, \mathcal{G}^{M,M}_{\nu}$. 
\subsection{Fluctuation-dissipation relations}
For a system in contact with an environment at equilibrium, the fluctuation dissipation relation (FDR) links the noise and the dissipation kernel  \cite{kubo1966fluctuation} respectively given by Eqs. \eqref{eq:corr1}, \eqref{eq:corr2}.
The relation can be proved directly from Eq. \eqref{eq:corr2}, since the derivative of $C^{(1)}_{\nu}$ is proportional to $C^{(2)}_{\nu}$ with a proportionality factor given by $\beta_{\nu}$.
An analogous symmetry exists in the quantum action (in the case with $\boldsymbol{\lambda_B} = 0$) for which we have (directly from Eqs. \eqref{eq:4GFs1.0})
\begin{align} \notag
 & \frac{d}{dt} G^{q,q}_{\nu}(\omega,t,t')  = - i \omega  \coth(\frac{\hbar \beta_{\nu} \omega}{2}) \\ & \times  [G^{q,cl}_{\nu}(\omega,t,t') - G^{cl,q}_{\nu}(\omega,t,t')]. \label{eq:finalFDR}
\end{align}
The equality above reduces to its classical counterpart in terms of $C_{\nu}^{(1)},
C_{\nu}^{(2)}$ in the classical limit.
The FDR is the consequence of a more fundamental symmetry of the Schwinger-Keldysh action that characterizes equilibrium systems \cite{aron2018non,yeo2019symmetry,sieberer2015thermodynamic}, see, for instance, Eq.~(69) in \cite{aron2018non}.
While the relation \eqref{eq:finalFDR} retains its validity for displaced reservoirs (in which the correlation functions are the same as at equilibrium), it is violated in the presence of squeezing \footnote{Displaced reservoirs are still out of equilibrium: the displacement induces an effective driving of the system Hamiltonian that goes beyond the noise and dissipation associated with a thermal reservoir. This effect is fully encoded in the effective driving term (see also Eq. \eqref{eq:motionnew}), while the noisy part of the evolution remains unchanged.}.
In the classical case, this can be easily checked by comparing $\mathcal{C}^{(2)}_{\nu}$ in Eq. \eqref{eq:modcorr2} with its equilibrium counterpart $C^{(2)}_{\nu}$ in Eq. \eqref{eq:corr2}.
A comment is mandatory about the relation between the FDR and FTs discussed in Sec. \ref{sec:fluctuationtheo}.
In classical systems the FDR is taken as a hypothesis to prove FT in a broad category of scenarios (among the others, the Langevin equation for a system in contact with a single ohmic reservoir, see, for example, \cite{kurchan1998fluctuation,seifert2012stochastic}).
The fact that in Sec. \ref{sec:fluctuationtheo} we demonstrated that the FT holds in the NECL while the FDR is violated might initially appear counterintuitive. However, it is important to note that in the scenarios mentioned above, heat accounts solely for the energy lost by the system to the reservoir (corresponding to 
$\Delta E_{\nu}$
  in Eq. \eqref{eq:heatclass}).
In contrast, our setting requires including the energy expended to drive the reservoir 
$\nu$ out of equilibrium in the heat calculation (as shown in Eqs. \eqref{eq:defheat} and \eqref{eq:momclass2}) to ensure compliance with the FT. Conversely, the statistics of 
$\Delta E_\nu $
  do not satisfy a FT  (see also \cite{yadalam2022counting}), precisely because the FDR fails to hold for squeezed reservoirs.
  We conclude by noting that, although the FDR does not hold for non-equilibrium reservoirs, it is still possible to derive exact inequalities linking the dissipation and noise kernels. These results lead to a fluctuation–dissipation inequality, which states that quantum fluctuations are bounded from below by quantum dissipation \cite{fleming2013nonequilibrium}.
\subsection{Other symmetries of the Green's function components}
The presence of the counting fields $\lambda_{\nu}$ determines the appearance of other GF components in addition to the one commonly considered in open quantum systems for equilibrium reservoirs $G^{-,-}_{\nu}$, $G^{+,-}_{\nu}$, $G^{-,+}_{\nu}$, $G^{+,+}_{\nu}$.
These new components are obtained by selecting the arguments of $G_{\nu}$ (Eq. \eqref{eq:GFmain}) on $\gamma_{\uparrow}^{\nu}, \gamma_{\downarrow}^{\nu}$, for instance, obtaining $G^{\pm,\uparrow}_{\nu}$, $G^{\pm, \downarrow}_{\nu}$, $G^{\uparrow,\pm}_{\nu}$, $G^{\uparrow,\downarrow}_{\nu}$, etc.
If the time duration of the protocol is a multiple of the characteristic period of motion of the oscillating mode of the reservoir, that is, $\tau = \frac{2 \pi}{\omega}$, we have
\begin{equation} \label{eq:periodsymm}
    G_{\nu}^{\uparrow,\uparrow}(\omega) + 
    G_{\nu}^{\downarrow,\downarrow}(\omega) - G_{\nu}^{\downarrow,\uparrow}(\omega)  - 
    G_{\nu}^{\uparrow,\downarrow}(\omega) = 0,
\end{equation}
where we omitted the time dependence for ease of notation. Eq. \eqref{eq:periodsymm} can be proved easily in the case without squeeze by doing the explicit calculations on the components \eqref{eq:components3}. Eq. \eqref{eq:periodsymm} has the same shape of another well-known symmetry that holds when we do not perform any measurement of the energy of the reservoir $\nu$ ($\lambda_{\nu}=0$) i.e. \cite{Kamenev}
\begin{equation} \label{eq:kamensymm}
    G_{\nu}^{+,+}(\omega) + 
    G_{\nu}^{-,-}(\omega) - G_{\nu}^{+,-}(\omega)  - G_{\nu}^{-,+}(\omega) = 0.
\end{equation}
This symmetry is related to the conservation of probability in the reservoirs, and it is broken when $\lambda_{\nu} \neq 0$ since the inside of the trace in Eq. \eqref{eq:genmom} is not normalized.
See Eq. \eqref{eq:classiGcl} for an explicit expression of $G_{\nu}^{cl,cl}$, the fact that $G_{\nu}^{cl,cl} =0$ for $\lambda_{\nu} = 0$ can be easily verified.

\section{Conclusions}
We studied the thermodynamics of the NECL, a non-equilibrium version of the CL model in which a particle is coupled to squeezed and displaced reservoirs.
This framework can be used to give a self-consistent description of work exchanges in driven thermal engines, by representing the external driving as a collection of displaced sources.
%
%
%A particularly relevant application concerns a refined description of energy consumption and its fluctuations in driven quantum-computing architectures. In these platforms, qubit control is typically implemented via semiclassical microwave pulses, which can be modeled as externally driven, out-of-equilibrium electromagnetic fields. 
Our approach provides a microscopic framework to quantify both the energetic cost of such driving protocols and the associated fluctuations, going beyond phenomenological descriptions.
We proved that, to conciliate the non-equilibrium nature of the reservoirs with the fluctuation theorem, the energy that is used to squeeze and displace the reservoirs has to be explicitly taken into account: the sole energy flows between the system and the reservoirs do not lead to a statistics satisfying the FT, and this is strictly connected with the violation of the fluctuation-dissipation relation for squeezed reservoirs. 
To compute the full energy statistics, we devised a new way to represent the energetic corrections due to squeezing and displacement by using generalized Hamiltonians on modified Keldysh contours.
This approach allows us to bridge the gap between the two-point measurement framework and the common techniques
used in stochastic thermodynamics for classical systems, where
the energy flows are defined at the trajectory level. 
Due to its generality, the framework presented here can be used as a paradigm to study
the fluctuating thermodynamics of several out of equilibrium setups in which the reservoirs are Gaussian.
From a practical point of view, this encompassess a variety of relevant physical platforms, including quantum optics (e.g., lasing cavities) and superconducting architectures driven by resonators \cite{blais2021circuit}.
Future applications could also include the study of fluctuating thermodynamics in quantum computing devices, in which gates are often realized by coupling the qubits to auxiliary non-equilibrium degrees of freedom \cite{blais2007quantum,liu2006scalable}.
By not relying on specific approximations, the present treatment extends beyond the conventional approaches based on weak coupling and secular regimes, and could be used as a starting point to assess the role of non-Markovian effects in non-equilibrium quantum thermodynamics.

\section*{Acknowledgments} VC acknowledges financial support by MUR (Ministero dell’Universit{\`a} e della Ricerca) through the PNRR MUR project PE0000023-NQSTI. 

\bibliographystyle{quantum}
\bibliography{bibliography}

\clearpage
\begin{appendix}
\begin{widetext}
\section{The action of displacement and squeezing operators} 
\label{app:squeez}
Here we summarize some of the properties of the displacement and squeezing operators that we are going to use to derive our results.
We start by computing the action of a displacement operator on the ladder operators $\hat{a}, \hat{a}^{\dag}$
\begin{equation}
\hat{D}^{\dagger}(\alpha) \hat{a} \hat{D}(\alpha) = \hat{a} + \alpha; \quad \quad  \hat{D}^{\dagger}(\alpha) \hat{a}^{\dag} \hat{D}(\alpha) = \hat{a} + \alpha^*,
\end{equation}
Given that, for a generic oscillator with frequency $\omega$ and mass $m$ we have $\hat{X} = \sqrt{\frac{\hbar}{2 m \omega}}(\hat{a} + \hat{a}^{\dagger}) $ and $\hat{P} = i \sqrt{\frac{\hbar m \omega}{2}} (\hat{a}^{\dagger} - \hat{a}) $ and we can derive
\begin{equation}
  \hat{D}^{\dagger}(\alpha) \hat{X} \hat{D}(\alpha)  = 
  \hat{X} + \sqrt{\frac{2\hbar}{ m \omega}} Re(\alpha); \quad \quad   \hat{D}^{\dagger}(\alpha) \hat{P} \hat{D}(\alpha)  = 
  \hat{P} + \sqrt{2\hbar m \omega} Im(\alpha).
\end{equation}
The relations above can be used to compute the effect of the displacement operator on the Hamiltonian of a harmonic oscillator
\begin{equation}
  \hbar \omega \hat{D}^{\dag}(\alpha) [\hat{a}^{\dag} \hat{a} +\frac{1}{2}] \hat{D}(\alpha) =    \hbar \omega [(\hat{a}^{\dag} + \alpha^*) (\hat{a} + \alpha) +\frac{1}{2}],
\end{equation}
while in terms of the $\hat{X},\hat{P}$ operators we have
\begin{equation}
\hat{D}^{\dag}(\alpha) [\frac{1}{2} m \omega^2 \hat{X}^2  + \frac{1}{2 m} \hat{P}^2] \hat{D}(\alpha)  = 
[\frac{1}{2} m \omega^2 (\hat{X} - L)^2  + \frac{1}{2 m} (\hat{P}+Y)^2]  \label{eq:dispapppp}
\end{equation}
where we introduced, as in the main text, $L = - \sqrt{\frac{2 \hbar}{m \omega}} Re(\alpha)$, and the momentum shift $Y =   \sqrt{2\hbar m \omega} Im(\alpha)$.
From the definition \eqref{eq:squeeze} with the squeeze parameter $r = |r| e^{i \theta}$ we obtain the transformation of the creation and annihilation operators under the action of a squeeze operator
\begin{equation}
    \hat{S}^{\dagger}(r) \hat{a} \hat{S}(r) = 
    \cosh(|r|) \hat{a} - e^{i \theta} \sinh(|r|) \hat{a}^{\dag} ; \quad 
     \hat{S}^{\dagger}(r) \hat{a}^{\dagger} \hat{S}(r) = 
    \cosh(|r|) \hat{a}^{\dagger} - e^{-i \theta} \sinh(|r|) \hat{a}.
\end{equation}
The equations above can be used to calculate the action of the squeeze operator on $\hat{X}$ and $\hat{P}$. Using the relations between $\hat{X}$, $\hat{P}$ and $\hat{a}$, $\hat{a}^{\dag}$ we can derive
\begin{align} \notag
  \hat{S}^{\dagger}(r) \hat{X} \hat{S}(r) & = \cosh(|r|) \hat{X} - \cos(\theta) \sinh(|r|) \hat{X} -
  \frac{1}{m\omega } \sin(\theta) \sinh(|r|) \hat{P}, \\
   \hat{S}^{\dagger}(r) \hat{P} \hat{S}(r) & = \cosh(|r|) \hat{P} + \cos(\theta) \sinh(|r|) \hat{P} - 
  m\omega \sin(\theta) \sinh(|r|) \hat{X},
\end{align}
The squeezed reservoir Hamiltonian is given by
\begin{align} \notag
 \hbar \omega & \hat{S}^{\dag}(r) [\hat{a}^{\dag} \hat{a} +\frac{1}{2}] \hat{S}(r) = 
 \hbar \omega \big\{ [\cosh(|r|) \hat{a}^{\dag} - e^{i\theta} \sinh(|r|) \hat{a}][\cosh(|r|) \hat{a} - e^{-i\theta} \sinh(|r|) \hat{a}^{\dag}] +\frac{1}{2} \big\} \\ \notag
 & = \hbar \omega \big[ \cosh(|r|)^2 \hat{a}^{\dag} \hat{a} + \sinh(|r|)^2 \hat{a} \hat{a}^{\dag} - e^{i\theta} \cosh(|r|) \sinh(|r|) \hat{a}^2 - e^{-i \theta} \cosh(|r|) \sinh(|r|) \hat{a}^{\dag \, 2} +\frac{1}{2}  \big] \\
 &  =
 \hbar \omega \big\{\cosh(2|r|) \hat{a}^{\dag} \hat{a} -\frac{1}{2} \sinh(2|r|)[e^{i\theta} \hat{a}^2 + e^{-i \theta} \hat{a}^{\dagger \, 2}]  +\frac{1}{2}\cosh(2|r|)\big\}.\label{app:squeezlast}
\end{align}
where we used $[\hat{a},\hat{a}^{\dag}]=1$ and the hyperbolic duplication formulas.
On many occasions in this manuscript, we will prefer the expression of the Hamiltonian in terms of $\hat{X}$ and $\hat{P}$. For this sake, consider the identities 
\begin{align} 
 \hbar \omega \big(\hat{a}^{\dag} \hat{a} +\frac{1}{2}\big) & = \frac{1}{2} m \omega^2 \hat{X}^2  + \frac{1}{2 m} \hat{P}^2, \quad \quad  \hat{a}^2 
 = \frac{m \omega}{2 \hbar} \hat{X}^2 -\frac{1}{2 \hbar m\omega} \hat{P}^2 + \frac{i}{2 \hbar}(\hat{X}\hat{P} +\hat{P}\hat{X}), 
\end{align}
that replaced in the last of equations \eqref{app:squeezlast} give
\begin{gather} \notag
 \hat{S}^{\dag}(r) [\frac{1}{2} m \omega^2 \hat{X}^2  + \frac{1}{2 m} \hat{P}^2] \hat{S}(r) = \cosh(2 |r|) [\frac{1}{2} m \omega^2 \hat{X}^2  + \frac{1}{2 m} \hat{P}^2] \\ + \sinh(2|r|) \{\cos(\theta)[\frac{1}{2m} \hat{P}^2 - \frac{1}{2} m \omega^2 \hat{X}^2] 
   + \frac{\omega}{2} \sin\theta(\hat{X}\hat{P} +\hat{P}\hat{X})\},
\end{gather}
Note that in the case of zero phase $\theta=0$ we have $|r|=r$ and the equation above simplifies to 
\begin{equation}
     \hat{S}^{\dag}(r) [\frac{1}{2} m \omega^2 \hat{P}^2  + \frac{1}{2 m} \hat{X}^2] \hat{S}(r) = e^{2r} \frac{1}{2 m} \hat{P}^2 + e^{-2r} \frac{m \omega^2}{2} \hat{X}^2 . \label{eq:squeezapppp}
\end{equation}
Finally, let us consider the case where we apply a squeezing operation followed by a displacement.
Using Eq. \eqref{eq:squeezapppp} and Eq. \eqref{eq:dispapppp} we have, for real $r$ and $\alpha$,
\begin{align} \notag
   \hat{D}^{\dag}(\alpha)  \hat{S}^{\dag}(r) [\frac{1}{2} m \omega^2 \hat{X}^2  + \frac{1}{2 m} \hat{P}^2] \hat{S}(r) \hat{D}(\alpha) & =  \hat{D}^{\dag}(\alpha)\big[ e^{2r} \frac{1}{2 m} \hat{P}^2 + e^{-2r} \frac{m \omega^2}{2}  \hat{X}^2 \big]  \hat{D}(\alpha) \\ 
   & = \big[ e^{2r} \frac{1}{2 m} \hat{P}^2 + e^{-2r} \frac{m \omega^2}{2}  (\hat{X}+L)^2 \big] 
   . 
   \label{eq:squeezedispapppp}
\end{align}

\section{Work source limit for a system in contact with a displaced reservoir mode} \label{app:balreservoir}
It is convenient to divide the Hamiltonian \eqref{eq:hampart}
into two contributions, one relative to the free evolution of the system and the reservoir, $\hat{H}_S + \hat{H}_{4,1}$, and one relative to the coupling between the two, $\chi_4(t) \hat{V}_{4,1}$.
Introducing the interaction picture relative to the coupling Hamiltonian, we have $
\hat{U}(t,0) = \hat{U}^{(0)}(t,0) \tilde{U}(t,0) $
    where 
    \begin{equation}
    \hat{U}^{(0)}(t,0) = e^{ - \frac{i t}{\hbar}\big(  \frac{m_{4,1} \omega_{4,1}^2 \hat{X}_{4,1}^2}{2} + \frac{\hat{P}_{4,1}^2}{2 m_{4,1}} + \hat{H}_S \big)   }
    \end{equation}
    is the unperturbed evolution operator and
$\tilde{U}(t,0)  = \mathcal{T}\{ e^{-\frac{i}{\hbar} \int_0^t \tilde{H}(s) ds}  \}$, with $\tilde{H}$ being the Hamiltonian in the interaction picture
\begin{equation}
\tilde{H}(t) = \chi_4(t) e^{\frac{i t  \hat{H}_S }{\hbar}} V_{4,1}(\hat{X})  e^{-\frac{i t  \hat{H}_S }{\hbar}} \big[ \hat{X}_{4,1} \cos \omega_{4,1} t - \frac{\hat{P}_{4,1}}{m_{4,1} \omega_{4,1} } \sin \omega_{4,1} t \big].
\end{equation}
 We are interested in the evolution of a system coupled to the displaced mode, i.e. 
 \begin{equation} \label{eq:appdispcalc} \hat{\rho}(t) = 
  \hat{U}^{(0)}(t,0) \tilde{U}(t,0) \hat{\rho}_S(0) \otimes \hat{D}^{\dagger}(\alpha_{4,1}) \hat{\rho}_{4,1}(0^-) \hat{D}(\alpha_{4,1})  \tilde{U}^{\dagger}   (t,0)\hat{U}^{(0) \,\dagger}(t,0)
  \end{equation}
  where $\hat{\rho}_{4,1}(0^-)$ is a Gibbs state of the mode, with inverse temperature $\beta_4$.
  Inserting $\hat{D}^{\dagger}(\alpha_{4,1}) \hat{D}(\alpha_{4,1})$ in between $\hat{U}^{(0)}$ and $\tilde{U}$ and at the beginning and at the end of Eq. \eqref{eq:appdispcalc} we can express the time evolution inside the trace in terms of a displaced Hamiltonian operator.
  We want the displacement to change the center of the Gaussian packet to $L_{4,1}$, so we choose $L_{4,1} =- \sqrt{\frac{2 \hbar}{m_{4,1} \omega_{4,1}}} Re \alpha_{4,1} $ and obtain
  \begin{equation} \label{eq:dispint}
\tilde{H}_D(t) = \hat{D}(\alpha_{4,1}) \tilde{H}(t) \hat{D}^{\dagger}(\alpha_{4,1}) = 
\chi_4(t) e^{\frac{i t  \hat{H}_S }{\hbar}} V_{4,1}(\hat{X})  e^{-\frac{i t  \hat{H}_S }{\hbar}} \big[ (\hat{X}_{4,1} + L_{4,1} ) \cos \omega_{4,1} t - \frac{\hat{P}_{4,1}}{m_{4,1} \omega_{4,1} } \sin \omega_{4,1} t \big].
  \end{equation}
  The corresponding unitary evolution operator $\tilde{U}_D(t,0)$ appears in Eq. \eqref{eq:appdispcalc} that (after taking the trace over the reservoir) reduces to
  \begin{align} \nonumber
  \hat{\rho}_S(t) & = Tr_{E}[   \hat{U}^{(0)}(t,0) \hat{D}^{\dagger}(\alpha_{4,1}) \tilde{U}_D(t,0) \hat{\rho}_S(0) \otimes  \hat{\rho}_{4,1}(0^-) \tilde{U}^{\dagger}_D(t,0)  \hat{D}(\alpha_{4,1})   \hat{U}^{(0) \,\dagger}(t,0)    ] \\ & = 
   \hat{U}^{(0)}_S(t,0) Tr_E[   \tilde{U}_D(t,0) \hat{\rho}_S(0) \otimes  \hat{\rho}_{4,1}(0^{-}) \tilde{U}^{\dagger}_D(t,0)   ]     \hat{U}^{(0) \,\dagger}_S(t,0), \label{eq:appdyn1}
  \end{align}
  where $\hat{U}^{(0)}_S(t,0)$ is the evolution operator generated by the sole system Hamiltonian.
The Hamiltonian $\tilde{H}_D(t)$ can be divided into two parts, the one proportional to $L_{4,1}$ and the rest. 
Taking this second part as a perturbation, we can further decompose the unitary evolution as $ \tilde{U}_D(t,0)  =  \tilde{U}_D^{(0)}(t,0)  \tilde{\tilde{U}}_D(t,0)$ where
\begin{equation}
\tilde{U}_D^{(0)}(t,0) = \mathcal{T} \big\{  e^{- \frac{i \chi_4 L_{4,1}}{\hbar} \int_0^t  e^{\frac{i s \hat{H}_S }{\hbar}} V_{4,1}(\hat{X})  e^{-\frac{i s  \hat{H}_S }{\hbar} }\cos \omega_{4,1} s
ds}  \big\}, \label{eq:UD0.1}
\end{equation}
where we also neglected the time dependence in the switching function.
After noting that $\tilde{U}_D^{(0)}(t)$ does not depend on reservoir operators, we can bring it outside of the trace, obtaining 
  \begin{align}  \label{eq:appdisp1}
  \hat{\rho}_S(t) =
   \hat{U}^{(0)}_S \tilde{U}^{(0)}_D Tr_E[   \tilde{\tilde{U}}_D \hat{\rho}_S(0) \otimes  \hat{\rho}_{4,1}(0^-) \tilde{\tilde{U}}^{\dagger}_D   ]    \tilde{U}^{(0)\,\dagger}_D \hat{U}^{(0) \,\dagger}_S \approx 
    \hat{U}^{(0)}_S \tilde{U}^{(0)}_D \hat{\rho}_S(0)   \tilde{U}^{(0)\,\dagger}_D \hat{U}^{(0) \,\dagger}_S ,
  \end{align}
where we omitted the time dependence inside the evolution operators (for ease of notation), and assumed that the contribution of $\tilde{\tilde{U}}_D(t,0) $ is small and kept only the zeroth order correction. 
To ensure that the force induced by the displaced reservoir is of the same order as the one generated by the free potential of the particle $V(\hat{X})$ we choose $\chi_4 L_{4,1} V_{4,1}(\hat{X}) \sim V(\hat{X})$ (where, maintaining the discussion at an informal level, with "$\sim$" we mean that the two operators have expectation values of the same order of magnitude on the vectors of the system Hilbert space of interest).
In this case, the evolution reduces to the one described in \eqref{eq:motionnew}.
Otherwise, note that $\tilde{\tilde{U}}_D(t,0)$ is generated by a Hamiltonian
\begin{equation}
 \tilde{\tilde{H}}_D(t) = 
\chi_4  \tilde{U}^{(0)\, \dagger}_D(t,0) e^{\frac{i t  \hat{H}_S }{\hbar}} V_{4,1}(\hat{X})  e^{-\frac{i t  \hat{H}_S }{\hbar}}  \tilde{U}^{(0)}_D(t,0) \big[ \hat{X}_{4,1}  \cos \omega_{4,1} t - \frac{\hat{P}_{4,1}}{m_{4,1} \omega_{4,1} } \sin \omega_{4,1} t \big], \label{eq:tilttilt}
\end{equation}
and if we want to compute further perturbative corrections we can keep the first terms in the expansion of $\tilde{\tilde{U}}_D(t,0)$ as a time-ordered exponential. 
Note that the first correction nullifies, since $Tr_E[\hat{X}_{4,1} \hat{\rho}_{4,1}(0^-)] = 
Tr_E[\hat{P}_{4,1}\hat{\rho}_{4,1}(0^-)] = 0$ because $\hat{\rho}_{4,1}(0^-)$ is quadratic in $\hat{X}_{4,1}$ and $\hat{P}_{4,1}$.
Using the formulas for the second moments, we find
\begin{align} \notag
& Tr [\hat{X}_{4,1}^2 \hat{\rho}_{4,1}(0^-) ]  = Tr [ \frac{\hat{P}_{4,1}^2}{\omega_{4,1}^2 m_{4,1}^2} \hat{\rho}_{4,1}(0^-) ]  =  \frac{\hbar}{2 m_{4,1} \omega_{4,1}} \coth \big(\frac{\beta_4 \hbar \omega_{4,1}}{2} \big); \\
& Tr [ \hat{X}_{4,1} \frac{\hat{P}_{4,1}}{m_{4,1} \omega_{4,1}} \hat{\rho}_{4,1}(0^-) ]  = \frac{i \hbar }{2  m_{4,1} \omega_{4,1}}. \label{eq:variances}
\end{align}
 For future convenience we define
$\tilde{\tilde{O}}_{4,1}(s) = \tilde{U}^{(0)\, \dagger}_D(s,0) e^{\frac{i s  \hat{H}_S }{\hbar}} V_{4,1}(\hat{X})  e^{-\frac{i s  \hat{H}_S }{\hbar}}  \tilde{U}^{(0)}_D(s,0)  $ 
and $\tilde{O}_{4,1}(s) =  e^{\frac{i s  \hat{H}_S }{\hbar}} V_{4,1}(\hat{X})  e^{-\frac{i s  \hat{H}_S }{\hbar}}  $ for the system operators appearing in Eq. \eqref{eq:tilttilt} and \eqref{eq:dispint}, respectively.
One of the contributions to the second-order correction in $\tilde{\tilde{H}}_D(t)$ containing two $\hat{X}_{4,1}$ operators reads

\begin{equation} \label{O!11}
\Delta^{(2)} = - \frac{\chi_4^2}{\hbar^2} 
\big[ \int_0^{t} \tilde{\tilde{O}}_{4,1}(s) \cos \omega_{4,1} s ds 
 \big ] \hat{\rho}_S(0) \big[ \int_0^{t} \tilde{\tilde{O}}_{4,1}(s') \cos \omega_{4,1} s'ds' \big] Tr[ \hat{X}_{4,1}^2 \hat{\rho}_{4,1}].
\end{equation}

We can compare the quantity above with the contribution proportional to $L_{4,1}$ of \eqref{eq:dispint}, which reads

\begin{equation} \label{O!1}
\Delta^{(1)} = - \frac{i \chi_{4} L_{4,1}}
{\hbar} \int_0^t ds [\tilde{O}_{4,1}(s), \hat{\rho}_S(0)] ds .
\end{equation}

%Recalling that $\chi_4 L_{4,1} V_{4,1}(\hat{X}) \sim V(\hat{X})$, we can prove that
%the ratio between the trace norm of the corrections becomes proportional to the ratio between the length scales of the fluctuations and $L_{4,1}$ 
%\begin{equation} 
%\frac{||\Delta^{(2)}||}{||\Delta^{(1)}||}  =
%f(t) \frac{ \hbar \coth\big( \frac{\beta_4 \hbar \omega_{4,1}}{2}\big)}{2 m_{4,1} \omega_{4,1} L_{4,1}^2 },
%\end{equation}
%where $f(t)$ is a dimensionless time-dependent function containing the contribution of the time integrals in Eqs. \eqref{O!1}. and \eqref{O!11}.
Comparing the standard deviation of the fluctuating coupling constant associated to Eq. \eqref{O!1} with the prefactor in Eq. \eqref{O!11} we obtain the condition \eqref{eq:bigbig}. We conclude that if the displacement is sufficiently large if compared to the range of the position fluctuations (induced by both thermal and quantum effects) of the Gaussian mode, we can neglect their noisy contribution to the dynamics of the system and we are left with an evolution that is effectively unitary.
This implies that $\Delta S_S$ is negligible during the process, as required in Section \ref{sec:avgtherm} for pure driving sources. To complete the discussion, we compute the energy and entropy variations of the reservoir mode.
For this sake, we derive the reduced density matrix of the reservoir mode, which reads 
\begin{align} \nonumber
  \hat{\rho}_E(t) & = Tr_{S}[   \hat{U}^{(0)} \hat{D}^{\dagger}(\alpha_{4,1}) \tilde{U}_D \hat{\rho}_S(0) \otimes  \hat{\rho}_{4,1}(0^-) \tilde{U}^{\dagger}_D  \hat{D}(\alpha_{4,1})   \hat{U}^{(0) \,\dagger}    ] \\ & = 
   \hat{U}^{(0)}_E \hat{D}^{\dagger}(\alpha_{4,1}) Tr_S[   \tilde{U}_D \hat{\rho}_S(0) \otimes  \hat{\rho}_{4,1}(0^-) \tilde{U}^{\dagger}_D   ]    \hat{D}(\alpha_{4,1})  \hat{U}^{(0) \,\dagger}_E 
  \notag \\ & =  \hat{U}^{(0)}_E \hat{D}^{\dagger}(\alpha_{4,1}) Tr_S[   \tilde{\tilde{U}}_D \hat{\rho}_S(0) \otimes  \hat{\rho}_{4,1}(0^-) \tilde{\tilde{U}}^{\dagger}_D   ]   \hat{D}(\alpha_{4,1})   \hat{U}^{(0) \,\dagger}_E 
   , \label{eq:appdyn2} \end{align} 
where $\hat{U}_E^{(0)}(t,0)$ is the propagator of the free evolution of the reservoir.
To compute the final value of the Von Neumann entropy, we can use its invariance under unitary transformations and write
\begin{equation} \label{app:entropy}
    S_E(t) = - Tr_E[  Tr_S[   \tilde{\tilde{U}}_D(t,0) \hat{\rho}_S(0) \otimes  \hat{\rho}_{4,1}(0^-) \tilde{\tilde{U}}^{\dagger}_D(t,0)   ]  \log  Tr_S[   \tilde{\tilde{U}}_D(t,0) \hat{\rho}_S(0) \otimes  \hat{\rho}_{4,1}(0^-) \tilde{\tilde{U}}^{\dagger}_D(t,0)   ]     ].
\end{equation}
The only effect that contributes to the Von Neumann entropy change is due to the coupling with the system and is induced by the Hamiltonian \eqref{eq:tilttilt}.
 At first order in the expansion of Eq. \eqref{app:entropy} we find
\begin{align} \notag
Tr_S[...] & = \hat{\rho}_{4,1}(0^-) - \frac{i}{\hbar} \int_0^t Tr_S\big\{ [ \chi_4  \tilde{U}^{(0)\, \dagger}_D(s,0) e^{\frac{i s  \hat{H}_S }{\hbar}} V_{4,1}(\hat{X})  e^{-\frac{i s  \hat{H}_S }{\hbar}}  \tilde{U}^{(0)}_D(s,0) \times \\ & \times \big( \hat{X}_{4,1}  \cos \omega_{4,1} s -  \frac{\hat{P}_{4,1}}{m_{4,1} \omega_{4,1} } \sin \omega_{4,1} s \big) , \hat{\rho}_S(0) \otimes  \hat{\rho}_{4,1}(0^-)  ] \big\}  \label{eq:pertappnuovo}
\end{align}
that can be seen as a perturbation of the reduced density matrix of the reservoir.
Our goal is to expand the Von Neumann entropy in the perturbation above; for this, we can use the equation 
\begin{equation} \label{eq:expentrop}
S(\hat{A} + \epsilon \hat{B}) 
= S(\hat{A}) - \epsilon Tr[\hat{B} \log \hat{A}] - \epsilon^2 \int_0^{\infty} Tr \big\{ s \big(\hat{A} +  s\big)^{-1}  \big[\hat{B} \big(\hat{A} +  s\big)^{-1}\big]^2
\big\} ds + O(\epsilon^3),
\end{equation}
where $\epsilon$ is a small parameter.
By replacing $\hat{A}$ and $\hat{B}$ in the equations above with $\hat{\rho}_{4,1}(0^-)$ and the perturbation (see Eq. \eqref{eq:pertappnuovo}), respectively, we find that the first correction to the Von Neumann entropy nullifies.

The variation of energy is, instead, order zero in the perturbation; indeed, we have
\begin{align} \notag
E_{4,1}(t) - E_{4,1}(0) & = Tr_E [ \hat{H}_{4,1} \hat{D}^{\dagger}(\alpha_{4,1}) Tr_S[   \tilde{\tilde{U}}_D(t,0) \hat{\rho}_S(0) \otimes  \hat{\rho}_{4,1}(0^-) \tilde{\tilde{U}}^{\dagger}_D(t,0)   ]   \hat{D}(\alpha_{4,1})  ] \\ & - 
Tr_E [ \hat{H}_{4,1} \hat{D}^{\dagger}(\alpha_{4,1}) \hat{\rho}_{4,1}(0^-) \hat{D}(\alpha_{4,1})  ].  \label{eq:energdispapp}
\end{align}
Remembering that
\begin{equation} \label{eq:disphamappfin}
     \hat{D}(\alpha_{4,1})  \hat{H}_{4,1} \hat{D}^{\dagger}(\alpha_{4,1})  = 
     \frac{\omega_{4,1}^2 m_{4,1}}{2} \big(\hat{X}_{4,1} + L_{4,1}\big)^2 +  \frac{1}{2 m_{4,1}} \hat{P}_{4,1}^2, 
\end{equation}
considering the contribution of the energy proportional to $ L_{4,1} \hat{X}_{4,1}$ and replacing it into the Eq. \eqref{eq:energdispapp} after expanding $\tilde{\tilde{U}}_D(t,0)$ we remain with a product $\chi_4 L_{4,1}$, which is therefore finite.
In this discussion, we did not consider the cost of switching on the coupling between the system and the reservoir. This coupling contribution can be safely disregarded since we are assuming $\chi_4$ to be very small.
 We conclude the section with a proof of Eq. \eqref{eq:phenpow}.
In the weak coupling limit we can expand Eq. \eqref{eq:energdispapp} at first order in $\chi_{4}$
\begin{equation} \label{eq:newworkfin}
E_{4,1}(t) - E_{4,1}(0) = -\frac{i}{\hbar} \int_0^t Tr\big[\hat{D}(\alpha_{4,1}) \hat{H}_{4,1} \hat{D}^{\dag}(\alpha_{4,1})   [ \tilde{\tilde {H} }_D(s), \hat{\rho}_S(0) \otimes  \hat{\rho}_{4,1}(0)   ]   \big] ds.
\end{equation}
For the displaced Hamiltonian, we use Eq. \eqref{eq:disphamappfin} and consider the dominant contribution (that is the one linear in $L_{4,1}$ since the quadratic one gives a $0$ contribution becoming the trace of a commutator in Eq. \eqref{eq:newworkfin}). We obtain
\begin{equation}
E_{4,1}(t) - E_{4,1}(0) = - \frac{i \omega^2_{4,1} m_{4,1} L_{4,1} }{\hbar} \int_0^t Tr\big[ \hat{X}_{4,1}  [ \tilde{\tilde {H} }_D(s), \hat{\rho}_S(0) \otimes  \hat{\rho}_{4,1}(0)   ]   \big] ds.
\end{equation}
The integrand of the equation above is the rate of the energy flow in the mode of the displaced reservoir, let us call it $\dot{E}_{4,1}$.
We now replace Eq. \eqref{eq:tilttilt} inside the equation above, obtaining
\begin{equation}
\dot{E}_{4,1} =   \frac{ \omega_{4,1}  \chi_4 L_{4,1} i}{\hbar} Tr\big[ \hat{X}_{4,1}  [ \tilde{U}^{(0)\, \dagger}_D(t,0) e^{\frac{i t  \hat{H}_S }{\hbar}} V_{4,1}(\hat{X})  e^{-\frac{i t  \hat{H}_S }{\hbar}}  \tilde{U}^{(0)}_D(t,0) \hat{P}_{4,1} , \hat{\rho}_S(0) \otimes  \hat{\rho}_{4,1}(0)   ]   \big] \sin \omega_{4,1} t,
\end{equation}
where we mantained only the contribution of Eq. \eqref{eq:tilttilt} proportional to $\hat{P}_{4,1}$ since the contribution of the position operator vanishes in the commutator. Using the third of Eqs. \eqref{eq:variances} we have
\begin{equation}
\dot{E}_{4,1} =   - \omega_{4,1}  \chi_4 L_{4,1} Tr_S [  \tilde{U}^{(0)\, \dagger}_D(t,0) e^{\frac{i t  \hat{H}_S }{\hbar}} V_{4,1}(\hat{X})  e^{-\frac{i t  \hat{H}_S }{\hbar}}  \tilde{U}^{(0)}_D(t,0) \hat{\rho}_S(0) ] \sin \omega_{4,1} t,
\end{equation}
that by rearranging the terms inside the trace reduces to \begin{equation}
\dot{E}_{4,1} =   - \omega_{4,1}  \chi_4 L_{4,1} Tr_S [   V_{4,1}(\hat{X})   \hat{\rho}_S(t) ] \sin \omega_{4,1} t,
\end{equation}
that is exactly Eq. \eqref{eq:phenpow} in the main text.
\section{Work source limit for a system in contact with a squeezed reservoir mode} \label{app:squeeztherm}
In the case of a squeezed reservoir mode, we can reason in an analogous way to what we did for the displaced case, but we have to replace the displacement with a squeeze. The interaction Hamiltonian (see also Eq. \eqref{eq:dispint}) in this case reads 
\begin{equation} \label{eq:squeezinter}
\tilde{H}_{SQ}(t) = \hat{S}(r_{3,1}) \tilde{H}(t) \hat{S}^{\dag}(r_{3,1}) = 
\chi_3(t) e^{\frac{i t}{\hbar} \hat{H}_S} 
 V_{3,1}(\hat{X}) e^{-\frac{i t}{\hbar} \hat{H}_S} [\hat{X}_{3,1} e^{+r_{3,1}} \cos \omega_{3,1} t -  \frac{\hat{P}_{3,1} e^{-r_{3,1}}}{m_{3,1} \omega_{3,1}} \sin \omega_{3,1} t ],
\end{equation}
in which we assumed $r_{3,1}$ to be real.
If the squeeze is sufficiently large, we can neglect the term proportional to $\hat{P}_{3,1}$ and write
\begin{equation}
\tilde{U}_{SQ}(t,0) = \hat{S}(r_{3,1}) \tilde{U}(t) \hat{S}^{\dag}(r_{3,1}) = 
\mathcal{T} \{ e^{  -\frac{i}{\hbar} \int_0^t V_{3,1}[\tilde{X}(s)] \hat{X}_{3,1} e^{r_{3,1}} \cos \omega_{3,1} s \, ds}  \},
\end{equation}
where $  V_{3,1}[\tilde{X}(s)]  =  e^{\frac{i t}{\hbar} \hat{H}_S} 
 V_{3,1}(\hat{X}) e^{-\frac{i t}{\hbar}\hat{H}_S}$.
The evolution of the system density matrix in the interaction picture is now given by
\begin{align} \nonumber
Tr_B[ \tilde{U}_{SQ}(t,0) \hat{\rho}_{S}(0) \otimes \hat{\rho}_{3,1}(0^-)\tilde{U}_{SQ}^{\dagger}(t,0)] \\ 
&\hspace{-4cm}= Tr_B \big[\big( \mathds{I} - \hat{X}_{3,1} \frac{i}{\hbar} \int_0^t\tilde{\Gamma}(s)  -\frac{\hat{X}_{3,1}^2}{2  \hbar^2} \iint_0^t \mathcal{T}\{\tilde{\Gamma}(s) \tilde{\Gamma}(s') \}  ds ds' + ...\big)  \hat{\rho}_{S}(0) \otimes \hat{\rho}_{3,1}(0^-) \nonumber\\
&\hspace{-4cm} \times\big(  \mathds{I} + \hat{X}_{3,1} \frac{i}{\hbar} \int_0^t\tilde{\Gamma}(s)  - \frac{\hat{X}_{3,1}^2}{2  \hbar^2} \iint_0^t \bar{\mathcal{T}}\{\tilde{\Gamma}(s) \tilde{\Gamma}(s') \}  ds ds' + ...    \big) \big], \label{eq:appexp}  \end{align}
where we introduced $\tilde{\Gamma}(s) = \chi_3(s)  V_{3,1}[\tilde{X}(s)] e^{r_{3,1}} \cos \omega_{3,1} s
$ and $\bar{\mathcal{T}}$ is the anti time-ordering operator.
The expansion above can be written more conveniently by using the Keldysh contour
\begin{align} \notag
Tr_B[ \tilde{U}_{SQ}(t,0) \hat{\rho}_{S}(0) \otimes \hat{\rho}_{3,1}(0^-)\tilde{U}_{SQ}^{\dagger}(t,0)]\\
&\hspace{-4cm} = Tr_B [\mathcal{T}_K \big\{ \big(  \mathds{I} - \hat{X}_{3,1} \frac{i}{\hbar} \int_{\gamma_K
  } \tilde{\Gamma}(s) ds  - \frac{\hat{X}_{3,1}^2}{2  \hbar^2} \iint_{\gamma_K} \tilde{\Gamma}(s) \tilde{\Gamma}(s') \, ds \, ds' + ...    \big) \hat{\rho}_{S}(0) \otimes \hat{\rho}_{3,1}(0^-) \big\}  ] \nonumber
  \\ &\hspace{-4cm} =
  Tr_B [\mathcal{T}_K \big\{ 
  \big( \sum_{k=0}^{\infty} \frac{(- \frac{i}{\hbar}  \int_{\gamma_K} \, ds \tilde{\Gamma}(s) \hat{X}_{3,1} )^k}{k!}  \big) \hat{\rho}_{S}(0) \otimes \hat{\rho}_{3,1}(0^-) \big\}  ], \label{eq:appexp2}
\end{align}
where $\mathcal{T}_K$ orders the operators according to their position on the Keldysh contour $\gamma_K$.
The averages of the powers of $\hat{X}_{3,1}$ over the initial state of the mode can be easily computed, and we have
\begin{equation}
Tr[\hat{X}_{3,1}^{2k} \hat{\rho}_{3,1}(0^-)] = (2k-1)!! \langle\hat{X}_{3,1}^2\rangle^k,
\end{equation}
where $!!$ denotes the double factorial and $\langle\hat{X}_{3,1}^2\rangle$ the expectation value in Eq. \eqref{eq:variances}.
The expectation value for odd values of $k$ is $0$.
Using $(2k-1)!! = \frac{(2k)!}{2^k k!} $ we obtain
\begin{align}  \notag
  Tr_B[ \tilde{U}_{SQ} \hat{\rho}_{S}(0) \otimes \hat{\rho}_{3,1}(0^-)\tilde{U}_{SQ}^{\dagger}]
 & = \mathcal{T}_K \big \{ \big[ \sum_{k=0}^{\infty}  \frac{  \big( -\int_{\gamma} \tilde{\Gamma}(s)   \, ds \big)^{2k} \langle \hat{X}_{3,1}^2 \rangle^k }{k! 2^k \hbar^{2k}}   \big] \hat{\rho}_{S}(0) \big\}
\\ & = \mathcal{T}_K \big \{  \exp\big[\frac{  \big( -\int_{\gamma} \tilde{\Gamma}(s)   ds \big)^2  \langle \hat{X}_{3,1}^2 \rangle} {2\hbar^{2}}   \big] \hat{\rho}_{S}(0) \big\}
 . \label{eq:bigexp}
 \end{align}
We note that this time  ordered exponential is the same we would obtain by considering the solution of the stochastic Von Neumann equation
\begin{equation}
    \dot{\hat{\rho}}^{(c)}_S(t) = - i\frac{\xi}{\hbar} [\tilde{\Gamma}(s),\hat{\rho}_S^{(c)}(t)],
\end{equation}
where the standard density matrix can be obtained by averaging over the noise $\langle \hat{\rho}_S^{(c)}(t) \rangle = \hat{\rho}_S(t) $ and the noise is Gaussian with
$\langle \xi^2 \rangle = \langle \hat{X}^2_{3,1} \rangle $.

To compute the energy and entropy variations, we can use a similar approach to the displaced case.
We calculate the evolution operator generated by $\tilde{H}_{SQ}(t)$ in \eqref{eq:squeezinter}, by decomposing it in a contribution relative to the dominant term, proportional to $\hat{X}_{3,1}$, and the rest
\begin{equation} \label{eq:squeezappcomput}
\tilde{U}_{SQ}(t,0) = \tilde{U}_{SQ}^{(0)}(t,0) 
\mathcal{T}\big\{ \exp \big[\frac{i}{\hbar} \int_0^t \chi_3(s) \tilde{U}_{SQ}^{(0) \, \dagger}(s,0)  
 \frac{\hat{P}_{3,1}}{m_{3,1} \omega_{3,1}} V_{3,1}[\tilde{X}(s)] e^{-r_{3,1}} \tilde{U}_{SQ}^{(0)}(s,0) \sin \omega_{3,1} s ds \big]\big\},
\end{equation}
 where $\tilde{U}_{SQ}^{(0)}(t,0)$ is the time-ordered exponential generated by $\tilde{\Gamma}(t) \hat{X}_{3,1}$.
Using that $[\hat{P}_{3,1}, f(\hat{X}_{3,1})] = - i \hbar f'(\hat{X}_{3,1})$ for every function $f$, we derive
\begin{equation}
\hat{P}_{3,1} \tilde{U}_{SQ}^{(0)}(t,0) = 
\tilde{U}_{SQ}^{(0)}(t,0) \hat{P}_{3,1} - i \hbar \frac{\partial}{\partial \hat{X}} \tilde{U}_{SQ}^{(0)}(t,0).
\end{equation}
The time-ordered exponential in \eqref{eq:squeezappcomput} becomes

\begin{align} \notag
\mathcal{T}\big\{ \exp \big[\frac{i}{\hbar} \int_0^t \chi_3(s) \tilde{U}_{SQ}^{(0) \, \dagger}(s,0)  
 \frac{\hat{P}_{3,1}}{m_{3,1} \omega_{3,1}} V_{3,1}[\tilde{X}(s)] e^{-r_{3,1}} \tilde{U}_{SQ}^{(0)}(s,0) \sin \omega_{3,1} s ds \big]\big\} \\ \notag =
 \mathcal{T}\big\{ \exp \big[\frac{i}{\hbar} \int_0^t \chi_3(s) \tilde{U}_{SQ}^{(0) \, \dagger}(s,0)  
  V_{3,1}[\tilde{X}(s)] e^{-r_{3,1}} \tilde{U}_{SQ}^{(0)}(s,0) \frac{\hat{P}_{3,1}}{m_{3,1} \omega_{3,1}} \sin \omega_{3,1} s ds 
   \\  
  + \frac{1}{m_{3,1} \omega_{3,1}} \int_0^t \chi_3(s) \tilde{U}_{SQ}^{(0) \, \dagger}(s,0)  
  V_{3,1}[\tilde{X}(s)] e^{-r_{3,1}}  \frac{\partial}{\partial \hat{X}} \tilde{U}_{SQ}^{(0)}(s,0)  \sin \omega_{3,1} s ds 
   \big]\big\}.
\end{align}
The time-ordered exponential above is sum of two contributions of order $\chi_{3} e^{- r_{3,1}}$, that is second order in the expansion parameter $\chi_3 \approx e^{- r_{3,1}}$.
We conclude that the contact with the system induces a variation of the state of the reservoir that is second order in the expansion parameter and thus contributes to the Von Neumann entropy only at second order (see Eq. \eqref{eq:expentrop}).
Conversely, the correction to the average final energy of the reservoir is order zero in the small parameter, since it is the expectation value of the Hamiltonian
\begin{equation}
     \hat{S}(r_{3,1})  \hat{H}_{3,1} \hat{S}^{\dagger}(r_{3,1})  = 
     \frac{ e^{- 2 r_{3,1}} \omega_{3,1}^2 m_{3,1}}{2} \hat{X}_{3,1}^2 +  \frac{e^{2 r_{3,1}}}{2 m_{3,1}} \hat{P}_{3,1}^2, 
\end{equation}
that diverges as $e^{2 r_{3,1}}$.

\section{Contour action for a system in contact with displaced and squeezed reservoirs}
\label{app:disp}
After summing over $\nu$ to consider the contributions of all the reservoirs and integrating by parts, the action \eqref{eq:quantact} with the replacements \eqref{eq:piecewisquee} and \eqref{eq:piecewisquee2} can be written as
 \begin{align} \notag
 \mathcal{S} & =  \int_{\gamma} \big\{ \frac{m}{2} \dot{x}^2(z) -V[x(z)] \big\} dz -\sum_{\nu,k} \int_{\gamma^{\nu}} \big\{ \chi^{(\gamma)}_{\nu}(z) x_{\nu,k}(z) V^{(\gamma)}_{\nu,k}[x(z)] dz  \\ & + \frac{m_{\nu,k}}{2} \ddot{x}_{\nu,k}(z) x_{\nu,k}(z) + \frac{m_{\nu,k} \omega_{\nu,k}^2}{2} x_{\nu,k}^2(z)\big\} dz +  i \hbar \sum_{\nu,k}(\beta_{\nu} + \lambda_{\nu})N_{\nu,k}, \label{eq:quantactapp}
 \end{align}
that can also be written as 
\begin{align} \notag
    \mathcal{S} &  =  \int_{\gamma} \big\{ \frac{m}{2} \dot{x}^2(z) -V[x(z)] \big\} dz -\sum_{\nu,k} \int_{\gamma^{\nu}} \big\{ \chi^{(\gamma)}_{\nu}(z) x_{\nu,k}(z) V^{(\gamma)}_{\nu,k}[x(z)] dz \\
    & + \frac{\omega_{\nu,k} m_{\nu,k}}{2} x_{\nu,k}(z)  G^{-1}_{\nu, k}(z,z') x_{\nu,k}(z') \big\} dz + i \hbar \sum_{\nu,k} (\beta_{\nu} + \lambda_{\nu})N_{\nu,k},  \label{eq:actiondispapp}
\end{align}
where $ \omega_{\nu,k} G^{-1}_{\nu, k}(z) = \partial_z^2 + \omega^2_{\nu,k}$.
To perform the path integration on the variables of the reservoirs, we multiply the action times $\frac{i}{\hbar}$ and then do the Gaussian path integral associated to such variables. Let us focus on the contribution of a single mode $\nu,k$, we obtain the following contribution to the action after the integration 
\begin{equation}
   \bar{\mathcal{S}}_{\nu,k} = \frac{1}{2 m_{\nu,k} \omega_{\nu,k}} \iint_{\gamma^{\nu}}   \chi^{(\gamma)}_{\nu}(z)V^{(\gamma)}_{\nu,k}[x(z)]   G_{\nu,k}(z,z')  \chi^{(\gamma)}_{\nu}(z') V^{(\gamma)}_{\nu,k}[x(z')] 
   dz dz'  + i \hbar (\beta_{\nu} + \lambda_{\nu}) N_{\nu,k}. \label{eq:effective}
\end{equation}
To understand the role of the different components in the path integration above, it is convenient to separate the contributions of the different branches of the contour. 
Let us remember that, for any given reservoir $\nu$ we have
\begin{align}
  \int_{\gamma^{\nu}} dz & = \int_{\gamma_-} dz + \int_{\gamma_+} dz + \int_{\gamma^{\nu}_{\uparrow}} dz + \int_{\gamma^{\nu}_{\downarrow}} dz + \int_{\gamma^{\nu}_M} dz.
\end{align}
Using this notation, we can transform the action \eqref{eq:effective} in
\begin{equation} \label{eq:effective2}
   \bar{\mathcal{S}}_{\nu,k} =  \frac{1}{2 m_{\nu,k} \omega_{\nu,k}} \sum_{l,j} \int_{\gamma^{\nu}_l} \int_{\gamma^{\nu}_j}   \chi^{(\gamma)}_{\nu}(z)V^{(\gamma)}_{\nu,k}[z]   G_{\nu,k}(z,z')  \chi^{(\gamma)}_{\nu}(z') V^{(\gamma)}_{\nu,k}[z']    dz dz' + i \hbar (\beta_{\nu} + \lambda_{\nu}) N_{\nu,k}
\end{equation}
where the indices $l,j$ run over the different components of the contour in Fig. \ref{fig:1}, that is $l,j = -,+,\uparrow, \downarrow,M$.
Following the main text, we assume that the only difference in the coupling potential between different modes of the same reservoir is a multiplicative factor $V_{\nu,k}[x(z)] = c_{\nu,k} V_{\nu}[x(z)]$. 
Isolating the contributions due to the forward and backward branches in Eq. \eqref{eq:effective2} and defining, as in the main text, $v_{\nu}^{j}(t) \equiv \chi_{\nu}^{j}(t)V_{\nu}[x_j(t)]  $, 
we finally obtain
\begin{align}  \notag
   \bar{\mathcal{S}}_{\nu,k} & =   \sum_{l,j=\pm} \int_{0}^{\tau}  \int_{0}^{\tau}   \frac{ c^2_{\nu,k}} {2 m_{\nu,k} \omega_{\nu,k}} v^l_{\nu}(t) \big[(l \times j) G^{l,j}_{\nu,k}(t,t') \big]  v^{j}_{\nu}(t') dt dt' \\  \notag &
    +i  \sum_{l=\pm} \int_{0}^{\tau} dt \int_{\hbar \lambda_{\nu}}^{0} d\zeta'    \frac{\omega_{\nu,k}  l L_{\nu,k} }{2} c_{\nu,k} v^l_{\nu}(t) \big[     G^{l,\downarrow}_{\nu,k}(t,\zeta')   + G^{\downarrow,l}_{\nu,k}(\zeta',t)  \big]  
   \\ \notag 
    & +i  \sum_{l=\pm} \int_{0}^{\tau} dt \int_{0}^{-\hbar \beta_{\nu}} d\zeta'    \frac{\omega_{\nu,k}  l L_{\nu,k} }{2}c_{\nu,k} v^l_{\nu}(t) 
    \big[     G^{l,M}_{\nu,k}(t,\zeta')   + G^{M,l}_{\nu,k}(\zeta',t)  \big] 
    \\ \notag
    & -  \int^{-\hbar \beta_{\nu}}_{\hbar \lambda_{\nu}} d\zeta \int^{-\hbar \beta_{\nu}}_{\hbar \lambda_{\nu}}d\zeta'    \frac{m_{\nu,k} \omega^3_{\nu,k}  L^2_{\nu,k} }{2} \big[    \Theta(\zeta) \Theta(\zeta')   G^{\downarrow,\downarrow}_{\nu,k}(\zeta,\zeta')   +   \Theta(-\zeta) \Theta(\zeta')   G^{M,\downarrow}_{\nu,k}(\zeta,\zeta') 
 \\ 
 & +  \Theta(-\zeta) \Theta(-\zeta')   G^{M,M}_{\nu,k}(\zeta,\zeta') 
    +   \Theta(\zeta) \Theta(-\zeta')   G^{\downarrow,M}_{\nu,k}(\zeta,\zeta')    
    \big] + i \hbar (\beta_{\nu} + \lambda_{\nu}) N_{\nu,k}.\label{eq:bigactdispapp}
%    \\
  %  & +   
% \int_{0}^{\hbar(\beta+\lambda)} \int_{0}^{\hbar(\beta+\lambda)}  d\zeta d\zeta'  \frac{m_{\nu,k} \omega_{\nu,k}^2 L^2_{\nu,k}}{2} g^{M,M}_{\nu,k}(\zeta,\zeta').
 \end{align}
 The last two lines of Eq. \eqref{eq:bigactdispapp} are equal to the factor $\mathcal{N}_{\nu}$ introduced in the main text.
 In the squeezed case the masses depend on the position of the variable $z$ on the contour. We do the same procedure as above for Eqs. \eqref{eq:dissipactsquee}, \eqref{eq:piecewisquee3} becomes
 \begin{align}  \notag
   \bar{\mathcal{S}}_{\nu,k} & =   \sum_{l,j=\pm} \int_{0}^{\tau}  \int_{0}^{\tau}   \frac{ c^2_{\nu,k}} {2 m_{\nu,k} \omega_{\nu,k}} v^l_{\nu}(t) \big[(l \times j) \mathcal{G}^{l,j}_{\nu,k}(t,t') \big]  v^{j}_{\nu}(t') dt dt' \\  \notag &
    +i  \sum_{l=\pm} \int_{0}^{\tau} dt \int_{\hbar \lambda_{\nu}}^{0} d\zeta'    \frac{\omega_{\nu,k}  l L_{\nu,k}    e^{- 2 r_{\nu,k}}}{2} c_{\nu,k} v^l_{\nu}(t) \big[   \mathcal{G}^{l,\downarrow}_{\nu,k}(t,\zeta')   +  \mathcal{G}^{\downarrow,l}_{\nu,k}(\zeta',t)  \big]  
   \\ \notag 
    & +i  \sum_{l=\pm} \int_{0}^{\tau} dt \int_{0}^{-\hbar \beta_{\nu}} d\zeta'    \frac{\omega_{\nu,k}  l L_{\nu,k} e^{- 2 r_{\nu,k}} }{2}c_{\nu,k} v^l_{\nu}(t) 
    \big[     \mathcal{G}^{l,M}_{\nu,k}(t,\zeta')   +     \mathcal{G}^{M,l}_{\nu,k}(\zeta',t)  \big] 
    \\ \notag
    & -  \int^{-\hbar \beta_{\nu}}_{\hbar \lambda_{\nu}} d\zeta \int^{-\hbar \beta_{\nu}}_{\hbar \lambda_{\nu}}d\zeta'    \frac{   e^{- 2 r_{\nu,k}}  m_{\nu,k} \omega^3_{\nu,k}  L^2_{\nu,k} }{2} \big[    \Theta(\zeta) \Theta(\zeta')  \mathcal{G}^{\downarrow,\downarrow}_{\nu,k}(\zeta,\zeta')   +   \Theta(-\zeta) \Theta(\zeta')   \mathcal{G}^{M,\downarrow}_{\nu,k}(\zeta,\zeta') 
 \\ 
 & +  \Theta(-\zeta) \Theta(-\zeta')   \mathcal{G}^{M,M}_{\nu,k}(\zeta,\zeta') 
    +   \Theta(\zeta) \Theta(-\zeta')   \mathcal{G}^{\downarrow,M}_{\nu,k}(\zeta,\zeta')    
    \big] + i \hbar (\beta_{\nu} + \lambda_{\nu})N'_{\nu,k}. \label{eq:bigactdispappsqueez}
%    \\
  %  & +   
% \int_{0}^{\hbar(\beta+\lambda)} \int_{0}^{\hbar(\beta+\lambda)}  d\zeta d\zeta'  \frac{m_{\nu,k} \omega_{\nu,k}^2 L^2_{\nu,k}}{2} g^{M,M}_{\nu,k}(\zeta,\zeta').
 \end{align}
 Equations \eqref{eq:bigactdisp} and \eqref{eq:bigactdisp22} are obtained from the ones above by using the symmetry of the Green's function $\mathcal{G}_{\nu,k}(z,z')= \mathcal{G}_{\nu,k}(z',z)$.
\section{Green's function of the harmonic oscillator on the contour} 
\label{app:Green}
The Green's function of the Harmonic oscillator with Hamiltonian $\hat{H} = \frac{1}{2 m} \hat{P}^2 + \frac{m \omega^2}{2} \hat{X}^2 $ with Kubo-Martin-Schwinger boundary conditions satisfies 
\begin{equation} \label{eq:kms}
    (\partial_z^2 + \omega^2) G(z,z') = \omega \delta(z-z'), \quad \quad 
 G(0,z')= G(-i \hbar \beta, z') , \quad \quad G'(0,z') =  G'(-i \hbar \beta, z'),    \quad \quad \end{equation}
where $G'$ denotes the derivative in the first argument of the Green's function and we dropped all the $\nu,k$ indices for ease of notation.
It is easy to verify that the solution is given by
Equation \eqref{eq:GFmain}, with $\omega_{\nu,k}$, $m_{\nu,k}$ replaced by $\omega,m$, where the domain $z,z'$ is given by the contour in Fig. \ref{fig:1}.

Using the theory of Sec. \ref{sec:greenreservoir} and Eq. \eqref{eq:defcompmain} we can extract the components of the Green's function \eqref{eq:GFmain}.
We stard from the components appearing in the standard Keldysh contour, by choosing $z \in  \gamma_{-}, \gamma_{+}$:
 \begin{align} \notag
 & G^{-,-}(t,t') = \frac{i}{2}\coth\big(\frac{\hbar \omega \beta}{2}\big) \cos\omega(t-t') +\big[\Theta(t-t') -\frac{1}{2} \big] \sin \omega(t-t'); \\ \notag
  & G^{+,+}(t,t') = \frac{i}{2}\coth\big(\frac{\hbar \omega \beta}{2}\big) \cos\omega(t-t') +\big[\Theta(t'-t) -\frac{1}{2} \big] \sin \omega(t-t'); \\ \notag
&  G^{+,-}(t,t') = \frac{i}{2}\coth\big(\frac{\hbar \omega \beta}{2}\big) \cos\omega(t-t' + i \hbar \lambda) + \frac{1}{2} \sin \omega(t-t'+i \hbar \lambda); 
\\ 
&  G^{-,+}(t,t') = \frac{i}{2}\coth\big(\frac{\hbar \omega \beta}{2}\big) \cos\omega(t-t' - i \hbar \lambda) - \frac{1}{2} \sin \omega(t-t'  - i \hbar \lambda). \label{eq:components1}
\end{align}
When the initial equilibrium state is represented using the Matsubara branch, that is, in contour of panel A in Fig, \ref{fig:5}, we can define Green's functions with arguments on $\gamma_M$.
The new Green's function components appearing in this case are
\begin{align} \notag
&  G^{-,M}(t,\zeta') = \frac{i}{2}\coth\big(\frac{\hbar \omega \beta}{2}\big) \cos\omega(t- i \zeta') -\frac{1}{2} \sin \omega(t- i \zeta');
\\ \notag
&  G^{M,-}(\zeta,t') = \frac{i}{2}\coth\big(\frac{\hbar \omega \beta}{2}\big) \cos\omega(i \zeta - t') +\frac{1}{2} \sin \omega(i \zeta - t');
\\ \notag
&  G^{+,M}(t,\zeta') = \frac{i}{2}\coth\big(\frac{\hbar \omega \beta}{2}\big) \cos\omega(t + i \hbar \lambda - i \zeta') -\frac{1}{2} \sin \omega(t + i \hbar \lambda - i \zeta');
\\  \notag
&  G^{M,+}(\zeta,t') = \frac{i}{2}\coth\big(\frac{\hbar \omega \beta}{2}\big) \cos\omega(i \zeta - t' - i \hbar \lambda) +\frac{1}{2} \sin \omega(i \zeta - t'- i \hbar \lambda);
\\ 
&  G^{M,M}(\zeta,\zeta') = \frac{i}{2}\coth\big(\frac{\hbar \omega \beta}{2}\big) \cos\omega(i \zeta - i\zeta') +\big[\Theta(\zeta'-\zeta) -\frac{1}{2} \big] \sin \omega(i \zeta - i\zeta'). \label{eq:components2}
\end{align}
Finally, in the presence of counting fields we add $\gamma_{\uparrow}$ and $\gamma_{\downarrow}$ to the picture (obtaining the contour in Fig. \ref{fig:4}) and we can define 
\begin{align} \notag
&  G^{M,\uparrow}(\zeta,\zeta') = \frac{i}{2}\coth\big(\frac{\hbar \omega \beta}{2}\big) \cos\omega(i \zeta - i\zeta' -\tau) +\frac{1}{2} \sin \omega(i \zeta - i\zeta' -\tau);
\\ \notag 
&  G^{\uparrow,\uparrow}(\zeta,\zeta') = \frac{i}{2}\coth\big(\frac{\hbar \omega \beta}{2}\big) \cos\omega(i \zeta - i\zeta') +\big[\Theta(\zeta-\zeta') -\frac{1}{2} \big] \sin \omega(i \zeta - i\zeta')
;
\\ \notag
&  G^{-,\uparrow}(t,\zeta) = \frac{i}{2}\coth\big(\frac{\hbar \omega \beta}{2}\big) \cos\omega(t- \tau - i \zeta)  -\frac{1}{2}  \sin \omega(t- \tau - i \zeta);
\\  \notag
&  G^{+,\uparrow}(t,\zeta) = \frac{i}{2}\coth\big(\frac{\hbar \omega \beta}{2}\big) \cos\omega(t + i \hbar \lambda - \tau - i \zeta)  +\frac{1}{2} \sin \omega(t + i \hbar \lambda- \tau - i \zeta);
\\ \notag
&  G^{-,\downarrow}(t,\zeta) = \frac{i}{2}\coth\big(\frac{\hbar \omega \beta}{2}\big) \cos\omega( t- i \zeta )  -\frac{1}{2} \sin \omega(t-i \zeta);
\\ 
&  G^{+,\downarrow}(t,\zeta) = \frac{i}{2}\coth\big(\frac{\hbar \omega \beta}{2}\big) \cos\omega(t+ i \hbar \lambda - i\zeta ) -\frac{1}{2} \sin \omega(t + i \hbar \lambda - i\zeta )
. \label{eq:components3}
 \end{align}
 In case we want to extract components of Eq. \eqref{eq:GFmain} we have to replace $\omega, m,\beta, \lambda$ in the list above respectively with $\omega_{\nu,k}, m_{\nu,k}, \beta_{\nu}, \lambda_{\nu}$.
\section{Corrections to Eq.~\eqref{eq:finalclass} due to initial displacement and squeezing}
\label{app:coordispsque}
The energy $\Delta E_{\nu}$ in Eq.~\eqref{eq:heatclass} does not take into account the contribution of initial displacement and squeezing.
The energy pumped into the reservoir mode $\nu,k $ by initially displacing a particle in position $x_{\nu,k}(0^-)$ is
\begin{equation} \label{eq:contendisp}
    \Delta E_{0,\nu}^{dp} = \frac{1}{2} m_{\nu,k} \omega_{\nu,k}^2 (\bar{x}_{\nu,k}(0^-) + L_{\nu,k})^2 -  \frac{1}{2} m_{\nu,k} \omega_{\nu,k}^2 \bar{x}_{\nu,k}^2(0^-)  = \frac{1}{2}   m_{\nu,k} \omega_{\nu,k}^2 L_{\nu,k}^2 + m_{\nu,k} \omega_{\nu,k}^2 \bar{x}_{\nu,k}(0^-) L_{\nu,k}.
\end{equation}
Differently from the quantum case, we are free to express the energy variation also in terms of the position of the reservoir mode {\it after} the displacement.
For every position $\bar{x}_{\nu,k}(0)$ resulting from the displacement, the original position was given by 
$\bar{x}_{\nu,k}(0) - L_{\nu,k}$, so that, in terms of this new variable, the energy variation reads
\begin{equation}
\Delta E_{0,\nu}^{dp} = 
\frac{1}{2} m_{\nu,k} \omega_{\nu,k}^2 
\bar{x}_{\nu,k}^2(0) 
-
 \frac{1}{2} m_{\nu,k} \omega^2_{\nu,k} (\bar{x}_{\nu,k}(0) - L_{\nu,k})^2 =
  -\frac{1}{2}   m_{\nu,k} \omega_{\nu,k}^2 L_{\nu,k}^2 + m_{\nu,k} \omega_{\nu,k}^2 \bar{x}_{\nu,k}(0) L_{\nu,k}.
\end{equation}
In the presence of squeezing, the equation above reduces to Eq. \eqref{eq:endispin} in the main text.
The squeeze operation is a quench of the reservoir massess. It can be implemented as the following transformation 
\begin{equation}
    \bar{x}_{\nu,k}(0) =  e^{ r_{\nu,k}} \bar{x}_{\nu,k}(0^-); \quad 
     \bar{p}_{\nu,k}(0) =  e^{- r_{\nu,k}} \bar{p}_{\nu,k}(0^-).
\end{equation}
This can be realized, for instance, by modulating the frequency of the reservoir mode during a transient (one possibility is to choose a new frequency $\omega_{\nu,k} e^{r_{\nu,k}}$ and let the mode evolve for a quarter of a period, then evolve with the normal frequency for three quarters of a period).
The energy exchanged during the process is 
\begin{equation}
    \Delta E_{0,\nu}^{sq} = \frac{1}{2} m_{\nu,k} \omega_{\nu,k}^2 \bar{x}^2_{\nu,k}(0) 
    + \frac{1}{2 m_{\nu,k}} \bar{p}^2_{\nu,k}(0) - 
    \frac{1}{2} m_{\nu,k} \omega_{\nu,k}^2 e^{- 2 r_{\nu,k}} \bar{x}^2_{\nu,k}(0) 
    -  \frac{1}{2 m_{\nu,k}} e^{2 r_{\nu,k}} \bar{p}^2_{\nu,k}(0).
\end{equation}
If squeezing and displacement are both present, 
the position of the particle before the quench is 
$x_{\nu,k}(0^-) = e^{- r_{\nu,k}} (x_{\nu,k}(0) - L_{\nu,k})$. Computing the energy difference between after and before the quench in this case we get Eqs. \eqref{eq:endispin}, \eqref{eq:ensquein}.
In the main text we found the action for the MGF \eqref{eq:momclass} and obtained the result in Eq. \eqref{eq:finalclass}.
To obtain an explicit result for \eqref{eq:momclass2}, we have to add the contributions due to the initial quench written above.
These contributions are expressed in terms of the initial positions and momenta of the reservoirs. We note that multiplying the MGF times the weight
$\exp(- \frac{1}{2} \lambda_{\nu} m_{\nu,k} \omega_{\nu,k}^2 L_{\nu,k}^2 + \lambda_{\nu} m_{\nu,k} \omega_{\nu,k}^2 \bar{x}_{\nu,k}(0) L_{\nu,k})$
is like modifying the initial displaced distribution as
\begin{align} \notag
& \exp \big[ - \frac{\beta_{\nu} m_{\nu,k} \omega^2_{\nu,k}}{2}  \big(\bar{x}_{\nu,k}(0) - L_{\nu,k} \big)^2 \big] \rightarrow \nonumber \\
&\hspace{2cm} \exp  \big[ - \frac{\beta_{\nu} m_{\nu,k} \omega^2_{\nu,k}}{2}  \big(\bar{x}_{\nu,k}(0) - L_{\nu,k} \big)^2 -\frac{\lambda_{\nu}}{2} m_{\nu,k} \omega_{\nu,k}^2 L_{\nu,k}^2 + \lambda_{\nu} m_{\nu,k} \omega_{\nu,k}^2 \bar{x}_{\nu,k}(0) L_{\nu,k} \big)\big] \nonumber
\\ &\hspace{2cm} =
\exp \big\{  -\frac{ \beta_{\nu} m_{\nu,k} \omega_{\nu,k}^2}{2} \big[\bar{x}_{\nu,k}(0) - \big(1+ \frac{\lambda_{\nu}}{\beta_{\nu}} \big) L_{\nu,k} \big]^2   \big\} \exp \big[
\frac{m_{\nu,k} \omega_{\nu,k}^2 L_{\nu,k}^2 }{2} \big( \lambda_{\nu} + \frac{ \lambda^2_{\nu}}{\beta_{\nu}} \big) \big]. \label{eq:reshift}
\end{align}
The exponent in the second exponential above is the term $\mathcal{M}_{\nu}$ in Eq. \eqref{eq:finalclass2}, while the first two lines in Eq. \eqref{eq:finalclass2} can be obtained by shifting $R_{\nu}' \rightarrow R_{\nu}' + \frac{\lambda_{\nu}}{\beta_{\nu}} F_{\nu,d}$ in the first of Eqs. \eqref{eq:finalclass}. This shift coincides with the one introduced in the last line of Eq. \eqref{eq:reshift}.
\section{Generalized Langevin equation}
\label{app:GLEQ}
In this appendix we study the energy statistics of a classical system coupled to an infinite amount of classical oscillators.
The Hamilton equations associated with the Hamiltonian \eqref{eq:hamclass} are
\begin{align}  \notag
\dot{p} &= - V'(x) - \sum_{\nu,k} \chi_{\nu} x_{\nu,k} V'_{\nu,k}(x), \quad \quad \dot{x} = \frac{p}{m},\\ 
\dot{p}_{\nu,k} & = - m_{\nu, k} \omega^2_{\nu,k} x_{\nu,k} - \chi_{\nu} V_{\nu,k}(x), \quad \quad \dot{x}_{\nu,k} = \frac{p_{\nu,k}}{m_{\nu,k}}. \label{eq:dynham}
\end{align}
The solution of the equations of motion for $p_{\nu,k}, x_{\nu,k}$ conditioned on a given trajectory of the system variables is given by Eqs. \eqref{eq:reservoirvar}. Therefore, we can replace these solutions within the first two equations in \eqref{eq:dynham}
to obtain the Langevin equation \eqref{eq:genlang}. 
The noise correlation function in Eq. \eqref{eq:corr1} can be obtained from the correlation between two environmental position variables
\begin{align} \notag
 &  \langle  \bar{x}_{\nu,k}(t) \bar{x}_{k',\nu}(t') \rangle \\ \notag & = \langle \sum_{k'} \sum_{k} \big[\bar{x}_{\nu,k}(0) \cos \omega_{\nu,k}t + \frac{\bar{p}_{\nu,k}(0)}{m_{\nu,k} \omega_{\nu,k}} \sin \omega_{\nu,k}t \big] \big[\bar{x}_{\nu,k'}(0) \cos \omega_{\nu,k'}t' + \frac{\bar{p}_{\nu,k'}(0)}{m_{\nu,k'} \omega_{\nu,k'}} \sin \omega_{\nu,k'}t' \big] \rangle  \\ & + \sum_{k'} \sum_{k} \int_{0}^{t'} ds' \int_{0}^{t} ds \frac{\chi_{\nu}(s)}{m_{\nu,k} \omega_{\nu,k}} V_{\nu,k}[x(s)] \sin{\omega_{k',\nu} (t-s) } \frac{\chi_{\nu}(s')}{m_{k',\nu} \omega_{k',\nu}} V_{\nu,k'}[x(s')] \sin{\omega_{k',\nu} (t'-s') } .
\end{align}
If the initial distribution of the reservoir is assumed to be a Gibbs state,
\begin{equation} \label{app:gibbs1}
    P_{\nu,k}(\bar{x}_{\nu,k}(0),\bar{p}_{\nu,k}(0))=\frac{1}{Z_{\nu,k}}e^{-\beta \big[ \frac{1}{2} m_{\nu,k} \omega_{\nu,k}^2 \bar{x}_{\nu,k}(0)^2 +  \frac{1}{2 m_{\nu,k}} \bar{p}_{\nu,k}(0)^2 \big]},
\end{equation}
we have
\begin{equation} \label{app:gibbs2}
\langle \bar{x}_{\nu,k}(0) \bar{x}_{\nu,k'}(0)  \rangle =  \frac{1}{m_{\nu,k} \omega_{\nu,k}^2 \beta_{\nu}} \delta_{k,k'} ,  \quad \quad
\langle \bar{p}_{\nu,k}(0) \bar{p}_{\nu,k'}(0)  \rangle =  \frac{m_{\nu,k}}{  \beta_{\nu}} \delta_{k,k'}.
\end{equation}
If we define new variables
\begin{equation}
    y_{\nu,k}(t) = \bar{x}_{\nu,k} + \int_{0}^{t} ds \frac{\chi_{\nu}(s)}{m_{\nu,k} \omega_{\nu,k}} V_{\nu,k}[x(s)] \sin{\omega_{\nu,k} (t-s) },
\end{equation}
and
\begin{equation}
    w_{\nu,k}(t) =  -\int_{0}^{t} ds \frac{\chi_{\nu}(s)}{m_{\nu,k} \omega_{\nu,k}} V_{\nu,k}[x(s)] \sin{\omega_{\nu,k} (t-s) },
\end{equation}
we have
\begin{equation}
  \langle  y_{\nu,k}(t) y_{\nu,k}(t') \rangle = 
 \sum_k \big[ \frac{\cos \omega_{\nu,k}t \cos \omega_{\nu,k}t'}{m_{\nu,k} \omega_{\nu,k}^2 \beta_{\nu}} +   \frac{\sin \omega_{\nu,k}t \sin \omega_{\nu,k}t' }{m_{\nu,k} \omega_{\nu,k}^2 \beta_{\nu}}  \big]= \sum_{k} \frac{\cos \omega_{\nu,k} (t-t') }{m_{\nu,k} \omega_{\nu,k}^2 \beta_{\nu}}.
\end{equation}

The force acting on the system from the reservoir $\nu$ after defining $V_{\nu,k}(x) = c_{\nu,k} V_{\nu}(x)$
\begin{equation}
F_{\nu,k}^{(S)} =
-\chi_{\nu} V'_{\nu}(x) \sum_k c_{\nu,k} (y_{\nu,k}(t) + w_{\nu,k}(t) ) 
\end{equation}
and defining 
$\xi_{\nu} = \sum_k c_{\nu,k} y_{\nu,k}$
we have 
\begin{equation}
    \langle \xi_{\nu}(t) \xi_{\nu}(t') \rangle = 
     \sum_{k} c^2_{\nu,k} \frac{\cos \omega_{\nu,k} (t-t') }{m_{\nu,k} \omega_{\nu,k}^2 \beta_{\nu}} .
\end{equation}
By introducing the spectral density defined in Eq. \eqref{spectraldensitymain}, the equation above becomes
%\begin{align} \label{spectral density 2}
 %   J_{\nu}(\omega)\equiv \frac{\pi}{2} \sum_k \frac{c_{\nu,k}^2}{m_{\nu,k} \omega_{\nu,k}} \delta\left(\omega-\omega_{\nu,k} \right),
%\end{align}
\begin{equation}
    \langle \xi_{\nu}(t) \xi_{\nu}(t') \rangle = 
  \int \frac{ 2 J_{\nu}(\omega) \cos \omega (t-t')}{\pi \beta_{\nu} \omega} d \omega,
\end{equation}
that is the classical noise kernel.
The force exerted by the reservoir $\nu$ on the system is then
\begin{equation}
F^{(S)}_{\nu,k} =
-\chi_{\nu} V'_{\nu}(x)\xi_{\nu} -\chi_{\nu} V'_{\nu}(x)R_{\nu},
\end{equation}
where $R_{\nu} = \sum_{k} c_{\nu,k} w_{\nu,k}$.
We have
\begin{equation}
R_{\nu} = - \sum_{k} c^2_{\nu,k} \int_{0}^{t} ds \frac{\chi_{\nu}(s)}{m_{\nu,k} \omega_{\nu,k}} V_{\nu}[x(s)] \sin{\omega_{\nu,k} (t-s) } = -\frac{2}{\pi} \int d\omega \int_{0}^t \chi_{\nu}(s) V_{\nu}[x(s)] J(\omega) \sin \omega(t-s).
\end{equation}
For the heat absorbed by the reservoir $\nu$ instead we have
\begin{align}
\int \sum_{k} F_{\nu,k}[\bar{x}(t), \boldsymbol{\bar{x}_B}(t)] \dot{\bar{x}}_{\nu,k}(t) 
&= \int -\chi_{\nu}(t) V_{\nu}[x(t)] \sum_{k} c_{\nu,k} (\dot{y}_{\nu,k}(t) + \dot{w}_{\nu,k}(t) ) \nonumber\\
&= + \int \frac{d}{dt} \{\chi_{\nu}(t) V_{\nu}[x(t)] \}\sum_{k} c_{\nu,k} ({y}_{\nu,k}(t) + w_{\nu,k}(t) ) 
 \end{align}
where we used the switching on/off property.
After using the definition of $\xi$ and $R_{\nu} = \sum_{k} c_{\nu,k} r_{\nu,k}$ we end up with
\begin{equation}
  \int \sum_{k} F^{ex}_{\nu,k}[\bar{x}(t), \boldsymbol{\bar{x}_B}(t)] \dot{\bar{x}}_{\nu,k}(t) =   +  \int  
  \frac{d}{dt} \{\chi_{\nu}(t) V_{\nu}[x(t)] \}(\xi_{\nu}(t) + R_{\nu}(t)).
\end{equation}
\subsection{Quenching the reservoirs masses}
\label{app:quencma}
We note that the strict equivalent of the quantum squeezing corresponds to a quench of the masses.
In this case, the masses appearing in equations \eqref{app:gibbs1} and \eqref{app:gibbs2} are different from the dynamical ones; we denote them by $M_{\nu,k} = e^{- 2 r_{\nu,k} } m_{\nu,k}$ and obtain
\begin{align} \notag
  \langle  y_{\nu,k}(t) y_{\nu,k}(t') \rangle & = 
 \sum_k \big[ \frac{\cos \omega_{\nu,k}t \cos \omega_{\nu,k}t' }{M_{\nu,k} \omega_{\nu,k}^2 \beta_{\nu}} +   \frac{M_{\nu,k}  \sin \omega_{\nu,k}t \sin \omega_{\nu,k}t' }{m^2_{\nu,k} \omega_{\nu,k}^2 \beta_{\nu}}  \big] \\ &= \sum_k \big[
\frac{M_{\nu,k} \cos \omega_{\nu,k} (t-t')}{m_{\nu,k}^2 \omega_{\nu,k}^2 \beta_{\nu}}  + \frac{M_{\nu,k} n_{\nu,k}}{ m^2_{\nu,k}  \omega_{\nu,k}^2 \beta_{\nu}}  \cos \omega_{\nu,k} t \cos \omega_{\nu,k} t'  \big]  \label{eq:classsqueezmass}  \end{align}
with 
$n_{\nu,k} = \big(\frac{m^2_{\nu,k}}{M^2_{\nu,k}} - 1 \big)
$.
We can define two modified spectral functions as follows
\begin{equation}
    \mathcal{J}_{\nu}(\omega)\equiv \frac{\pi}{2} \sum_k \frac{M_{\nu,k} c_{\nu,k}^2}{m^2_{\nu,k} \omega_{\nu,k}} \delta\left(\omega-\omega_{\nu,k} \right),
\end{equation}

\begin{equation}
    \Delta \mathcal{J}_{\nu}(\omega)\equiv \frac{\pi}{2} \sum_k n_{\nu,k} \frac{M_{\nu,k} c_{\nu,k}^2}{m^2_{\nu,k} \omega_{\nu,k}} \delta\left(\omega-\omega_{\nu,k} \right),
\end{equation}
and finally obtain the expression for the squeezed correlation function
\begin{equation}
\mathcal{C}_{\nu}^{(2)}(t,t') = 
\frac{2}{\pi} \int_{-\infty}^{\infty} \frac{1}{\beta \omega} \big[  \mathcal{J}_{\nu}(\omega) \cos \omega (t-t')+  \Delta \mathcal{J}_{\nu}(\omega) \cos \omega t \cos \omega t' \big] d\omega,
\end{equation}
that coincides with Eq. \eqref{eq:modcorr2}.
\section{Path integral and energy statistics for classical stochastic equations}
\label{app:classpath}
Let us consider a system coupled to many thermal reservoirs evolving under the action of Eq.~\eqref{eq:genlang}.
The corresponding Hamilton equations write
\begin{align}
m \dot{\bar{x}}(t) = \bar{p}(t), \quad \dot{\bar{p}}(t) = \mathcal{F}[\bar{x}(t),\boldsymbol{\xi_B}(t)],
\end{align}
 where $\mathcal{F}$ are the total forces acting on the system (that is, the right hand side of Eq. \eqref{eq:genlang}) and we used the overbar $\bar{\cdot}$ to denote a variable that satisfies the equations of motion.
After introducing a time slicing in $L$ intervals such that $t_n = \frac{n}{L} \tau = n \Delta $ with $n=0,...L-1$, we can discretize the Hamilton equations as 
\begin{gather} 
 \bar{x}(t_{n+1}) =  \bar{x}(t_{n}) + \frac{\bar{p}(t_n)}{m} \Delta, \quad \quad \bar{p}(t_{n+1}) = \bar{p}(t_n) + \mathcal{F}[\bar{x}(t_n),\boldsymbol{\xi_B}(t_n)] \Delta.
\label{eq:discretham} \end{gather}
To simplify the notation, we can introduce a phase space vector $\boldsymbol{q} = (x,p)$ and reduce the set of equations \eqref{eq:discretham} to a compact $  \boldsymbol{\bar{q}}(t_{n+1}) = 
\boldsymbol{\bar{q}}(t_{n}) + \Delta \mathcal{L}[\boldsymbol{\bar{q}}(t_{n})]  $
where $\mathcal{L}$ is a linear operator acting on $\boldsymbol{q}$ defined using Eq. \eqref{eq:discretham}.
The next step is to write the equations of motion as a constraint in an integral over a set of free fields
\begin{equation}
  f[\boldsymbol{\bar{q}}(t_{n+1})]= \int_{-\infty}^{\infty} d \boldsymbol{q}  
   \, f[\boldsymbol{q}] \delta\{
\boldsymbol{q} - \boldsymbol{\bar{q}}(t_n)  - \Delta \mathcal{L}[\boldsymbol{\bar{q}}(t_n)] \},
\end{equation}

where $f$ denotes a generic function of the trajectory.
Using the integral representation of the Dirac delta function, the equation above can also be expressed in terms of an auxiliary field $\boldsymbol{\eta}$ (up to a normalization factor that is irrelevant for our purposes) as
\begin{equation} \label{eq:phasedelta}
   f[\boldsymbol{\bar{q}}(t_{n+1})]= \int_{-\infty}^{\infty} \int_{-\infty}^{\infty} d\boldsymbol{\eta} \, d \boldsymbol{q}  
   \, f[\boldsymbol{q}] e^{i \boldsymbol{\eta} \cdot \{ [\boldsymbol{q}- \boldsymbol{\bar{q}}(t_n) ] - \Delta \mathcal{L}[\boldsymbol{\bar{q}}(t_n)] \} }.   
\end{equation}
If we have a functional of the whole trajectory, $f[\{\boldsymbol{\bar{q}}(t_n)\}_{n=0}^{L-1}]$
the above procedure can be iterated.
By definition, we have
\begin{equation}
    f[\{\boldsymbol{\bar{q}}(t_n)\}_{n=0}^{L-1}]= \int_{-\infty}^{\infty} \prod_{n=0}^{L-1} d \boldsymbol{q}_n  \delta[\boldsymbol{\bar{q}}(t_n)- \boldsymbol{q}_n] f[\{\boldsymbol{q}_n\}_{n=0}^{L-1}].
\end{equation}
The first Dirac delta ($n=0$) gives the initial condition, while the others ($1\leq n \leq L-1$) can be expressed in terms of the decomposition \eqref{eq:phasedelta}, obtaining
\begin{equation} f[\{\boldsymbol{\bar{q}}(t_n)\}_{n=0}^{L-1}]  =
\int_{-\infty}^{\infty} \int_{-\infty}^{\infty} \prod_{n=0}^{L-1} \big[d\boldsymbol{\eta}_n \, d \boldsymbol{q}_n  \big]
   \, e^{i  \sum_{n=0}^{L-1} \boldsymbol{\eta}_n \cdot \{ [\boldsymbol{q}_n- \boldsymbol{q}_{n-1} ] - \Delta \mathcal{L}[\boldsymbol{q}_{n-1}] \} }  \delta[\boldsymbol{\bar{q}}(0)- \boldsymbol{q}_0] f[\{\boldsymbol{q}_n\}_{n=0}^{L-1}].
\end{equation}
In the continuous limit, the finite differences are replaced by derivatives $ \boldsymbol{q}_n- \boldsymbol{q}_{n-1}  \approx \dot{\boldsymbol{q}}(t_n) \Delta$ and the sum over $n$ is replaced by an integral
\begin{equation} \label{eq:trajdelta}
   f[\{\boldsymbol{\bar{q}}(t)\}_t] = \int \int \mathcal{D} \boldsymbol{\eta} \mathcal{D} \boldsymbol{q}
   \, e^{i \int_0^{\tau} \boldsymbol{\eta}(t) \cdot \{ \dot{\boldsymbol{q}}(t) -  \mathcal{L}[\boldsymbol{q}(t)] \} dt }  \delta[\boldsymbol{\bar{q}}(0)- \boldsymbol{q}(0)] f[\{\boldsymbol{q}(t)\}_{t=0}^{\tau}]
, 
\end{equation}
where $\mathcal{D} \boldsymbol{\eta} \mathcal{D} \boldsymbol{q}$ is the measure of integration over the trajectories in $\boldsymbol{\eta}$ and $\boldsymbol{q}$.
Replacing $\boldsymbol{q}$ with the original fields $x,p$ and introducing the following action
\begin{align}
    \mathcal{S}^{(cl)}[\{x(t), p(t), \eta_x(t), \eta_p(t), \boldsymbol{\xi_B}(t) \}_{t=0}^{\tau}]& = \int_0^{\tau} \big(\eta_x(t) \{\dot{p}(t) - \mathcal{F}[x(t),\boldsymbol{\xi_B}(t)]\} +  \eta_p(t) [\dot{x}(t) - \frac{p(t)}{m}]\big), 
\end{align}
equation \eqref{eq:trajdelta} assumes the simple form
\begin{equation} \label{eq:trajdelta2}
   f[\{\boldsymbol{\bar{q}}(t)\}_t] = \int \int \mathcal{D} \boldsymbol{\eta} \mathcal{D} \boldsymbol{q}
   \, e^{i \mathcal{S}^{(cl)}}  \delta[\boldsymbol{\bar{q}}(0)- \boldsymbol{q}(0)] f[\{\boldsymbol{q}(t)\}_{t=0}^{\tau}]
.
\end{equation}
After doing the path integration in the momentum variables and performing an integration by parts at the exponent, $\mathcal{S}^{(cl)}$ reduces to the expression given in Eq. \eqref{eq:claclos}.
Note that the term inside the average in the generating function \eqref{eq:momclass} can be written as in \eqref{eq:genen}, \eqref{eq:genheat} and is a functional of the noises and of the trajectories of the system: 
\begin{equation} \label{eq:funcnew22}
     e^{\lambda_S \Delta E_S(0,\tau) + \boldsymbol{\lambda_B}\cdot \boldsymbol{\Delta E_B}(0,\tau)} =
    e^{- \lambda_S \int_0^{\tau}  \sum_{\nu}\chi_{\nu}(t) V'_{\nu}(\bar{x}) \dot{\bar{x}} (\xi_{\nu}(t) + R_{\nu}(t)) dt + \sum_{\nu} \lambda_{\nu}\int_0^{\tau}  
  \frac{d}{dt} [\chi_{\nu}(t) V_{\nu}(\bar{x}) ] (\xi_{\nu}(t) + R_{\nu}(t)) dt },
\end{equation}
thus it can be written in the representation \eqref{eq:trajdelta2} (replacing the functional $f \equiv f[\{x(t) ,\boldsymbol{\xi}_B \}_{t=0}^{\tau}]$ with the functional in Eq. \eqref{eq:funcnew22}). We have 
\begin{align}  \notag
 e^{\lambda_S \Delta E_S(0,\tau) + \boldsymbol{\lambda_B}\cdot \boldsymbol{\Delta E_B}(0,\tau)}  
& =  \int \int \mathcal{D} \eta
 \mathcal{D}  x
   \, e^{i \mathcal{S}^{(cl)} } 
    e^{- \lambda_S \int_0^{\tau}  \sum_{\nu} \chi_{\nu}(t) V'_{\nu}(x) \dot{x} (\xi_{\nu}(t) + R_{\nu}(t)) dt}  \\ \times & e^{\sum_{\nu} \lambda_{\nu}\int_0^{\tau}  
  \frac{d}{dt} [\chi_{\nu}(t) V_{\nu}(x) ] (\xi_{\nu}(t) + R_{\nu}(t)) dt } P(\boldsymbol{x}(0), \dot{x}(0))
,  \label{app:mgf}
\end{align}
where $P$ is the probability distribution of the initial system positions and velocities.
The MGF is obtained by averaging Eq. \eqref{app:mgf} over the noises $\boldsymbol{\xi_B}$.
From Sec. \ref{sec:redusist} we know that the autocorrelation of the noise is given by Eq. \eqref{eq:corr1}, so that the averaging procedure corresponds to a path integration over the variables $\xi_{\nu}$ with a Gaussian weight depending on $C^{(2)}_{\nu}$ for each one of the noise sources (up to a normalization factor).

\begin{equation}
\langle f[...] \rangle = \int \prod_{\nu} \mathcal{D} \xi_{\nu} e^{-\frac{1}{2}\iint_{0}^{\tau} C^{(2)\, -1}_{\nu}(t-t') \xi_{\nu}(t) \xi_{\nu} (t')}  f[ \{x(t), \boldsymbol{\xi}_B(t)\}_{t=0}^{\tau}].
 \end{equation}
After carrying the integration above for the function in Eq. \eqref{app:mgf} we obtain a MSRJD path integral expression with an action given by Eq. \eqref{eq:finalclass}.
\section{Keldysh rotation}
\label{app:keld}
 To identify the noise and dissipation contribution to the dynamics, it is convenient to introduce the Keldysh rotation \cite{caldeira1983path, Kamenev} i.e. the change of variables
 \begin{equation}
     x_- = x_{cl} + \frac{x_q}{2}, \quad \quad   x_+ = x_{cl} - \frac{x_q}{2}.
 \end{equation}
 The change of variable is linear, so that we can write
 \begin{equation}
     \left(\begin{array}{cc}  x_- \\ x_+    
     \end{array} \right)  = M  \left(\begin{array}{cc}  x_{cl} \\ x_q     \end{array} \right); \quad \quad M = \left(  \begin{array}{cc} 1  & \frac{1}{2} \\ 
    1 &  - \frac{1}{2} \end{array} \right) .
     \end{equation}
Let us assume, for ease of calculation, that $V_{\nu}[x]= x$, that is, $v_{\nu}(s)=\chi_{\nu}(s) x(s)$.
In this case, the integrand in the first line of Eq. \eqref{eq:bigactdispapp} attains the form
\begin{equation}
   \left(\begin{array}{cc}  x_- & x_+    
     \end{array} \right)   \hat{G}_{\nu,k}   \left(\begin{array}{c}  x_- \\ x_+  
     \end{array} \right)   =  \left(\begin{array}{cc}  x_{cl} & x_q    
     \end{array} \right)  \hat{G'}_{\nu,k}\left(\begin{array}{c}  x_{cl} \\ x_q    
     \end{array} \right) ,
\end{equation}
where
\begin{align}
 \hat{G}_{\nu,k} =\left(  \begin{array}{cc} G_{\nu,k}^{-,-}  & - G_{\nu,k}^{-,+}  \\ 
    - G_{\nu,k}^{+,-}  & G_{\nu,k}^{+,+}  \end{array} \right)  
\end{align}
and 
\begin{align} \notag
\hat{G'}_{\nu,k} & \equiv  M^t \hat{G}_{\nu,k}   M  =  \left(  \begin{array}{cc} 1  & 1 \\ 
    \frac{1}{2} &  - \frac{1}{2} \end{array} \right) \left(  \begin{array}{cc} G_{\nu,k}^{-,-}  & - G_{\nu,k}^{-,+}  \\ 
    - G_{\nu,k}^{+,-}  & G_{\nu,k}^{+,+}  \end{array} \right)   \left(  \begin{array}{cc} 1  & \frac{1}{2} \\ 
    1 &  - \frac{1}{2} \end{array} \right) 
    \\ & =  \left[  \begin{array}{cc} G_{\nu,k}^{-,-} - G_{\nu,k}^{+,-} - G_{\nu,k}^{-,+} + G_{\nu,k}^{+,+}  & \frac{1}{2} \big( G_{\nu,k}^{-,-} - G_{\nu,k}^{+,-} + G_{\nu,k}^{-,+} - G_{\nu,k}^{+,+} \big) \\ 
    \frac{1}{2} \big(G_{\nu,k}^{-,-} + G_{\nu,k}^{+,-} - G_{\nu,k}^{-,+} - G_{\nu,k}^{+,+}\big) &  \frac{1}{4}\big(G_{\nu,k}^{-,-} + G_{\nu,k}^{+,-} + G_{\nu,k}^{-,+} + G_{\nu,k}^{+,+} \big) \end{array} \right]. \label{eq:matGFs}
\end{align}
This leads to the introduction of the following definitions 
\begin{align}  \notag
G^{cl,cl}(t,t')  & = i \coth\big( \frac{\hbar \omega \beta}{2} \big)
\cos \omega(t-t')  \big[ 1 - \cos i \omega \hbar \lambda  \big] -  \cos \omega (t-t') \sin i \omega \hbar \lambda,  \\ \notag
G^{cl,q}(t,t')  & = \big[\frac{1}{2}\sign(t-t') - \frac{1}{2} \cos i \omega \hbar \lambda \big] \sin \omega (t-t') +\frac{i}{2} \coth\big(\frac{\hbar \omega \beta}{2} \big)  \sin \omega (t-t') \sin i \omega \hbar \lambda, \\ 
G^{q,cl}(t,t') & =  \big[\frac{1}{2}\sign(t-t') + \frac{1}{2} \cos i \omega \hbar \lambda \big] \sin \omega (t-t') - \frac{i}{2} \coth\big(\frac{\hbar \omega \beta}{2} \big)  \sin \omega (t-t') \sin i \omega \hbar \lambda ,  \notag \\
G^{q,q}(t,t') & = \frac{i}{4}  \coth\big( \frac{\hbar \omega \beta}{2} \big)
\cos \omega(t-t')  \big[ 1 + \cos i \omega \hbar \lambda  \big] +  \frac{1}{4}  \cos \omega (t-t') \sin i \omega \hbar \lambda, 
\label{eq:4GFs}
\end{align}
where as we omitted all the indices $k,\nu$, for ease of notation, we stress again that to obtain $G^{cl,cl}_{\nu,k}, G^{q,q}_{\nu,k}, G^{q,cl}_{\nu,k}, G^{cl,q}_{\nu,k}$ we have to replace $\omega, m,\beta, \lambda$ respectively with $\omega_{\nu,k}, m_{\nu,k}, \beta_{\nu}, \lambda_{\nu}$.
Note that for $\lambda=0$ the GF $G_{\nu,k}^{cl,cl}$ identically nullifies, this is a well-known symmetry of the dissipative action without counting fields \cite{Kamenev}.
Finally, we have to study the effect of the Keldysh rotation on the second term of Eq. \eqref{eq:bigactdispapp}.
Note that in \eqref{eq:bigactdispapp} we have the terms of the form
\begin{align}
 - \big[G^{-,M}(t,\zeta) v^-(t) - G^{+,M}(t,\zeta) v^+(t) \big] = - \big[ v^{cl}(t) G^{cl,M}(t,\zeta) + G^{q,M}(t,\zeta) v^q(s) \big],
\end{align}
where we introduced 
\begin{align} \notag
    G^{cl,M}( t, \zeta) & =  G^{-,M}(t,\zeta)  -G^{+,M}(t, \zeta)  \\  \notag & = \frac{i}{2} \coth\big(\frac{\hbar \omega \beta}{2}\big) \big[\cos \omega (t - i \zeta) -\cos \omega (t+ i \hbar \lambda - i \zeta)\big] +  \frac{1}{2}\big[ \sin \omega(i \zeta - t') -  \sin \omega(i \zeta - t'- i \hbar \lambda) \big], \\  \label{eq:matsucq}
     G^{q,M}( t, \zeta) & = \frac{1}{2} \big[ G^{-,M}(t,\zeta) +  G^{+,M}(t, \zeta) \big] \\ & = \frac{i}{4} \coth\big(\frac{\hbar \omega \beta}{2}\big) \big[\cos \omega (t - i \zeta) + \cos \omega (t+ i \hbar \lambda - i \zeta)\big] + \frac{1}{4} \big[ \sin \omega(i \zeta - t') +  \sin \omega(i \zeta - t'- i \hbar \lambda) \big]. \notag
\end{align}
A similar definition holds for $G^{q, \uparrow}, G^{cl, \uparrow}, G^{q, \downarrow}, G^{cl, \downarrow}$.
Using all the definitions introduced until now, Equation \eqref{eq:bigactdispapp} simplifies to 

\begin{align}  \notag
   \bar{\mathcal{S}}_{\nu,k} & =   \sum_{l,j=cl,q} \int_{0}^{\tau}  \int_{0}^{\tau}   \frac{ \alpha^2_{\nu,k}} {2 m_{\nu,k} \omega_{\nu,k}} v^l_{\nu}(t)  G^{l,j}_{\nu,k}(t,t')  v^{j}_{\nu}[s'] dt dt'
    - i  \sum_{l=cl,q} \int_{0}^{\tau} dt \int_{\hbar \lambda_{\nu}}^{0} d\zeta'    \omega_{\nu,k}   L_{\nu,k} \alpha_{\nu,k} v^l_{\nu}(t)      G^{l,\downarrow}_{\nu,k}(t,\zeta')   
   \\  
    & - i  \sum_{l=cl,q} \int_{0}^{\tau} dt \int_{0}^{-\hbar \beta_{\nu}} d\zeta'    \omega_{\nu,k}   L_{\nu,k} \alpha_{\nu,k} v^l_{\nu}(t) 
       G^{l,M}_{\nu,k}(t,\zeta')  
    + \mathcal{N}_{\nu,k}.\label{eq:bigactdispapp23}
%    \\
  %  & +   
% \int_{0}^{\hbar(\beta+\lambda)} \int_{0}^{\hbar(\beta+\lambda)}  d\zeta d\zeta'  \frac{m_{\nu,k} \omega_{\nu,k}^2 L^2_{\nu,k}}{2} g^{M,M}_{\nu,k}(\zeta,\zeta').
 \end{align}
Note that the structure of the integrands described above is the same as in Eq.~\eqref{eq:bigactdispapp} but the indices, and thus the components of Green's function, are different (above we have $cl,q$ instead of $-,+$).
After introducing the density functions $J_{\nu}(\omega)$, $K_{\nu}(\omega)$ as in the main text (Eqs. \eqref{spectraldensitymain} and \eqref{eq:Kdensity}), the sum of all contributions within a given reservoir becomes
\begin{align}  \notag
  \sum_{k} \bar{\mathcal{S}}_{\nu,k} & =   \sum_{l,j=cl,q} \int d\omega \int_{0}^{\tau}  \int_{0}^{\tau}   \frac{J_{\nu}(\omega)}{\pi} v^l_{\nu}(s)  G^{l,j}_{\nu}(\omega,t,t')  v^{j}_{\nu}[t'] dt dt' 
   \notag \\ & - i  \sum_{l=cl,q}  \int d\omega \int_{0}^{\tau} dt \int_{\hbar \lambda_{\nu}}^{0} d\zeta'  \frac{2 K_{\nu}(\omega) \omega}{\pi} v^l_{\nu}(t)      G^{l,\downarrow}_{\nu}(\omega,t,\zeta')   
  \notag \\  
    & -i  \sum_{l=cl,q}  \int d\omega  \int_{0}^{\tau} dt \int_{0}^{-\hbar \beta_{\nu}} d\zeta'     \frac{2 K_{\nu}(\omega) \omega }{\pi} v^l_{\nu} 
       G^{l,M}_{\nu,k}(\omega,t,\zeta') + \mathcal{N}_{\nu},\label{eq:bigactdispcont}
    \end{align}
    where since we are considering the case without squeezing, we have to use $r_{\nu,k} =0$ in Eq. \eqref{eq:Kdensity}.
\section{Classical limit of the quantum action}
\subsection{Equilibrium reservoirs in the strong coupling regime}
\label{app:classqdyn}
Here we show that the case without displacement reproduces the energy statistics of the Caldeira-Leggett model with time-dependent strong coupling \cite{FunoQuanPRE2018, Aurell2017}.
Setting $K_{\nu}(\omega) = 0$ and $\mathcal{N}_{\nu}=0$ in Eq. \eqref{eq:bigactdispcont} and summing over all thermal reservoirs we have
\begin{equation} \label{eq:appCL}
 \sum_{\nu,k} \bar{\mathcal{S}}_{\nu,k}  =   \sum_{l,j=cl,q} \int d\omega \int_{0}^{\tau}  \int_{0}^{\tau}   \frac{J_{\nu}(\omega)}{\pi} v^l_{\nu}(t)  G^{l,j}_{\nu}(\omega,t,t')  v^{j}_{\nu}(t') dt dt'.
\end{equation}
Let us start by considering only the dynamics, so we set $\lambda_{\nu}=0$ for all the thermal reservoirs
\begin{align} \notag
G_{\nu}^{q,q}(\omega,t,t') & = \frac{i}{2} \coth\big( \frac{\hbar \omega \beta_{\nu}}{2} \big)
\cos \omega(t-t'), \quad \quad 
G_{\nu}^{cl,cl}(\omega,t,t')  = 0, \\ \notag
G_{\nu}^{cl,q}(\omega,t,t')  & =   \frac{1}{2} \big[\sign(t-t') -1\big] \sin \omega (t-t'), \\ 
G_{\nu}^{q,cl}(\omega,t,t') & =  \frac{1}{2} \big[\sign(t-t') + 1 \big] \sin \omega (t-t'). \label{eq:4GFs1.0}
\end{align}
Equation \eqref{eq:appCL} can now be written as
\begin{align} \notag
    \sum_{\nu,k} \bar{\mathcal{S}}_{\nu,k} & =    \sum_{\nu} \int d \omega  \frac{J_{\nu}(\omega)}{2 \pi}  \big\{\int_{0}^{\tau} \int_{0}^{\tau} dt dt'  \chi(t) \chi(t') \big[  i \coth \big(\frac{\hbar \omega \beta_{\nu}}{2}\big)  \cos \omega (t-t') V_{\nu}^q(t) V_{\nu}^q(t') \\
   & + 2 (\sign(t-t') + 1) \sin \omega (t-t') V_{\nu}^{q}(t) V^{cl}_{\nu}(t') \big] \big\}, \label{eq:appCL20} 
\end{align}
we recognize here the memory kernels of noise and dissipation \cite{caldeira1983path,leggett1987dynamics}
\begin{align} \label{eq:Lfunctions}
    L_1^{(\nu)}(t-t')  & = \int d \omega  J_{\nu}(\omega) \sin \omega (t-t'),   \\ 
    L_2^{(\nu)}(t-t')  & = \int d \omega  J_{\nu}(\omega) \coth\big( \frac{\beta_{\nu} \hbar \omega}{2} \big) \cos \omega (t-t')   ,
\end{align}
that allow us to write the action as
\begin{align} \notag
 \sum_{\nu,k} \bar{\mathcal{S}}_{\nu,k}  =  \frac{1}{2 \pi} & \sum_{\nu} \int_{0}^{\tau} \int_{0}^{\tau} dt dt' \chi(t) \chi(t')  \big[ i L_2^{(\nu)}(t-t')  V_{\nu}^q(t) V_{\nu}^{q}(t') 
    \\ & + 2 (\sign(t-t') +1)  L_1^{(\nu)}(t-t')  V_{\nu}^{cl}(t) V_{\nu}^{q}(t') \big]. \label{eq:appCL2} 
\end{align}
Let us turn our attention to the semiclassical limit in the
statistics of heat in the case without squeezing and displacement. 
After expanding the hyperbolic cotangent in Eq. \eqref{eq:appCL2} and using the definitions \eqref{eq:corr1}, \eqref{eq:corr2} we find 
\begin{align} \notag
    \sum_{\nu} \bar{\mathcal{S}}_{\nu,k} & =   \frac{1}{2} \sum_{\nu}    \big\{\int_{0}^{\tau} \int_{0}^{\tau} dt dt'  \chi(t) \chi(t') \big[    \frac{i}{\hbar} C_{\nu}^{(2)} (t-t') V_{\nu}^q(t) V_{\nu}^q(t') \\
   & +  (\sign(t-t') + 1) C^{(1)}_{\nu}(t-t') V_{\nu}^{q}(t) V^{cl}_{\nu}(t') \big] \big\}. \label{eq:appCL3} 
\end{align}
After expanding $V_{\nu}^q(t) \approx V'[x_{cl}(t)] x_{q} (t)$ and replacing $x_q \rightarrow - \hbar \eta_x$ the first line of Equation \eqref{eq:appCL3} reduces to the quadratic term of Eq. \eqref{eq:finalclass}.
The second contribution can be rewritten as
\begin{equation}
    - \int_0^{\tau} \int_0^{\tau} dt dt' \Theta(t-t') \chi(t) \chi(t')  C^{(1)}_{\nu}(t-t') V'[x_{cl}(t)] x_q(t) V[x_{cl}(t')]  =  \int_0^{\tau} dt \chi(t)  R_{\nu}(t) V'[x_{cl}(t)] x_q(t)
\end{equation}
that is equal to the last term in the first line of Eq. \eqref{eq:finalclass}. 
To focus on the role of $\lambda_{\nu}$ in the classical expansion, we can expand Eq. \eqref{eq:4GFs} for small values of $\hbar$ and obtain
\begin{align} \notag
G^{q,q}_{\nu}(\omega,t,t') & \approx \frac{i}{ \hbar \omega \beta_{\nu}} 
\cos \omega(t-t')   , \\ \notag
G^{cl,cl}_{\nu}(\omega,t,t')  & \approx
%-\frac{2 i}{\hbar \omega \beta}
%\cos \omega(t-t')  \frac{1}{2}\big(\hbar \omega \lambda\big)^2 + \cos \omega (t-t') \omega \hbar \lambda  = 
-\frac{i \hbar \omega \lambda_{\nu}^2}{\beta_{\nu}}
\cos \omega(t-t')  - i \cos \omega (t-t') \omega \hbar \lambda_{\nu}
,  \\ \notag
G^{cl,q}_{\nu}(\omega,t,t')  & \approx \big[\frac{1}{2}\sign(t-t') - \frac{1}{2}  \big] \sin \omega (t-t') - \frac{ \lambda_{\nu}}{\beta_{\nu}} \sin \omega (t-t') , \\ 
G^{q,cl}_{\nu}(\omega,t,t') & \approx \big[\frac{1}{2}\sign(t-t') + \frac{1}{2} \big] \sin \omega (t-t') +  \frac{ \lambda_{\nu}}{\beta_{\nu}} \sin \omega (t-t') . \label{eq:4GFs2}
\end{align}
We already met the parts independent by $\lambda_{\nu}$ in the first half of this section, 
if we focus on the $\lambda_{\nu}$ dependent parts of \eqref{eq:appCL}, that we call $\bar{\mathcal{S}}_{\nu,k}^{\lambda_{\nu}}$, we are left with
\begin{align}
  \sum_{\nu,k} \bar{\mathcal{S}}^{(\lambda_{\nu})}_{\nu,k} =   \sum_{\nu} \int d\omega \int_{0}^{\tau}  \int_{0}^{\tau}   & \frac{J_{\nu}(\omega)}{\pi} \big[ v^{cl}_{\nu}(t)   v^{cl}_{\nu} (t') \cos \omega (t-t') \big(-i\omega \hbar \lambda_{\nu} - \frac{i \hbar \omega \lambda^2_{\nu}}{\beta_{\nu} }\big)  \notag \\ &  + 2  \frac{\lambda_{\nu}}{\beta_{\nu}} v^{cl}_{\nu}(t')  v^{q}_{\nu} (t)  \sin \omega (t-t') \big]dt dt'.
\end{align}
It is convenient to integrate by parts and obtain
\begin{align} \notag
  \sum_{\nu,k} \bar{\mathcal{S}}^{(\lambda_{\nu})}_{\nu,k} & =  \sum_{\nu} \int d\omega \int_{0}^{\tau}  \int_{0}^{\tau}   \frac{J_{\nu}(\omega)}{\pi} \big[- \dot{v}^{cl}_{\nu}(t)   \dot{v}^{cl}_{\nu} (t') \cos \omega (t-t') \times \frac{i \hbar  \lambda^2_{\nu}}{\beta_{\nu} \omega} \\ & + i \hbar \lambda_{\nu}   \dot{v}^{cl}_{\nu}(t)   v^{cl}_{\nu} (t') \sin \omega (t-t')  -2  \frac{\lambda_{\nu}}{\omega \beta_{\nu}} \dot{v}^{cl}_{\nu}(t')  v^{q}_{\nu} (t)  \cos \omega (t-t') \big]dt dt'.
\end{align}
Using the definitions introduced in the classical case, we can rewrite
\begin{align} \notag
  \sum_{\nu,k} \bar{\mathcal{S}}^{(\lambda_{\nu})}_{\nu,k} & =  \sum_{\nu}  \int_{0}^{\tau}  \int_{0}^{\tau}   \big[- \dot{v}^{cl}_{\nu}(t)   \dot{v}^{cl}_{\nu} (t') \frac{i}{2} \hbar  \lambda_{\nu}^2 C^{(2)}_{\nu}(t-t') \\ & + \frac{i}{2} \hbar \lambda_{\nu}   \dot{v}^{cl}_{\nu}(t)   v^{cl}_{\nu} (t') C^{(1)}_{\nu}(t-t')  -  \lambda_{\nu} \dot{v}^{cl}_{\nu}(s')  v^{q}_{\nu} (t)  C^{(2)}_{\nu}(t-t') \big]dt dt'. \label{eq:lambdacompare}
\end{align}
Let us compare the equation above term by term with the $\lambda_{\nu}$-dependent parts of Eq. \eqref{eq:finalclass}.
First of all, we use the definition \eqref{eq:potrot} to prove that  $v^{cl}_{\nu}(t) \approx \chi_{\nu}(t) V_{\nu}[x_{cl}(t)] + O(x_q^2)$.
With this, we can easily check that the first line of Eq. \eqref{eq:lambdacompare} is exactly the term quadratic in $\lambda_{\nu}$ in Eq. \eqref{eq:finalclass}.
Instead, the first term of the second line in \eqref{eq:finalclass} coincides with the last term in Eq. \eqref{eq:lambdacompare}.
The last term to check is the one containing $C^{(1)}_{\nu}(t-t')$, this term is equal to the first term in the second line of Eq. \eqref{eq:finalclass}.
Indeed, we can rewrite 
\begin{align} \notag
  \frac{i}{2} & \int_0^{\tau} \int_0^{\tau}  \hbar \lambda_{\nu}   \dot{v}^{cl}_{\nu}(t)   v^{cl}_{\nu} (t') C^{(1)}_{\nu}(t-t') dt dt' = 
  i \int_0^{\tau} dt \int_0^{t} dt' \hbar \lambda_{\nu} \frac{d}{dt}[ \chi_{\nu}(t) V^{cl}_{\nu}(t)  ] V^{cl}_{\nu} (t')  \chi_{\nu}(t') C^{(1)}_{\nu}(t-t') \\ & 
  = 
 -i \int_0^{\tau} dt  \hbar \lambda_{\nu} \frac{d}{dt}[ \chi_{\nu}(t) V^{cl}_{\nu}(t)  ] R_{\nu}(t).
\end{align}
\subsection{Displaced work sources and the path integral approach}
\label{app:worksource}
As discussed in Sec. \ref{sec:worksour} the contribution of noise and dissipation becomes negligible in the weak coupling and very large displacement limit.
In this case we can neglect the first addend in Equation \eqref{eq:bigactdispcont}, if we now set $\lambda_{\nu}=0$ (and ignore $\mathcal{N}_{\nu}$, we will come back to it at the end of the section) we are left with 
\begin{align}
 \sum_{k} \bar{\mathcal{S}}_{\nu,k} = -i  \sum_{l=cl,q}  \int d\omega  \int_{0}^{\tau} dt \int_{0}^{-\hbar \beta_{\nu}} d\zeta'     \frac{2 K_{\nu}(\omega) \omega}{\pi} v^l_{\nu} 
       G^{l,M}_{\nu,k}(\omega,t,\zeta')  
    .\label{eq:forza}
    \end{align}
In the classical limit, we have
    \begin{equation}  \label{eq:matsusem}
    G_{\nu,k}^{cl,M}(\omega, t, \zeta) \approx 0, \quad \quad 
     G_{\nu,k}^{q,M}(\omega, t, \zeta) \approx \frac{i}{ \hbar \beta_{\nu} \omega} \cos \omega (t - i \zeta) .
\end{equation}
so that Eq. \eqref{eq:forza} becomes
\begin{equation}
   \sum_{k} \bar{\mathcal{S}}_{\nu,k} =   \int d\omega  \int_{0}^{\tau} ds \int_{0}^{-\hbar \beta_{\nu}} d\zeta'     \frac{2 K_{\nu}(\omega)}{\pi} \frac{\cos \omega (t-i\zeta)}{\hbar \beta_{\nu} \omega} v^q_{\nu}(t) \approx  - \int d\omega  \int_{0}^{\tau} dt   \frac{2 K_{\nu}(\omega)}{\pi} \frac{\cos \omega t}{\omega} v^q_{\nu}(t).
\end{equation}
Following the literature on the subject \cite{Kamenev,Polkovnikovreview}, we expand for small $x_q$ 
\begin{equation} \label{eq:Vexpand}
 V^{q}_{\nu} = V_{\nu}[x_{cl}(t) +\frac{x_q(t)}{2}] - V_{\nu}[x_{cl}(t) - \frac{x_q(t)}{2}]  =   
V'_{\nu}[x_{cl}(t)] x_q(t) + O[x_q^3] 
\end{equation}
and we finally obtain
\begin{equation}
   \sum_{k} \bar{\mathcal{S}}_{\nu,k} =  - \int d\omega  \int_{0}^{\tau} dt  \chi_{\nu}(t) \frac{2 K_{\nu}(\omega)}{\pi} \cos \omega t V_{\nu}'[x_{cl}(t)] x_q(t).
\end{equation}
After introducing
$F_{\nu,d}$ like in Eq. \eqref{eq:avgforce} and replacing $ x_q(t) \rightarrow - \hbar \eta_{x}(t) $ the action becomes the same as the contribution associated with the force on the system induced by the displacement of the reservoirs in the first line of Eq. \eqref{eq:finalclass} , that is the contribution of $F_{\nu,d}$ contained in $R_{\nu}'$.
Let us now consider the generic case in which $\lambda_{\nu} \neq 0$, in this case also the GF components $G_{\nu}^{cl,\downarrow}, G_{\nu}^{q,\downarrow}$ (and, by completeness, the components $ G_{\nu}^{cl,\uparrow}, G_{\nu}^{q,\uparrow}$) enter in the calculation of Eq. \eqref{eq:bigactdispcont}. Writing these components explicitly (starting from Eq. \eqref{eq:components3}) we have
\begin{align} \notag
G_{\nu}^{cl,\uparrow}(\omega,t,\zeta) & = G_{\nu}^{-,\uparrow}(\omega,t,\zeta) 
- G_{\nu}^{+. \uparrow}(\omega,t,\zeta) \\ \notag& = 
 \frac{i}{2} \coth\big(\frac{\hbar \omega \beta_{\nu}}{2}\big) \big[\cos\omega(t- \tau - i \zeta)  - \cos\omega(t + i \hbar \lambda_{\nu} - \tau - i \zeta)\big] \\ \notag  & -\frac{1}{2}  \big[\sin \omega(t- \tau - i \zeta) + \sin \omega(t + i \hbar \lambda_{\nu}- \tau - i \zeta)\big], \notag
\\  \notag
G_{\nu}^{q,\uparrow}(\omega,t,\zeta) & =  \frac{1}{2} \big[ G_{\nu}^{-,\uparrow}(\omega,t,\zeta) 
+ G_{\nu}^{+. \uparrow}(\omega,t,\zeta) \big]
\\ \notag  &= \frac{i}{4} \coth\big(\frac{\hbar \omega \beta_{\nu}}{2}\big) \big[\cos\omega(t- \tau - i \zeta)  + \cos\omega(t + i \hbar \lambda_{\nu} - \tau - i \zeta)\big]  \\ \notag  &  - \frac{1}{4}  \big[\sin \omega(t- \tau - i \zeta) - \sin \omega(t + i \hbar \lambda_{\nu}- \tau - i \zeta)\big], \notag
\\ \notag
 G_{\nu}^{cl,\downarrow}(\omega, t, \zeta) & =  G_{\nu}^{-,\downarrow}(\omega,t,\zeta)  -G_{\nu}^{+,\downarrow}(\omega,t, \zeta)  \\  \notag &  = \frac{i}{2} \coth\big(\frac{\hbar \omega \beta_{\nu}}{2}\big) \big[\cos \omega (t - i \zeta) -\cos \omega (t+ i \hbar \lambda_{\nu} - i \zeta)\big] - \\ \notag  &  \frac{1}{2}\big[ \sin \omega(t-i \zeta) -  \sin \omega(t' -i \zeta + i \hbar \lambda_{\nu}) \big], 
 \\  \notag
G_{\nu}^{q,\downarrow}(\omega, t, \zeta) & = \frac{1}{2} \big[ G_{\nu}^{-,\downarrow}(\omega,t,\zeta)    + G_{\nu}^{+,\downarrow}(\omega,t, \zeta) \big] \\ &  = \frac{i}{4} \coth\big(\frac{\hbar \omega \beta_{\nu}}{2}\big) \big[\cos \omega (t - i \zeta) + \cos \omega (t+ i \hbar \lambda_{\nu} - i \zeta)\big]\notag   \\  &  - \frac{1}{4} \big[ \sin \omega(t-i \zeta) +  \sin \omega(t'-i \zeta+i \hbar \lambda_{\nu}) \big]. \label{eq:compclassvert}
\end{align}

In the limit $\hbar \rightarrow 0$ we can write

\begin{equation}
G_{\nu}^{q,\downarrow} (\omega,t,\zeta) \approx 
\frac{i}{\hbar \omega \beta_{\nu} } \cos \omega (t - i \zeta), \quad \quad 
G_{\nu}^{cl,\downarrow} (\omega,t,\zeta) \approx 
- \frac{\lambda_{\nu}}{ \beta_{\nu} } \sin \omega (t - i \zeta) .
\end{equation}
The second line of Eq. \eqref{eq:bigactdispcont} can be expanded as 
\begin{align} \notag
- i &  \sum_{l=cl,q}  \int d\omega \int_{0}^{\tau} dt \int_{\hbar \lambda_{\nu}}^{0} d\zeta'  \frac{2 K_{\nu}(\omega) \omega}{\pi} v^l_{\nu}(t)      G^{l,\downarrow}_{\nu}(\omega,t,\zeta')  
 \\ \approx  - i &   \int d\omega \int_{0}^{\tau} dt \int_{\hbar \lambda_{\nu}}^{0} d\zeta' \Big[ \frac{2 i K_{\nu}(\omega) }{\pi \hbar \beta_{\nu}} v^q_{\nu}(t)   \cos \omega (t - i \zeta) - \frac{2 K_{\nu}(\omega) \omega \lambda_{\nu}}{\pi \beta_{\nu}} v^{cl}_{\nu}(t)  
 \sin \omega (t - i \zeta)  \Big]
 \notag \\ \approx -  i &   \int d\omega \int_{0}^{\tau} dt \int_{\hbar \lambda_{\nu}}^{0} d\zeta' \Big\{ \frac{2 i K_{\nu}(\omega) }{\pi \hbar \beta_{\nu}} \chi_{\nu}(t) V_{\nu}'[x_{cl}(t)] x_q(t)   \cos \omega (t - i \zeta) \notag \\&  \quad  \quad  \quad  \quad  \quad  \quad  \quad  \quad -  \frac{2 K_{\nu}(\omega) \omega \lambda_{\nu}}{\pi \beta_{\nu}} \chi_{\nu}(t)  
V_{\nu}[x_{cl}(t)] 
 \sin \omega (t - i \zeta)  \Big\}
  \label{eq:finalwquantdis}\\ = - i &   \int d\omega \int_{0}^{\tau} dt \Big\{-  \lambda_{\nu} \frac{2 i K_{\nu}(\omega) }{\pi \beta_{\nu}} \chi_{\nu}(t) V_{\nu}'[x_{cl}(t)] x_q(t)   \cos \omega t  + \hbar   \frac{2 K_{\nu}(\omega) \omega \lambda^2_{\nu}}{\pi \beta_{\nu}} \chi_{\nu}(t) 
V_{\nu}[x_{cl}(t)] 
 \sin \omega t  \Big\} \notag
\end{align}
where in the second equality we used Eq. \eqref{eq:Vexpand}.\ and
\begin{equation} \label{eq:Vexpand2}
 V^{cl}_{\nu}(s) = \frac{1}{2} \{ V_{\nu}[x_{cl}(s) +\frac{x_q(s)}{2}] + V_{\nu}[x_{cl}(s) - \frac{x_q(s)}{2}] \}  = 
V_{\nu}[x_{cl}(s)] + O[x_q^2]. 
\end{equation}
The two terms in the last line of Eq. \eqref{eq:finalwquantdis} match with the terms in Eq. \eqref{eq:finalclass2} after replacing $x_q \rightarrow - \hbar \eta_x$ and multiplying by $\frac{i}{\hbar}$. 
The classical limit of $\mathcal{N}_{\nu}$ does not involve the Keldysh rotation and is easy to compute. It is also not necessary to go to the continuum limit, so we will keep the notation with $\nu$ and $k$ when calculating this term.
Since $\hbar$ is small, the two integrals in the last two lines of Eq. \eqref{eq:bigactdispapp} reduce to multiplications by $\hbar (\beta_{\nu} + \lambda_{\nu}) $, and keeping only the first non zero order in the cosines in \eqref{eq:components3} we obtain
\begin{align} \notag
    \mathcal{N}_{\nu} & \approx - \hbar^2(\beta_{\nu} + \lambda_{\nu})^2 \times \sum_k \frac{m_{\nu,k} \omega^3_{\nu,k} L_{\nu,k}^2}{2} \times \frac{i}{\hbar \beta_{\nu} \omega_{\nu,k}} + i  (\hbar\beta_{\nu} + \hbar\lambda_{\nu})  \sum_k \frac{m_{\nu,k} \omega_{\nu,k}^2 L_{\nu,k}^2 }{2}
    \\ = &  -i \hbar \big( \lambda_{\nu} + \frac{\lambda_{\nu}^2}{\beta_{\nu}} \big) \sum_k \frac{m_{\nu,k} \omega_{\nu,k}^2 L_{\nu,k}^2}{2} .
\end{align}
After multiplying with $\hbar^{-1}$ this coincides with $\mathcal{M}_{\nu}$ in  Eq. \eqref{eq:finalclass2}.
\section{Approximate calculation of the squeezed Green's function}
\label{sec:squeezKeld}
In this appendix we study the solution of Eq. \eqref{eq:squeeqg} using a perturbative approach.  
As a first step, it is convenient to note that a quench of the masses is equivalent (i.e. produces the same MGF) of a quench of $\beta_{\nu}, \lambda_{\nu}$ alongside a quench in the frequencies.
This is clear from the formula of the squeezed Hamiltonian \eqref{eq:squeezapppp}: the squeezed thermal state can be obtained by replacing $m_{\nu,k} \rightarrow e^{-2r_{\nu,k}} m_{\nu,k}$ but also by replacing $\beta_{\nu} \rightarrow \beta_{\nu} e^{2 r_{\nu,k}} $ and $\omega_{\nu,k} \rightarrow \omega_{\nu,k} e^{- 2 r_{\nu,k}} $ for every mode of the reservoir. 
Using this property, we can represent the quench in the masses as a combination of a quench in the temperatures and the frequencies, and turn our attention to the calculation of the Green's function for an on oscillator with a varying frequency on the contour
  \begin{align}
\omega_{\nu,k}(z) = \begin{cases}
   \omega_{\nu,k} &  \text{for } z =t \in \gamma_{\pm}, \gamma^{\nu}_{\uparrow} , \\
  e^{-2 r_{\nu,k}} \omega_{\nu,k}  &  \text{for } z \in \gamma_{M}, \gamma^{\nu}_{\downarrow}.
  \end{cases}    \label{eq:piecewisqueesqapp}
\end{align}

In this case, the Green's function will satisfy 

\begin{equation} \label{eq:newsqueezzzz}
\big[\frac{\partial^2}{\partial z^2} + \omega_{\nu,k}(z)^2 \big] \mathcal{G}_{\nu,k}(z,z') = \delta(z-z') \omega_{\nu,k}
\end{equation}
and the domain of integration is given by a "quenched" version of the contour $\gamma^{\nu \star}$ in which $\beta_{\nu} \rightarrow \beta_{\nu} e^{2 r_{\nu,k}} $  and $\lambda_{\nu} \rightarrow \lambda_{\nu} e^{2 r_{\nu,k}} $ .
If we introduce the quenched value of the frequency $\Omega_{\nu,k} \equiv \omega_{\nu,k} e^{- 2 r_{\nu,k}}$ and the function
  \begin{align}
f(z) = \begin{cases}
   0 &  \text{for } z \in \gamma_{\pm} , \\
  1  &  \text{for } z \in \gamma^{\nu \star}_{M}, \gamma^{\nu \star}_{\downarrow}, \gamma^{\nu \star}_{\uparrow},
  \end{cases}    \label{eq:piecewisquee3app}
\end{align}
the Eq. \eqref{eq:newsqueezzzz} becomes
\begin{equation} \label{eq:appsqGF}
\Big(    \frac{\partial^2}{\partial z^2} + \omega_{\nu,k}^2 \Big) \mathcal{G}_{\nu,k}(z,z')
+ f(z) (\Omega_{\nu,k}^2-\omega_{\nu,k}^2) \mathcal{G}_{\nu,k}(z,z') = \omega_{\nu,k}  \delta(z-z').
\end{equation}
Eq. \eqref{eq:dissipactsquee} can also be written as
\begin{align}    \label{eq:dissipactsqueenew} \bar{\mathcal{S}}_{\nu,k} =\iint_{\gamma^{\nu \star}} \frac{dz  dz'}{2 m_{\nu,k} \omega_{\nu,k}} \mathcal{G}_{\nu,k}(z,z') \chi^{(\gamma)}_{\nu}(z) \chi^{(\gamma)}_{\nu}(z') V^{(\gamma)}_{\nu,k}[x(z)] V^{(\gamma)}_{\nu,k}[x(z')] + i \hbar (\beta_{\nu} + \lambda_{\nu}) N'_{\nu} 
\end{align}
where $N'_{\nu} = \sum_k \frac{m_{\nu,k} e^{- 2 r_{\nu,k}} \omega_{\nu,k}^2 L_{\nu,k}}{2}$ and we replaced $\gamma^{\nu}$ with $\gamma^{\nu \star}$.
Assuming a perturbative expansion in $\varepsilon_{\nu,k} = \Omega_{\nu,k}^2 - \omega_{\nu,k}^2$ that is
$\mathcal{G}_{\nu,k}(z,z') = \mathcal{G}^{(0)}_{\nu,k}(z,z') + \varepsilon_{\nu,k} \mathcal{G}^{(1)}_{\nu,k}(z,z') + O(\varepsilon_{\nu,k}^2)$ we have
\begin{align} \label{eq:0ord}
  \Big(  \frac{\partial^2}{\partial z^2} +  \omega_{\nu,k}^2\Big) \mathcal{G}_{\nu,k}^{(0)}(z,z') &= \omega_{\nu,k}  \delta(z-z'), \\
   \Big(  \frac{\partial^2}{\partial z^2} +  \omega_{\nu,k}^2\Big) \mathcal{G}^{(1)}_{\nu,k}(z,z') &= - f(z) \mathcal{G}^{(0)}_{\nu,k}(z,z'). \label{eq:1stcorr}
\end{align}
It is clear from the system of equations above that $\mathcal{G}_{\nu,k}^{(0)}(z,z')$ follows the same equations of the Green's function \eqref{eq:GFmain}, thus we have %with the difference that the coefficient of the delta function is time dependent.
%It is easy to check that the solution of \eqref{eq:0ord} is given by
\begin{equation}
    \mathcal{G}_{\nu,k}^{(0)}(z,z') = G_{\nu,k}(z,z'),
\end{equation}
where $G_{\nu,k}(z,z')$ is given in Eq.~\eqref{eq:GFmain}.
The first correction instead can be solved with standard Green's function techniques. Since the Green's function of the differential equation \eqref{eq:1stcorr} is just the Eq. \eqref{eq:GFmain} divided by $\omega_{\nu,k}$, the correction takes the form of
\begin{equation}
  \mathcal{G}_{\nu,k}^{(1)}(z,z')   = -
  \frac{1}{\omega_{\nu,k}}\int_{\gamma_{\nu}^*} G_{\nu,k}(z,\bar{z}) f(\bar{z})  \mathcal{G}_{\nu,k}^{(0)}(\bar{z},z') d \bar{z}= 
  -
  \frac{1}{\omega_{\nu,k}}\int_{\gamma_{\nu}^*} G_{\nu,k}(z,\bar{z}) f(\bar{z})  G_{\nu,k}(\bar{z},z') d \bar{z} .\label{eq:perturb}
\end{equation}
We can start by looking only at the contribution of the squeezing to the dynamical part of the action by setting $\lambda_{\nu}=0$. 
In this case, the only nonzero contribution of the integral in Equation \eqref{eq:perturb} is on $\gamma^{M \star}$. 
We are interested in the correlation function, that can be obtained by looking at the quantum component of $\mathcal{G}_{\nu,k}$. 
Following equations \eqref{eq:rotgf} the quantum part of $\mathcal{G}_{\nu,k}^{(1)}(z,z')$ is obtained by combining its components as
\begin{equation} \label{eq:4contsqueez}
\mathcal{G}_{\nu,k}^{(1),q,q}(t,t') = \frac{1}{4} [\mathcal{G}_{\nu,k}^{(1),+,+}(t,t')+ \mathcal{G}_{\nu,k}^{(1),-,-}(t,t')+ \mathcal{G}_{\nu,k}^{(1),+,-}(t,t')+ \mathcal{G}_{\nu,k}^{(1),-,+}(t,t') ].
\end{equation}
Let us compute the first contribution to the equation above
\begin{equation} \label{eq:intmatsusque}
\mathcal{G}_{\nu,k}^{(1),+,+}(t,t') = - \frac{i}{\omega_{\nu,k}}
     \int_{0}^{-\hbar \beta_{\nu} e^{2 r_{\nu,k}}} G_{\nu,k}^{+,M}(t,\zeta) G_{\nu,k}^{M,+}(\zeta,t')d \zeta.
\end{equation}
Directly from \eqref{eq:components2} we can check that, for what concerns the term in the integral above, the four contributions \eqref{eq:4contsqueez} are the same, so that we obtain, at first order in the small parameter
\begin{equation}
\mathcal{G}_{\nu,k}^{q,q}(t,t') = G_{\nu,k}^{q,q}(t,t') 
     - \frac{i \epsilon_{\nu,k}}{\omega_{\nu,k}}
     \int_{0}^{-\hbar \beta_{\nu} e^{2 r_{\nu,k}}} G_{\nu,k}^{+,M}(t,\zeta) G_{\nu,k}^{M,+}(\zeta,t')d \zeta.
\end{equation}
For $\lambda_{\nu,k}=0$ and $\hbar \rightarrow 0$, keeping the first order in $\hbar$ we have, from Eqs. \eqref{eq:4GFs2} and \eqref{eq:matsusem}, that $G_{\nu,k}^{q,q}(t,t') \approx \frac{i}{\hbar e^{2 r_{\nu,k}} \beta_{\nu} \omega_{\nu,k}} \cos 
 \omega_{\nu,k}(t-t')$ and $G_{\nu,k}^{+,M}(\zeta,t) \approx G_{\nu,k}^{M,+}(t,\zeta) \approx \frac{i }{\hbar e^{2 r_{\nu,k}} \beta_{\nu} \omega_{\nu,k}} \cos  \omega_{\nu,k}(\zeta-t)$ so that we obtain
\begin{equation}
    G^{q,q}_{\nu,k}(z,z') = 
    \frac{i}{ \hbar e^{2 r_{\nu,k}}\beta_{\nu} \omega_{\nu,k}} \cos \omega_{\nu,k} (t-t') - \Big(\frac{\Omega_{\nu,k}^2}{\omega_{\nu,k}^2} -1 \Big) \frac{i }{\hbar e^{2 r_{\nu,k}}\beta_{\nu} \omega_{\nu,k} }  \cos \omega_{\nu,k} t \cos \omega_{\nu,k} t'. 
    \label{eq:appbalballa}
\end{equation}

After dividing by $\omega_{\nu,k} m_{\nu,k}$ as prescribed by Eq. \eqref{eq:dissipactsqueenew}, we obtain Eq. \eqref{eq:classsqueezmass}.
To prove that the two equations are matching, we use that in Eq. \eqref{eq:classsqueezmass} the prefactor of the second term is
\begin{equation}
\frac{M_{\nu,k}}{  \omega_{\nu,k}^2 m_{\nu,k}^2} n_{\nu,k} =
\frac{M_{\nu,k}}{  \omega_{\nu,k}^2 m_{\nu,k}^2} \big(\frac{m^2_{\nu,k}}{M^2_{\nu,k}} - 1 \big)=  
\frac{e^{-2 r_{\nu,k}}}{ \omega_{\nu,k}^2 m_{\nu,k}} \big(e^{4 r_{\nu,k}} - 1 \big), 
\end{equation}
while in the Eq. \eqref{eq:appbalballa} we have, after dividing by $m_{\nu,k} \omega_{\nu,k}$
\begin{equation} 
-\frac{1}{e^{2 r_{\nu,k}} m_{\nu,k} \omega^2_{\nu,k}}\Big(\frac{\Omega_{\nu,k}^2}{\omega_{\nu,k}^2}  -  1\Big) =
\frac{1}{e^{2 r_{\nu,k}} m_{\nu,k} \omega^2_{\nu,k}}\Big(1 - e^{-4 r_{\nu,k}} \Big) .
\end{equation}
The two quantities are equal at the first order in $\epsilon_{\nu,k}/\omega_{\nu,k}^2$.
This perturbative approach can also be used to study the matching between the $\lambda_{\nu}$ dependent terms in Eqs. \eqref{eq:finalclass} and \eqref{eq:bigactdispcontmain}. In such a case the integral in Eq. \eqref{eq:intmatsusque} has to be performed on the branch $\gamma^{\nu \star}_{\uparrow}$, that is obtained by dilatating the branch $\gamma_{\nu}^{\uparrow}$ by a factor $e^{2 r_{\nu,k}}$.

\section{Fluctuation theorem for squeezed and displaced reservoirs} 
\label{app:fluctsqdisp}
Let us consider the generating function of internal energy and heat for a system coupled with many squeezed and displaced harmonic modes.
Equations \eqref{eq:genmom} and \eqref{eq:Utilt} applied to the protocol described in Section \ref{sec:fluct} allow us to write the following compact relation for the MGF
\begin{align} \notag
M(&\lambda_S,\boldsymbol{\lambda_B},\tau,\boldsymbol{\alpha}, \boldsymbol{r}) = Tr[e^{\lambda_S \hat{H}_S + \boldsymbol{\lambda_B} \cdot \boldsymbol{\hat{H}_B}} \hat{U}(\tau,0) \\ \label{eq:genappf} & \times
\prod_{\nu=1}^N  \hat{D}^{\dagger}(\alpha_{\nu}) \hat{S}^{\dagger}(r_{\nu}) e^{-\lambda_S \hat{H}_S -\boldsymbol{\lambda_B} \cdot \boldsymbol{\hat{H}_B}} \hat{\rho}(0^-) 
\prod_{\nu=1}^N   \hat{S}(r_{\nu})  \hat{D}(\alpha_{\nu}) \hat{U}^{\dagger}(\tau,0) ],
\end{align}
where we made the dependence from the squeezing and displacement parameters explicit and used the shorthand notation $\hat{D}(\alpha_{\nu}) = \prod_k \hat{D}(\alpha_{\nu,k})$,
$\hat{S}(r_{\nu}) = \prod_k \hat{S}(r_{\nu,k})$.
As in Sec.~\ref{sec:fluct}, we have chosen $\tau_0 = 0^-$ to denote the initial time in which squeezing and displacement have not yet been performed.
In the time reversal protocol, any time-dependent external driving is inverted in time.
In our case, this corresponds to doing the following substitution
\begin{equation}
\hat{U}(\tau,0) 
\prod_{\nu=1}^N  \hat{D}^{\dagger}(\alpha_{\nu}) \hat{S}^{\dagger}(r_{\nu})  \rightarrow 
\prod_{\nu=1}^N  \hat{S}^{\dagger}(r_{\nu}) \hat{D}^{\dagger}(\alpha_{\nu})   \hat{U}_{rev}(\tau,0) 
\end{equation}
where $\hat{U}_{rev}(\tau,0)$ represents the unitary evolution in which any time dependence in the Hamiltonian has been reversed $\hat{H}_{rev}(s) = \hat{H}(\tau-s)$. The idea behind the transformation above is that the initial squeezing and displacement are part of the "evolution" of the system and reservoir, and as such, since the time evolution is inverted, their contribution now comes {\it after} the propagator $\hat{U}_{rev}(\tau,0)$. 
The time reversal operation is the combination of the replacement above with an initial time reversal applied to the quantum state, thus we finally obtain
\begin{align} \notag
M^R(& \lambda_S,\boldsymbol{\lambda_B},\tau,\boldsymbol{\alpha}, \boldsymbol{r})  = Tr[e^{\lambda_S \hat{H}_S + \boldsymbol{\lambda_B} \cdot \boldsymbol{\hat{H}_B}}\prod_{\nu=1}^N  \hat{S}^{\dagger}(r_{\nu}) \hat{D}^{\dagger}(\alpha_{\nu})  \\ & \times \hat{U}_{rev}(\tau,0)  e^{-\lambda_S \hat{H}_S - \boldsymbol{\lambda_B} \cdot \boldsymbol{\hat{H}_B}} \hat{\Theta} \hat{\rho}(0^-) \hat{\Theta}
\hat{U}^{\dagger}_{rev}(\tau,0) \prod_{\nu=1}^N     \hat{D}(\alpha_{\nu})\hat{S}(r_{\nu}) ],
\end{align}
where $\hat{\Theta}$ is the time reversal operator such that $\hat{\Theta} i  = - i \hat{\Theta}$ and $\hat{\Theta} \hat{\Theta} = \mathds{1}$.
To proceed we have to recall the properties of the displacement and squeezing under time reversal
\begin{equation}
\hat{\Theta} \hat{D}(\alpha_{\nu}) \hat{\Theta}  = \hat{D}(\alpha^*_{\nu}), \quad \quad \hat{\Theta} \hat{S}(r_{\nu}) \hat{\Theta} = \hat{S}(r_{\nu}^*),
\end{equation}
that together with $\hat{\Theta} \hat{U}_{rev}(\tau,0) \hat{\Theta} = \hat{U}^{\dag}(\tau,0)$ allow us to write 
\begin{align} \notag
M^R(&\lambda_S,\boldsymbol{\lambda_B},\tau,\boldsymbol{\alpha}, \boldsymbol{r})= Tr[e^{\lambda_S \hat{H}_S + \boldsymbol{\lambda_B} \cdot \boldsymbol{\hat{H}_B}}\prod_{\nu=1}^N  \hat{S}^{\dagger}(r^*_{\nu}) \hat{D}^{\dagger}(\alpha^*_{\nu})  \\ & \times \hat{U}^{\dagger}(\tau,0)  e^{-\lambda_S \hat{H}_S - \boldsymbol{\lambda_B} \cdot \boldsymbol{\hat{H}_B}} \hat{\rho}(0^-)
\hat{U}(\tau,0) \prod_{\nu=1}^N     \hat{D}(\alpha^*_{\nu})\hat{S}(r^*_{\nu}) ],
\end{align}
\newline
under the assumption that $\hat{H}_S$ is time reversal invariant.
To conclude the derivation of Eq. \eqref{eq:ftsqdis} we do the replacement $\lambda_{\nu} \rightarrow - \lambda_{\nu} - \beta_{\nu}$, $\lambda_S \rightarrow - \lambda_S- \beta_S$ and invert the signs of the real parts of $\alpha_{\nu}$ and $r_{\nu}$
\begin{align} \nonumber
M^R(-\lambda_S- \beta_S,-\boldsymbol{\lambda_B}- \boldsymbol{\beta_B},&\tau,-\boldsymbol{\alpha}^*, -\boldsymbol{r}^*)=  \frac{1}{Z_S(0)}Tr[e^{-(\lambda_S+ \beta_S) \hat{H}_S - (\boldsymbol{\lambda_B} + \boldsymbol{\beta_B}) \cdot \boldsymbol{\hat{H}_B}}\prod_{\nu=1}^N  \hat{S}(r_{\nu}) \hat{D}(\alpha_{\nu})   \hat{U}^{\dagger}(\tau,0)  \\ & \times e^{+(\lambda_S + \beta_S) \hat{H}_S + (\boldsymbol{\lambda_B} + \boldsymbol{\beta_B}) \cdot \boldsymbol{\hat{H}_B}} \hat{\rho}(0^-)
\hat{U}(\tau,0) \prod_{\nu=1}^N     \hat{D}^{\dagger}(\alpha_{\nu})\hat{S}^{\dagger}(r_{\nu}) ].
\end{align}
\newline
Using that the system and the reservoir are initially prepared in a Gibbs state and rearranging the terms, we obtain that the r.h.s of the equation above is equal to $1/Z_S(\tau)$ times the r.h.s. of \eqref{eq:genappf}, thus proving the FT in Eq.~\eqref{eq:ftsqdis}.
\end{widetext}
\end{appendix}
\par
\vfill
\null
\clearpage
\end{document}